\documentclass{article}
\usepackage{PRIMEarxiv}
\usepackage[utf8]{inputenc} 
\usepackage[T1]{fontenc}    
\usepackage{url}            
\usepackage{booktabs}       
\usepackage{amsfonts}       
\usepackage{nicefrac}       
\usepackage{fourier,amssymb,bbm,amsmath,gensymb}
\usepackage{microtype}      
\usepackage{lipsum}
\usepackage{fancyhdr}       
\usepackage{graphicx}       
\graphicspath{{media/}}     

\usepackage{booktabs}
\usepackage{textcomp,gensymb}
\usepackage[flushleft]{threeparttable} 
\usepackage{lscape}
\usepackage{multirow} 
\usepackage{graphicx} 
\usepackage{hyperref}
\hypersetup{
    colorlinks=true,
    linkcolor=blue,
    filecolor=magenta,      
    urlcolor=cyan}
\usepackage{natbib} 
\usepackage[nottoc]{tocbibind} 
\usepackage{tablefootnote}
\usepackage[table]{xcolor}

\usepackage{siunitx} 
\usepackage{chngcntr} 
\usepackage{copyrightbox} 
\usepackage{chngpage} 

\title{Heterogeneous earning responses to inheritance: new event-study evidence from Norway
}

\author{
  Xiaoguang Ling \\
  Department of Economics \\
  University of Oslo \\
  Oslo, Norway\\
  \texttt{lingxiaoguang@gmail.com} \\
}
\begin{document}

\maketitle

\begin{abstract}


It has long been assumed that inheritances, particularly large ones, have a negative effect on the labor supply of inheritors. Using Norwegian registry data, I examine the inheritance-induced decline in inheritors' wages and occupational income. In contrast to prior research, my estimates allow the dynamic effect of inheritances on labor supply to vary among inheritor cohorts. The estimation approach adopted and the 25-year long panel data make it possible to trace the dynamics of the effect for at least 20 years, which is twice as long as the study period in previous studies. Since all observations in the sample are inheritors, I avoid the selection problem arising in studies employing non-inheritors as controls. I find that large parental inheritances (more than one million Norwegian kroner) reduce annual wage and occupational income by, at most, 4.3 \%, which is about half the decrease previously identified. The magnitude of the effect increases with the size of the inheritance. Large inheritances also increase the probability of being self-employed by more than 1\%, although entrepreneurship may be dampened by inheritances that are excessively large. The inheritance effect lasts for up to 10 years and is heterogeneous across sexes and age groups. Male heirs are more likely to reduce their labor supply after receiving the transfer. Young heirs are more likely to be self-employed, and their annual occupational income is, therefore, less affected by inheritances in the long run; for the very young inheriting large amounts of wealth from their grandparents, the probability of their attaining a post-secondary education declines by 2\%.


\end{abstract}


\section{Introduction}
\label{sec:intro}
Inheritances have complex impacts on intergenerational mobility \citep{Bowles2002,Erikson2002,Kohli2004,Tomes1981}, tax revenue, and the labor choices of heirs. 
\footnote{Inheritance tax generates 2 billion Norwegian Kroner (about 300 million US dollar) in fiscal revenue in 2013, which accounts for around one percent of the total tax revenue. Inheritance tax was abolished in Norway in 2014 but the abolition is controversial \citep{dackling2020arveavgiften,pedersen2021arveavgift}}.
This paper focuses on the effect of inheritances on labor market performance. A sudden windfall of non-earned income will - as long as leisure is a normal good - induce recipients to reduce their own earnings, either by exiting the labor market altogether or by reducing the hours they work. Andrew Carnegie, the American industrialist and philanthropist, saw this as the downside of leaving a bequest to children \citep{carnegie2006gospel}; hence the common reference in the literature to the reduction of labor supply due to inheritances as the ``Carnegie effect''. This study investigates the effect's empirical importance by considering the impact of inheritances on earnings in Norway from 1993 to 2017.
\footnote{Working hour is a direct approximation of labor supply, however the quality of the Norwegian registration data for working hours prior to 2015 is insufficient. I therefore utilize earnings as the proxy for labor supply in this paper.}

A flaw in Carnegie's contention is that when there is no credit constraint and the size and timing of inheritances are fully foreseeable by a risk-neutral agent, that agent will smooth their labor supply and consumption across time. As such, inheritances should have no effect on earnings at the time that wealth is transferred. In reality, however, there are many factors that will prevent the perfect smoothing of labor supply. Children do not know with certainty how much or when they will inherit from their parents. Credit constraints and risk aversion may also alter an heir's behavior contrary to perfect smoothing \citep{doorley2016labour}. This research examines the effect of inheritances on earnings prior to and following the wealth transfer. An empirical investigation of the impact of inheritances on labor supply has important implications for wealth and estate taxes, intergenerational equality, and social welfare. The findings are likely of particular interest in developed countries in which there is a labor shortage. 

Despite the importance of this topic, there is limited and inconclusive evidence of the effect of inheritance on earning. Exploiting survey data detailing expected and unexpected inheritances, \cite{Joulfaian1994} finds that inheritances, whether expected or unexpected, have no significant effect on the labor supply of men and married women. Similarly, \cite{doorley2016labour} finds that inheritances have little effect on men's earnings and, especially when unexpected, can reduce the probability of female heirs working full-time by approximately 5\% and decrease their desired and actual working time by 1 to 2 hours per week. \cite{Brown2010} find that inheritances are also associated with a significant increase in the likelihood of retirement, particularly when they are unexpected. In these various studies, the unforeseen inheritances recorded in the micro-data function as exogenous shocks disentangling the effect of inheritances from unobservable confounders. As \cite{Boe2019} points out, however, because of the limited size of sample, the survey data may not adequately represent large inheritances, leading to inconclusive results.

\cite{Boe2019} makes use of tax registry data for the entire Norwegian population to comprehensively examine the effect of inheritances on earnings. They use the propensity score matching method to construct a counterfactual group of inheritors based on a rich set of observable characteristics for 3 years prior to receiving inheritances. The findings show that inheritors who received more than NOK 300,000 (at constant 2015 prices) earn, on average, approximately 7\% less in wage and occupational income (business and wage income) than the matched non-inheritors in the 5 years following the inheritance; for those who inherited less than NOK 300,000, the wage gap between the two groups is small and insignificant. In addition, the study reveals that the effect is heterogeneous across sex and age groups. It is stronger for women and individuals close to retirement age. Like other empirical work relying on the conditional independence assumption, the study assumes the same potential outcomes for matched inheritors and non-inheritors; this assumption might be invalid when selection to a group is based on unobservable characteristics, in which case individuals who never receive an inheritance may intrinsically differ from those who do. On the other hand, due to the length of the longitudinal study, individuals who did not inherit anything during the study period may have already received an inheritance prior to the study's start, rendering the control group inappropriate.

\cite{Elinder2012} circumvent the problem of selection on unobservable factors by focusing on the outcome for heirs who inherited in 2005, applying a static two-way-fixed-effect(TWFE) model to Swedish tax registry data. They find that receiving an inheritance decreases labor supply by around 8\%. For those aged between 50 and 59, inheritances reduce labor supply by as much as 11\%. Both static and dynamic TWFE models are utilized in the study by \cite{doorley2016labour} mentioned above. Although the TWFE model is often employed in event studies, the identification strategy may be invalid for so-called ``staggered'' treatment that affect the population in stages, making it 
difficult to find a proper control group. 
\footnote{The drawbacks of the traditional TWFE model are discussed in greater detail in Appendix Section \ref{asubsec:twfe}.}
Inheritances in my study are staggered, which means that the population receives inheritances cohorts by cohorts, and the entire population are treated in the end.

Another contribution made by this paper is its empirical method for analyzing labor effects of inheritances. I apply the ``staggered'' event study strategy established in recent years - that is, using only the not-yet-treated cohorts as the control group - thereby avoiding the danger present in the conventional TWFE model of an inappropriate control group when the treatment effect is heterogeneous across cohorts \citep{Callaway2020,de2020two,Sun2020}. Because the method allows the treatment effect of inheritances to vary across cohorts, I am able to disentangle treatment effect's dynamics from the effect due to cohort heterogeneity.
\footnote{\cite{Wooldridge2021} proposes a method employing a saturated (across all cohorts) TWFE model with the same intuition.} 
I find large bequests from parents have a negative effect on labor supply that is moderate (up to -4.3\%) but long-lasting (as long as 10 years) and much smaller than found in previous studies using Nordic administrative data. The effect is stronger among men than women, as suggested by the literature. 

Furthermore, large inheritances increase the probability of self-employment by around 1\%, but those that are excessively large inhibit entrepreneurship to some degree. Additionally, inheritances have a greater effect on youthful heirs; young heirs are more likely to have lower wage income and a higher likelihood of self-employment than their older counterparts. Very young inheritors who receiving large bequests from their grandparents exhibit an wage decrease of as much as 17\% and a 14\% increase in the probability of being self-employed. The effect of inheritances on the probability of self-employment that I identify is in line with the literature on liquidity constraint and entrepreneurship \citep{Earle2000,Fairlie2012,HoltzEakin1993,HoltzEakin1994,Hurst2004,Johansson2000,Laferrere2001,Lindh1996}. Interestingly, the probability of attaining upper secondary education decreases by 2\% for very young heirs.

My research is related to a body of literature using quasi-experiments, for example, stock or real-estate market fluctuations \citep{Coile2006,Disney2018,Li2020}, lottery win \citep{Golosov2021,Cesarini2017,imbens2001estimating} and policy reform \citep{Krueger1992} to study the effect of idiosyncratic income on labor supply. However, these experimental designs are not suitable for studying intergenerational wealth transfer through inheritances. Inheritances affect a larger proportion of the population than a lottery or stock market win, and, more importantly, inheriting after the death of a relative is not a self-selected behavior.


The remainder of the paper is structured as follows. Section \ref{sec:data} outlines the data and study sample. Section \ref{sec:model} explains the identification strategy and baseline model specification. The results and robustness check are contained in sections \ref{sec:results} and \ref{sec:robust}, respectively. In Section \ref{sec:hetero}, I explore the heterogeneity across sexes and age groups of the effect of inheritance on earnings; in this section, examining inheritances from grandparents affords me the opportunity to observe the earnings change of very young heirs. Section \ref{sec:coclude} concludes the study.
\section{Sample selection and data description}
\label{sec:data}
I utilize longitudinal data spanning the period 1993 to 2017 for the entire Norwegian population from the Norwegian taxation authority and Statistic Norway. Earnings. the main outcome variable in my study, is derived from tax records, which provide information about the source of the earnings, namely whether it is wage income or earnings from self-employment. It is interesting to identify whether inheritances affect the different types of earnings differently. I thus construct two earnings variables; (i) wage income and (ii) occupational income, which is the sum of wage and business (self-employment) income. I also define a binary indicator variable for those with non-zero business income as a proxy for self-employment. Note that being self-employed and having wage income are not mutually exclusive.

Inheritances, the treatment, are also derived from tax data; I observe when an individual receives a transfer from another individual. This transfer could be either an inheritance or a gift; the latter includes advances (gifts given during donors' lifetime, usually when the donors anticipate their own death). Gifts received from a person well in advance of their death can be endogenous since they and the recipient are able to plan the timing and extent of the gift. By contrast, inheritances triggered by death are arguably exogenous in nature because the timing is difficult to predict and because, by law, children of the deceased are entitled to at least two thirds of the inheritances.
\footnote{According to Norwegian inheritance law,
At least a minimal amount of the inheritances should be left to the surviving spouse. In 2015, the minimal amount is about 360 thousand NOK. The minimal amount is adjusted annually in accordance with the general wage increase. If the inheritances are less than or equal to the minimal amount, the surviving spouse becomes the sole heir. If the inheritances surpass the minimum amount, the alive spouse and the children of the deceased have the right to inherit a quarter and two thirds, respectively, of the inheritances. As a result of the law, the majority of children will receive inheritances at the death of the last remaining parent, unless they waive their inheritance rights. advances shortly before parental death are also highly prevalent, possibly because the parents are gravely ill and begin transferring their wealth before they die. See also (in Nowegian): \hyperlink{https://www.regjeringen.no/no/tema/lov-og-rett/innsikt/arv/arv-etter-foreldre/id2006132/}{https://www.regjeringen.no/no/tema/lov-og-rett/innsikt/arv/arv-etter-foreldre/id2006132/}.}  
In my study, I define transfers as inheritances from parents if they fall within a short interval right before and after the parent's death, which I also observe in my data.
\footnote{The impact of bequests from grandparents, which are defined as transfers that occur close to the death of a grandparent with no parental death during the same period, is examined in Section \ref{subsec:grand}. These heirs account for only 10 \% of the sample, and their characteristics differ from the baseline sample.}
In my baseline analysis, I select an interval of 7 years (the inheritance year, the 3 years prior, and the 3 years following the death of a parent) to include more observations.
\footnote{In the robustness check Section \ref{subsec:pdeath}, I further restrict the transfers to be either within 1 year or in the same year of parental death, which excludes advances. The estimates are mostly unchanged.} 


Based on previous findings, how inheritance affects labor supply is proportional to its size. In the analysis, I separate inheritances into four categories
by its size relative to the average annual wage income in Norway: designate the amount of an inheritance that happens in year $t$ as $I^t$, and the average annual wage income in Norway for the same year as $W^t$. $I^t$ belongs to one of the following four groups: \{I1: $I^t \in (0,\tfrac{1}{2}W^t]$\}, \{I2: $I^t \in (\tfrac{1}{2}W^t,W^t]$\}, \{I3: $I^t \in (W^t,2W^t]$\}, \{I4: $I^t \in (2W^t,\infty]$\}.
\footnote{The relative size of the inheritance to the mean national wage income functions similarly to using the national annual wage growth rate as a deflator. I use the relative size of inheritance because in my long study period between 1993 and 2017, the economic volume in Norway tripled and life quality improved dramatically. The same amount of purchasing power signifies different things over time. Comparing the size of an inheritance with the expense of maintaining an average living standard for a year, instead of only valuing purchasing power, is more likely to be how individuals perceive the magnitude of inheritances as well. The average annual wage in Norway was approximately 500,000 NOK (around 60,000 USD) in 2015. In practice, I also attempted to use the absolute size of an inheritance to classify it; however, when population size and death rate are considered, the number of large inheritance (CPI-deflated) beneficiaries in early cohorts is disproportionately smaller than in recent years. It indicates that it is easier to inherit the same amount (worth in purchasing power) today than it was in the past.} 
This study focuses on large inheritances those greater than the average national annual wage income (categories I3 and I4), since the effects of inheritances I1 and I2 are small and less significant (estimation results shown in Appendix Section \ref{asubsec:small}). More importantly, large transfers taking place close to the time of a parent's death are more likely to be inheritances or advances on inheritances that are directly induced by that death, as opposed to gifts obtained by the recipients for unknown reasons. This is depicted more clearly in Table \ref{tab:pct_pd_gpd}.

Table \ref{tab:pct_pd_gpd} displays the proportion of inheritances/gifts that fall within the time interval $[d-\Delta,d+\Delta]$, where $d$ is a year in which at least one parent dies, $\Delta \in \{0,1,3,4,7\}$. Note that a very small proportion (e.g. 0.75\% when $\Delta=3$) of the transfers occur closely to the death of both (at least) one parent and one grandparent.
\footnote{I include recipients whose parent and grandparent die in the same time interval because children are prior to grandchildren in the order of succession in Norway in the absence of a will. I thus assume the bequest from parents rather than grandparents. The effect of grandparents' bequest is discussed in Section \ref{subsec:grand}.} 
As evident in Table \ref{tab:pct_pd_gpd}, when the time interval is extended from the first to the fifth row of the table, more recipients are covered, but there is a diminishing marginal gain (numbers in parenthesis). In general, a large proportion of inheritances or gifts occur close to the time of parental death. A comparison of the columns in Table \ref{tab:pct_pd_gpd} indicates that, in a given interval, the larger the transfers, the more probable it is that they occurred close to the parent's death.
This is consistent with the intuition that inheritances and advances are usually substantial in size. According to Table \ref{tab:pct_pd_gpd}, 54\% and 65\% of the recipients receive categories I3 and I4 transfers within 3 years of their parents' death, separately, who constitute my baseline sample.



\begin{table}[htbp]
\centering
\scalebox{1}{
\begin{threeparttable}
\caption{Proportion of inheritances/gifts close to parental death (\%) \vspace{-10pt}}
\label{tab:pct_pd_gpd}
\begin{tabular}{l*{5}{c}}
\toprule
\multicolumn{1}{c}{$\Delta$} &All&I1&I2&I3&I4 \\
\midrule
0 year   & 22  & 16 & 20 & 28  & 37  \\
	     &(+0)   &(+0)  &(+0)  &(+0)   &(+0)   \\
1 year   & 36  & 26 & 34 & 45  & 57  \\
	     &(+14)   &(+10)  &(+14)  &(+17)   &(+20)   \\
3 years  & 45  & 34 & 43 & 54  & 65  \\
	     &(+9)   &(+8)  &(+9)  &(+9)   &(+8)   \\
5 years  & 50  & 41 & 49 & 59  & 69  \\
	     &(+5)   &(+7)  &(+6)  &(+5)   &(+4)   \\
7 years  & 55  & 45 & 53 & 63  & 72  \\
	     &(+5)   &(+4)  &(+4)  &(+4)   &(+3)   \\


\midrule
\multicolumn{1}{l}{no. of indiv.}  & 313,419& 124,551&80,459&63,872 &44,537 \\
\bottomrule
\end{tabular}
\begin{tablenotes}
      \footnotesize
      \item Note: (1) Column 1 is the one-side interval width $\Delta$. Transfers in each row fall in $[d-\Delta,d+\Delta]$, where $d$ is the year at least 1 parent/grandparent died; the second column is for all the recipients and columns 3-5 correspond to transfer size I1-I4. (2) The numbers in parentheses show how many more percentage transfers fall within the time interval as $\Delta$ increases relative to $\Delta=0$. (3) Less than 1.7\% of the transfers in Panel A are around the death of both parents and grandparents; while transfers in Panel B happened with no parental death in the same time interval. 
    \end{tablenotes}
\end{threeparttable}}
\end{table}
 

Implementing this event study requires that my baseline sample include only individuals whose age makes them eligible to remain in the labor market for the whole study period. Otherwise, I would, for example, wrongly compare inheritors of working age to those who have already retired or are yet to be born. I, therefore, restrict my sample to individuals aged 18 to 66 years old and alive for the entire study period 1993 to 2017 - that is, those born between 1951 and 1975.
\footnote{The normal retirement age in Norway is 67, however it is possible to draw retirement pension voluntarily at age 62 with a reduction in benefits. Earning reduction owing to early retirement is thus considered in my study. See also:  \hyperlink{https://www.nav.no/no/person/pensjon/alderspensjon/relatert-informasjon/kombinere-jobb-og-alderspensjon}{https://www.nav.no/no/person/pensjon/alderspensjon/relatert-informasjon/kombinere-jobb-og-alderspensjon}.}
\footnote{I eased the birth-year constraint (at the expense of panel length) to be $[1944,1976]$ and $[1948,1978]$ as a robustness check in Section \ref{subsec:birth}.}
My sample thus comprises panel data (25-year long) instead of pooled cross-section data. As a result of the constraint on birth year, the average age of inheritors (in the year of inheritances) varies over cohorts. If the
potential labor supply of heirs is affected differently depending on the life stage at which they inherit, the cohort-specific treatment effect is heterogeneous across cohorts, which is permissible in my research.  
\footnote{Age profiles of the individuals are also included in my regression to control for the effect of age and working experience on earnings, like in the classical Mincer equation \citep{mincer1974schooling}.}

\begin{figure}[htbp]
    \centering
    \copyrightbox[b]{\includegraphics[width=1\textwidth]{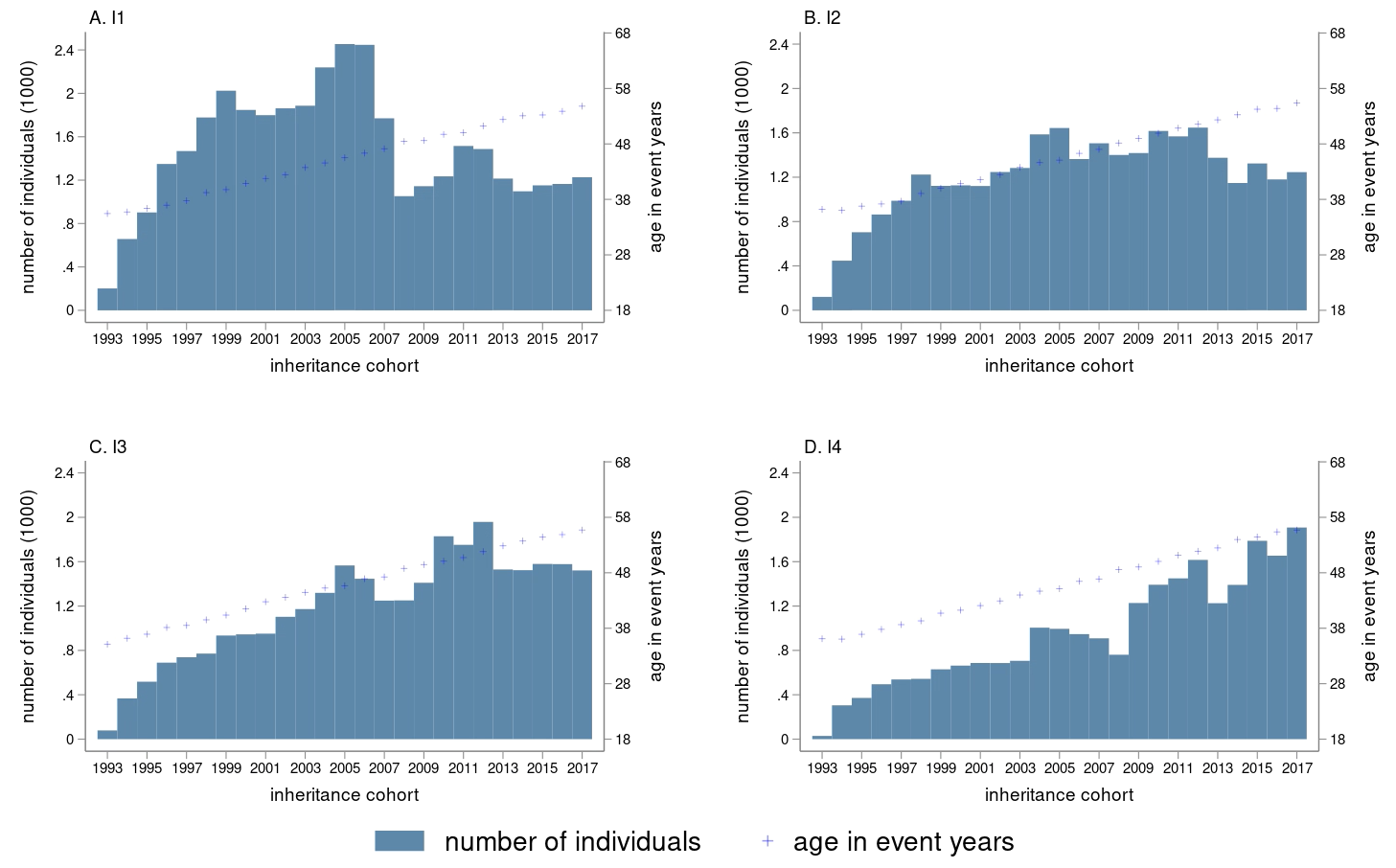}}{\scriptsize Note: $I1 \in (0,\tfrac{1}{2}W^t]$, $I2 \in (\tfrac{1}{2}W^t,W^t]$, $I_3 \in (W^t,2W^t]$, $I_4 \in (2W^t,\infty]$, $W^t$ is the mean annual wage in Norway in year $t$.}
    \caption{Sample size and average age of each cohort by size of transfers}
    \label{fig:bar}
\end{figure}

Figure \ref{fig:bar} depicts the number of individuals (cyan bars) and their average age (light-blue cross) in the year of transfer in each cohort by transfer size. A cohort is defined as individuals who receive transfers in the same year, and there are a total of 25 cohorts in my sample.
\footnote{Due to my identification strategy and assumptions introduced in Section \ref{sec:model}, cohorts before 1996 and after 2014 are excluded from my benchmark regression.} 
Part A of Figure \ref{fig:bar} indicates that, on average, recipients receive transfers that are less than half the average annual wage (category I1) when they are under 48 years old. Transfers exceeding half the average annual wage happen more frequently as beneficiaries age. This tendency is particularly obvious when transfers are larger than the average national annual wage in parts C and D of Figure \ref{fig:bar}.


Of the 1.42 million Norwegians born between 1951 and 1975, 22\% received an inheritance or gift once between 1993 and 2017, and 9\% were recipients on more than one occasion in the study period. For simplicity, I study only the ``oneoff'' inheritances in my baseline analysis and exclude individuals who inherited more than once over the study period.
\footnote{Individuals who receive small amount transfers more than once are examined in the robustness check Section \ref{subsec:recur}.}
Table \ref{tab:des_inheri} details the characteristics of the sample based on the aforementioned sample selection criteria by the size of the transfers. Panel A of Table \ref{tab:des_inheri} depicts the features of the sample in the event year. Column 2 sets out all inheritors and columns 3-6 detail recipients of transfers in categories I1 through I4. My baseline sample is provided in columns 5 and 6 and comprises 25-year longitudinal data on around 54,000 individuals. 
As seen in Table \ref{tab:des_inheri}, individuals receiving larger transfers are, on average, wealthier, have a higher income and level of education, and are older at time of transfers. Consistent with the average age, inheritors of larger bequest are more likely to be married and less likely to have living parents. Interestingly, the better-off have fewer children on average. Though there may be large wealth gaps between inheritors, the wage income distribution is relatively equal, which is typical in Nordic countries. In Panel B of Table \ref{tab:des_inheri}, I calculate the average income and wealth of the heirs during the whole study period $t\in[1993,2017]$. Since the average age in the year of receiving an inheritance is 47 in my baseline sample, and my study extends for 25 years, Panel B can be seen as an approximate of the annual average income and wealth in the middle-late stage of life cycle. As the inheritance size increases, the same pattern emerges in Panel B in terms of income and wealth. 

\begin{table}[htbp]
\centering
\caption{Baseline sample mean by inheritance size \label{tab:des_inheri}}  \vspace{-4pt}
\begin{threeparttable}
\begin{tabular}{l*{5}{c}}
\toprule
\textbf{Inheritance Size:} &\textbf{All}&\textbf{I1}&\textbf{I2}&\textbf{I3}&\textbf{I4} \\
\midrule
\multicolumn{6}{l}{\textbf{A.Sample mean in event year $E_i$}} \\
average cohort &         2007&    2005&    2006&    2008&    2009\\
inheritance amount&      568&      102&      272&      548&    1,691\\
wage income&            398&      348&      382&      416&      470\\
occupational income&     430&      379&      410&      447&      513\\
Pr(self-employment)&        0.14&         0.15&         0.13&       0.13&        0.14\\
capital income&           36&       17&       22&       36&       83\\
gross wealth&          1,146&      638&      817&    1,129&    2,369\\
gross debt  &      65 &      556&      608 &      687 &      829 \\
age         &             47&       45&       46&       48&       49\\
male        &              0.47&        0.47&        0.47&        0.47&        0.49\\
education level  &              4.24&        4.00&        4.12&        4.28&        4.69\\
married     &              0.78&        0.75&        0.77&        0.80&        0.80\\
number of kids&            0.99&        1.11&        0.99&        0.90&        0.91\\
paternal death&            0.79&        0.74&        0.77&        0.82&        0.86\\
maternal death&            0.62&        0.53&        0.58&        0.67&        0.75\\
\midrule
\multicolumn{6}{l}{\textbf{B.Sample mean in study period $t\in[1993,2017]$}} \\
wage income&        379&      343&      363&      385&      426\\
occupational income&  406&      374&      391&      414&      466 \\
entrepreneurship&   0.13 &  0.14    &  0.12    &    0.12 &  0.13    \\
capital income&       29&       18&       20&       27&       60\\
transfer    &         40&       42&       41&       39&       35\\
gross wealth&        836&      634&      676&      811&    1,384\\
gross debt  &        602&      549&      564&      608&      721\\
\midrule
\multicolumn{1}{l}{Number of individuals}& 120,952& 36,972&30,273&29,783 & 23,924 \\
\bottomrule
\end{tabular}
\begin{tablenotes}
      \small
      \item Note: (1) $I1 \in (0,\tfrac{1}{2}W^t]$, $I2 \in (\tfrac{1}{2}W^t,W^t]$, $I_3 \in (W^t,2W^t]$, $I_4 \in (2W^t,\infty]$, $W^t$ is the mean annual wage in Norway in year $t$. (2) Income and wealth are in 1,000 Norwegian kroner (NOK) at 2015 price. (3) ``occupational income'' = wage income + business income; entrepreneurship=1 if business income $\ne$ 0; ``gross wealth'' includes debt; ``education level'' is an ordered 0-8 categorical variable as defined by SSB: \href{https://www.ssb.no/klass/klassifikasjoner/36/}{https://www.ssb.no/klass/klassifikasjoner/36/}, e.g. education=4 for upper secondary education, education=5 for Post-secondary non-tertiary education; ``number of kids'' is the number of underage children; parental death indicators equal to 1 if the corresponding parents are dead.
    \end{tablenotes}
\end{threeparttable}
\end{table}
 

\section{Identification and model specification}
\label{sec:model}

An exogenous shock that affects part of the population will naturally create a treatment group and a control group. Under the (conditional) parallel trend assumption, canonical (fully dynamic) difference-in-differences (DiD) or, more generally, two-way fixed-effect (TWFE), models specify the dynamic average treatment effect ($ATE_t$) \citep{Angrist2008}. However, inheritances in the whole recipient population take place cohort by cohort until, eventually, every cohort is treated. Once transfers occur in a cohort, recipients are subject to a wealth shock that may affect their labor supply permanently. Given these features, inheritances are a form of ``staggered treatment'', as described in recent literature. When the treatment effect is heterogeneous across cohorts, the conventional TWFE estimators are biased because, intuitively, the control group includes cohorts of individuals that had already received an inheritance \citep{Borusyak2021,GoodmanBacon2021,Sun2020}.
\footnote{Further information regarding the bias of conventional TWFE estimators is introduced in Appendix part \ref{asubsec:twfe}.}


To avoid the pitfall of conventional TWFE, I employ the identification strategy proposed in the recent literature on staggered DiD (or TWFE) models, by \cite{Callaway2020}, \cite{de2020two} and \cite{Sun2020}. I first estimate each cohort-year-specific average treatment effect separately, using not-yet-treated cohorts as the control group, and then aggregate the effect by relative event time to study its dynamics.
Initially, I introduce the notations to explain the identification strategy for the cohort-year specific average treatment effect. Define inheritance status indicator $D_{it}$ and cohort indicator $G_{ig}$. $G_{ig}=1$ if $i$ inherits in year $g$. Assume one cannot anticipate inheritance precisely. In calendar year $t$, $i$ is treated as long as $t \ge g$: 

\begin{equation}
D_{it} =
\begin{cases}
1 \text{\ if \ } G_{ig}=1, t \ge g   \\
0 \text{\ if \ } G_{ig}=1, t \le g - 1 
\end{cases}
    \label{eq:D1} 
\end{equation}

Note that ``anticipation'' in this context refers exclusively to individuals behaving differently than they would have prior to receiving a bequest (potential outcome). When the precise timing and extent of a bequest are unknown, individuals can have a plan in place to smooth their labor supply since they have vague expectations. When they can precisely anticipate the bequest, they may adjust their plan accordingly. This adjustment is what I want to specify in this research.

In the baseline analysis, I assume individuals exhibit limited anticipation of their forthcoming inheritances, reflecting the fact that inheritances can, to some extent, be anticipate according to signs like parents becoming critically ill. If $i$ can anticipate forthcoming inheritances $\delta \ge 0$ years before it occurs, inheritance status $D_{it}$ becomes the following:

\begin{equation}
D_{it} =
\begin{cases}
1 \text{\ if \ } G_{ig}=1, t \ge g - \delta  \\
0 \text{\ if \ } G_{ig}=1, t \le g - \delta - 1 
\end{cases}
    \label{eq:D2} 
\end{equation}

How individuals anticipate inheritances determines when they are first exposed to the treatment and which years are pre-treatment, which is essential for my identification. By default, I assume $\delta=2$, which implies one cannot anticipate inheritances precisely 3 years in advance.
\footnote{In the robustness Section \ref{subsec:pdeath} part I set $\delta$ to be 4 to examine if the default setting is appropriate.} 

Denote the cohort-time specific average treatment effect as $\tau_{gt}$, where $t$ is calendar year and $g$ is cohort. $\tau_{gt}$ can evolve along the calendar year $t$ and vary among inheritance cohorts $g$. The temporal variation of $\tau_{gt}$ depicts the dynamics of the average treatment effect, like what the traditional dynamic TWFE model estimates. Allowing for heterogeneity in the average treatment effect among cohorts $g$ is necessary because the cohorts' age composition differs, and individuals in each cohort receive inheritances at different stages of their lives, as stated in Section \ref{sec:data}. How an individual disposes of the idiosyncratic income may depend on their phase of life. For example, young heirs may still drop out of school and change their careers easily, whereas heirs who are about to retire may lack the opportunity to adjust. This heterogeneity will be neglected if the canonical DID/TWFE method were applied to the whole sample. The benefits of allowing heterogeneous effects are amplified when the study period is extended and the sample includes individuals at different phases of life.

Let $Y^D_{it}$ represent one's potential earnings given their inheritance status $D_{it}$. The cohort-year specific average treatment effect $\tau_{gt}$ can be expressed as $\tau_{gt}=E[Y^1_{t} - Y^0_{t}|G_{g}=1]$. For convenience, the subscript $i$ is omitted at the cohort level. Conditional on treatment cohort $G_{g}=1$ and only ``not-yet-treated'' cohorts are used as controls, the heterogeneous treatment effect, $\tau_{gt}$, can be appropriately estimated using a traditional $2\times 2$ DiD model:

\begin{equation}
Y_{gt} = \beta_{t} d_{t} + \beta_{g}  G_{g} + \beta_{gt} D_{gt}  + \gamma \cdot age_{t} + \epsilon_{t}
\label{eq:spec_base} 
\end{equation}

Where $Y_{gt}$ is the earning of any individual in cohort $g$ in year $t$.
\footnote{I don't apply the logarithmic transformation on the outcome variable because a considerable proportion of observations leave the labor market after the transfers happen and the earnings, therefore, become zero, and $Pr(Y=0)$ could be an important response margin.} 
$d_t =\mathbbm{1}\{t \ge g-\delta\}$ is a binary variable indicating whether year $t$ is before or after the treatment, $G_{g}$ is the cohort indicator, $D_{t}$ is the treatment status indicator as defined in equation \ref{eq:D2}, and $D_{gt} = d_{t} \cdot G_{g}$. The coefficient on the interaction term, $\beta_{gt}$, is the estimator for the average treatment effect $\tau_{gt}$. Covariate $age_{t}$ is a collection of binary variables representing the age profile. The same as the canonical DiD method, the model relies on conditional parallel trend assumption:

\begin{equation}
E[Y^0_{g,t} - Y^0_{g,g-3} |age_t, G_{g} = 1] = E[Y^0_{g',t} - Y^0_{g',g-3} |age_t, D_{t} = 0]
\label{eq:para} 
\end{equation}

Where $Y^0$ is the potential earnings if individual do not receive inheritances, $g'$ represents the not-yet-treated cohorts, and $t=g-3$ is 3 years ``before the treatment'' used as the reference period.



Each cohort-year specific estimation has limited power for inference since the size of each cohort is relatively small. The heterogeneity among cohorts also makes it complex to explain $\beta_{gt}$ consistently. To study the dynamics of the treatment effect, I aggregate the cohort-year specific average treatment effect estimates $\beta_{gt}$ by time. Following the conventional event study method, define relative event time $s_{g} = t-g$. The cohort-calendar-year specific ATT estimates $\beta_{gt}$ can firstly be rewritten into cohort-relative-event-time specific estimates $\beta_{gs}$ and then be aggregated into relative time specific estimates $\beta_{s}$ using the following weighting scheme:

\begin{equation}
\beta_s = \sum^{G_s}_{g}  \frac{N_{gs}}{N_s}  \beta_{gs}
\label{eq:agg} 
\end{equation}

Where $N_{gs}$ represents the numbers of individuals of cohort $g$ in year $t$ such that $s=t-g$. Since the panel data in my study is strongly balanced, $N_{gs}=N_{g}, \forall s$. The denominator $N_{s}$ represents the number of individuals in all cohorts who are in the same relative event time $s$. A concrete example of the aggregation procedure is illustrated in the following section. The inference for the aggregated estimates is based on bootstrap.
\footnote{A refined aggregation scheme taking cohort heterogeneity (age composition) in to account will be discussed in the robustness check Section \ref{subsec:balance}.}

\section{Results}
\label{sec:results}
In this section, I first I give an example of the cohort-year specific treatment effect using the estimator in equation \ref{eq:spec_base}. Thereafter I illustrate the aggregation method given by equation \ref{eq:agg}.
 

\begin{figure}[htbp]
    \centering
\includegraphics[width=1\textwidth]{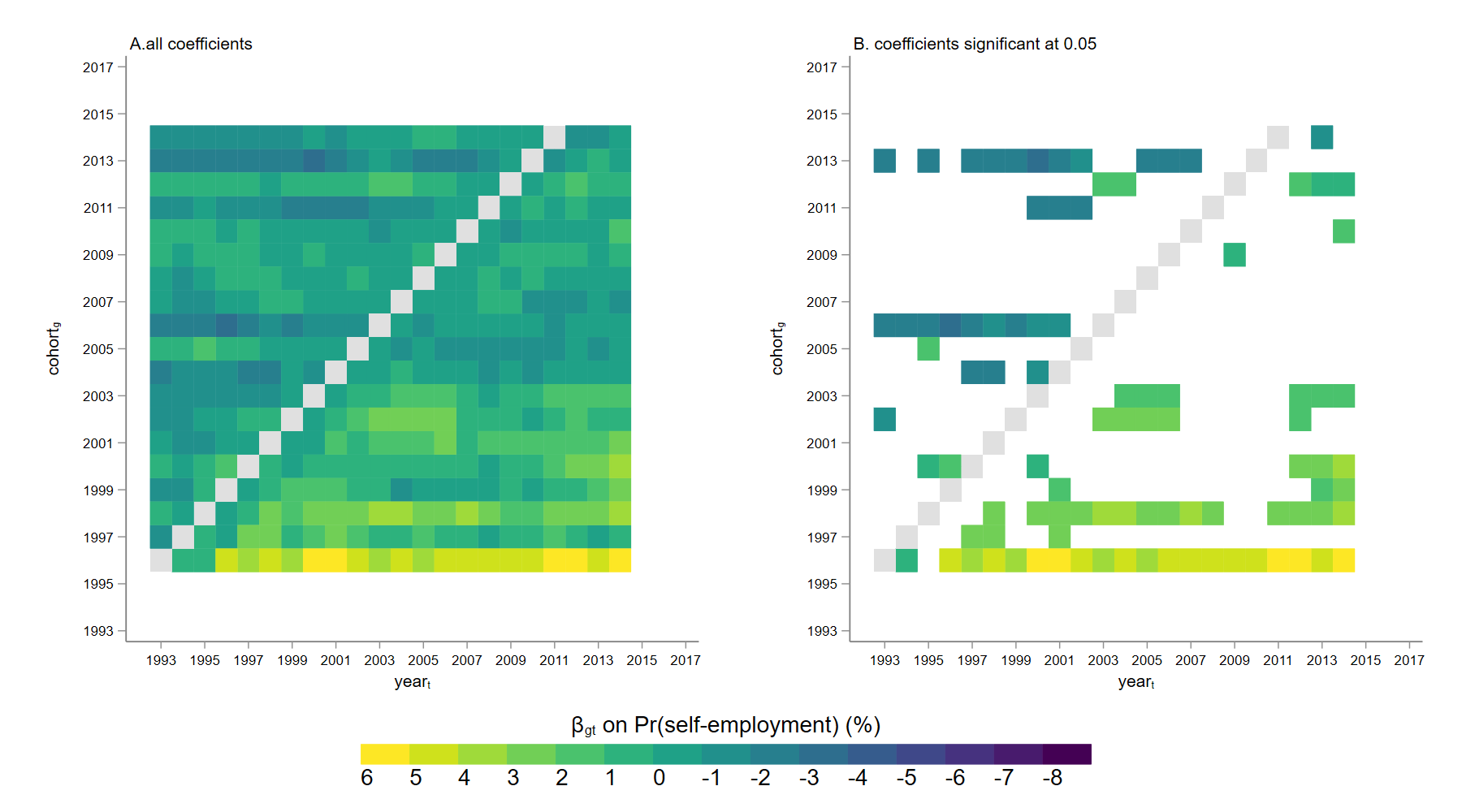}
    \caption{$\beta_{gt}$ of size $I3$ inheritances on Pr(self-employment)}
    \label{fig:heat_I3_attgt_ratio_p3_bq3_etpn}
\end{figure}
Figure \ref{fig:heat_I3_attgt_ratio_p3_bq3_etpn} is a pedagogical example of the regression result for heirs of inheritance size $I3$, i.e. $I_i \in$ \{I3: $I^t \in (W^t,2W^t)\}$ and the dependent variable is the probability of being self-employed, Pr(self-employment).
\footnote{I choose these estimates as an example because they show clear heterogeneity among cohorts. The other estimation results with annual wage and occupational income as outcome and with inheritance greater than $2W^t$ (I4) as treatment are in Appendix Section \ref{asubsec:heat}.} 
Every grid in Figure \ref{fig:heat_I3_attgt_ratio_p3_bq3_etpn} specified by cohort (y-axis) and calendar year (x-axis) in the heatmap corresponds to one cohort-year specific estimation, where cohort $g$ is the treated group, $t$ is the period of interest, year $g-3$ is the reference period, and cohorts that are not-yet-treated in year $t$ (and not-yet-treated in years before the reference year $g-3$) work as control group.
\footnote{The number of observations in each regression varies with the calendar year $t$ and the cohort $g$ because (i) the sample sizes varies by cohorts, and (ii) the composition of the control (not-yet-treated) group varies. As $g$ and $t$ increase, there are fewer not-yet-treated cohorts available as controls. This will be further discussed in Section \ref{subsec:near}.} 
As noted above, under the 2-year limited anticipation assumption, the 1993-1995 cohorts are not included as a treatment group since their reference periods, cohorts 1990-1992, are not observable. Similarly, the 2015-2017 cohorts are not included in the baseline regression because individuals in these cohorts may have inherited during the unobservable years 2018-2020, which, if included, would contaminate the not-yet-treated group. There are thus 418 cohort-year-specific estimates, $\beta_{gt}$, for the average treatment effect (19 cohorts $\times$ 22 calendar years), with the magnitude of effect represented by color intensity in the heatmap.

In part A of Figure \ref{fig:heat_I3_attgt_ratio_p3_bq3_etpn}, the lighter the color of the cell, the greater the average treatment effect estimates $\beta_{gt}$. The gray cells close to the diagonal of the heatmap matrix represent the reference year $t = g-3$ and are thus normalized as zero. To have an intuitive view of the estimates, I display only the results significant at the 0.05 level in portion B of Figure \ref{fig:heat_I3_attgt_ratio_p3_bq3_etpn}. Although we cannot claim joint significance here because the cohort-year specific treatment effect $\tau_{gt}$ are estimated separately, part B of Figure \ref{fig:heat_I3_attgt_ratio_p3_bq3_etpn} implies that the probability of being self-employed is likely to increase after the inheritance year, which suggests receiving a bequest increases the probability of being self-employed. Moreover, the positive effect of inheritances on Pr(self-employment) appears to be concentrated in the 1996-2003 cohorts, in which the heirs are relatively young when they receive their inheritances. This indicates that the treatment effect may be heterogeneous across cohorts and that younger heirs have a more pronounced response to receiving an inheritance.

\begin{figure}[htbp]
  \centering
  \copyrightbox[b]{ \includegraphics[width=1\textwidth]{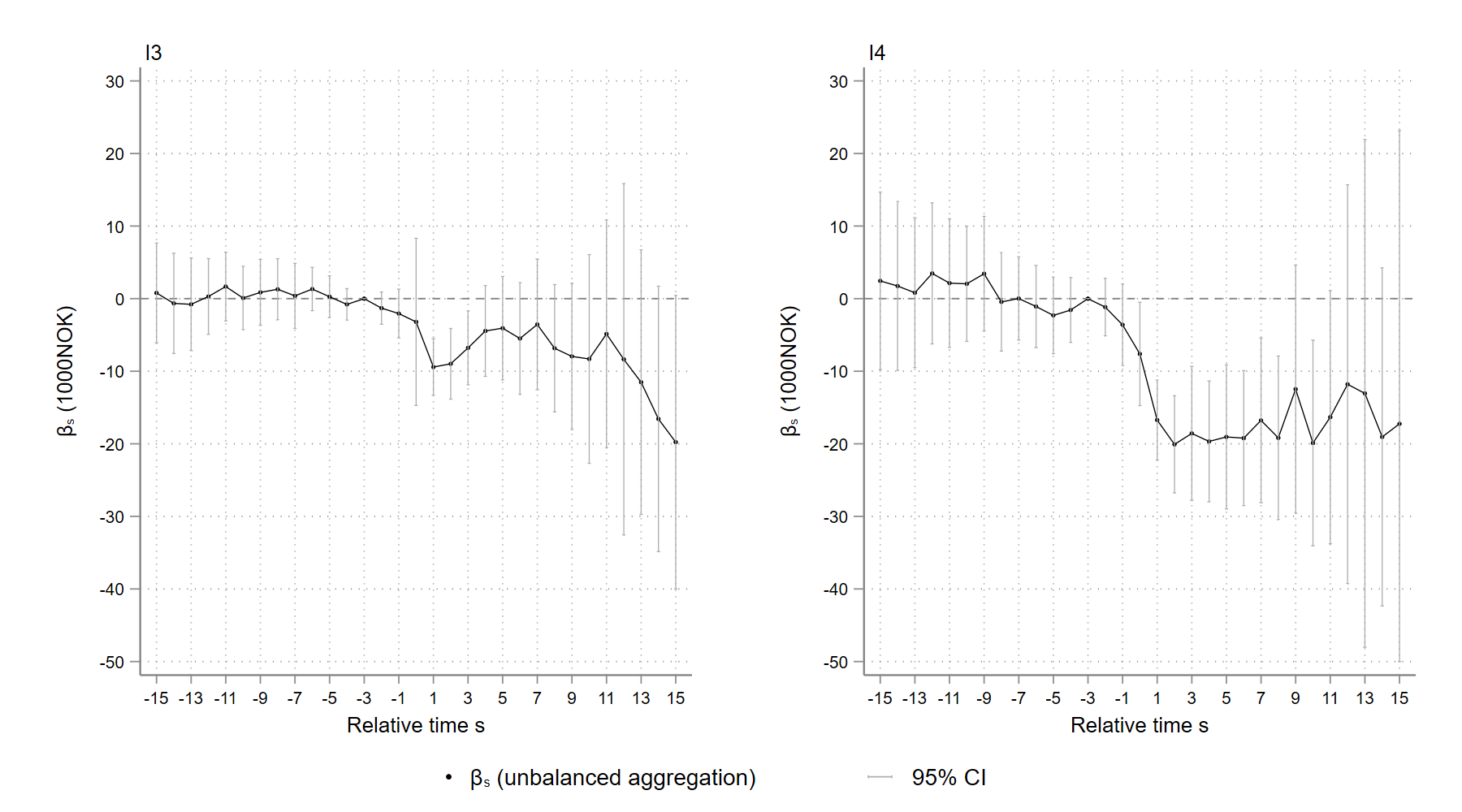}}{\scriptsize Note: $I_3 \in (W^t,2W^t]$, $I_4 \in (2W^t,\infty]$, $W^t$ is the mean annual wage in Norway in year $t$.}
  \caption{$\hat{ATT_s}$ on annual wage income}
  \label{fig:combine2_basic_wage_p3_bq3}
\end{figure}

\begin{figure}[htbp]
  \centering
  \copyrightbox[b]{ \includegraphics[width=1\textwidth]{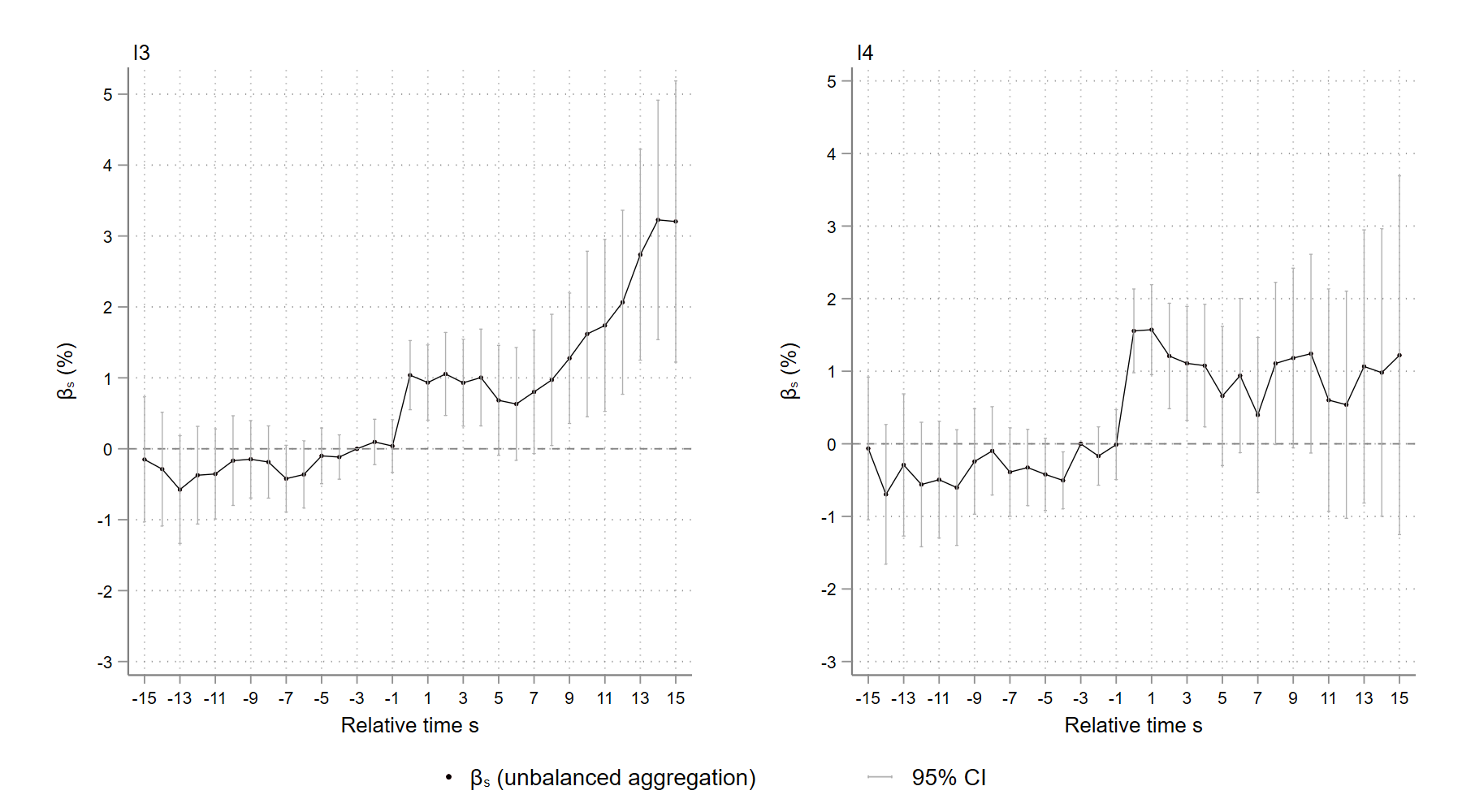}}{\scriptsize Note: $I_3 \in (W^t,2W^t]$, $I_4 \in (2W^t,\infty]$, $W^t$ is the mean annual wage in Norway in year $t$.}
  \caption{$\hat{ATT_s}$ on entrepreneurship}
  \label{fig:combine2_basic_etpn_p3_bq3}
\end{figure}

\begin{figure}[htbp]
  \centering
  \copyrightbox[b]{ \includegraphics[width=1\textwidth]{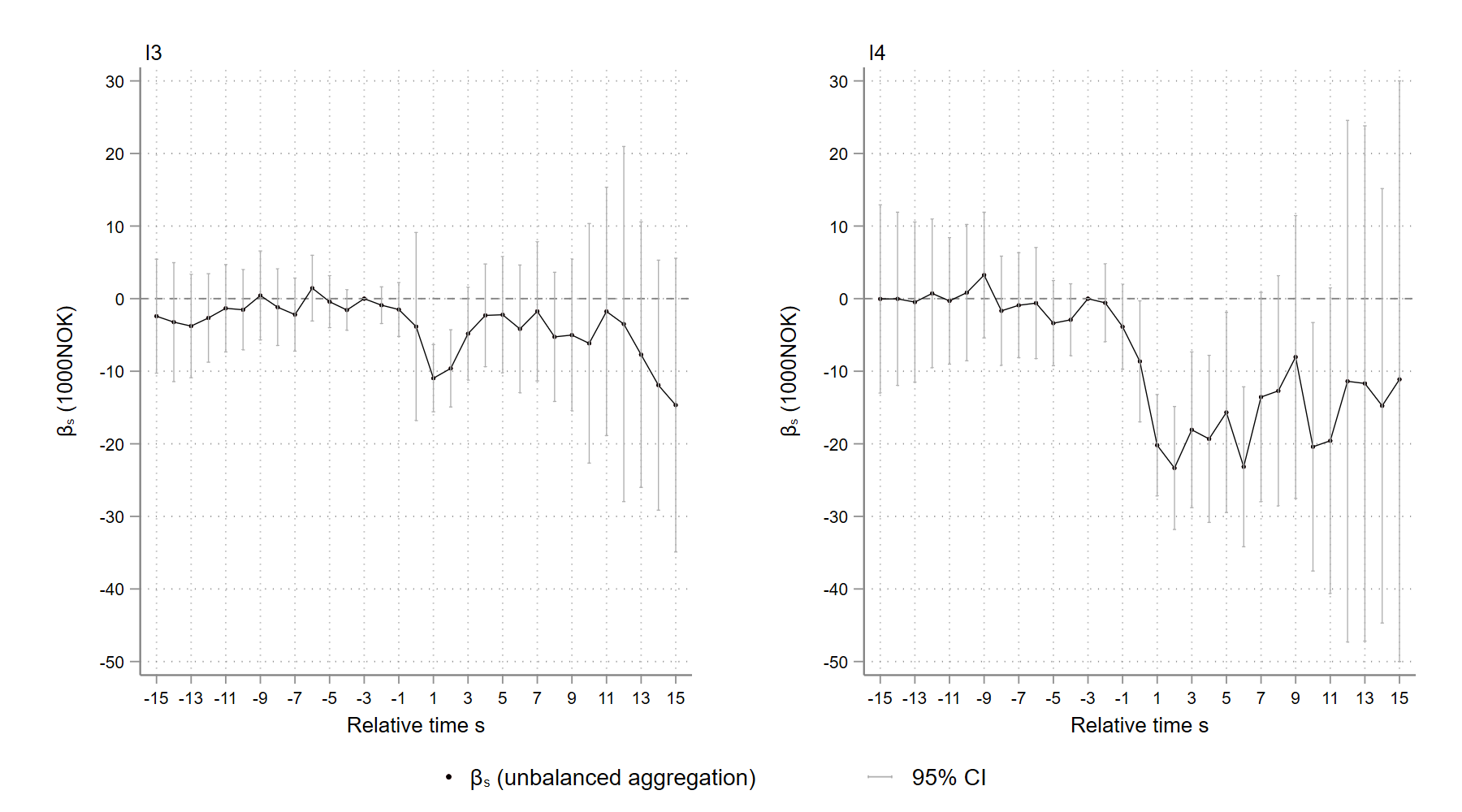}}{\scriptsize Note: $I_3 \in (W^t,2W^t]$, $I_4 \in (2W^t,\infty]$, $W^t$ is the mean annual wage in Norway in year $t$.}
  \caption{$\hat{ATT_s}$ on annual occupational income}
  \label{fig:combine2_basic_yrkinnt_p3_bq3}
\end{figure}

I then aggregate $\beta_{gs}$ ($s=g-t$) into relative-event-time specific ATT estimates $\beta_s$ according to weighting scheme \ref{eq:agg}. For instance, in Figure \ref{fig:heat_I3_attgt_ratio_p3_bq3_etpn}, cells with the same perpendicular distance to the grey cells (reference period) have the same relative event time and are thus aggregated into the same $\beta_s$. $\beta_s$ and bootstrap-derived confidence intervals at the significance level of 95\% constitute an event plot that resembles the conventional event study method. Figures \ref{fig:combine2_basic_wage_p3_bq3} - \ref{fig:combine2_basic_yrkinnt_p3_bq3} are event plots with annual wage income, probability of being self-employed, and annual occupational income as outcomes. Because there are fewer and fewer estimates $\beta_{gs}$ to be aggregated when $s$ increases, confidence intervals are too long to make the event plot readable when $s>15$. More importantly, the composition of the cohorts to be aggregated varies a lot when $s$ is too large, which will be discussed in detail in Section \ref{subsec:balance}. For these reasons, I only display the event plots with $|s| \le 15$. 

According to Figure \ref{fig:combine2_basic_wage_p3_bq3}, large bequests I3 and I4 decrease wage income by NOK 10,000 and 20,000 immediately after transfer, accounting for 2.5\% and 4.3\% of the average annual wage income of heirs in each group in the event year. The effect is about half that described in previous studies based on Nordic registry data \citep{Boe2019,Elinder2012}.
\footnote{In Appendix \ref{asubsec:boe}, I replicate the research of \cite{Boe2019} using my sample and obtain comparable results.} 
There are no estimates significant at the 0.05 level when $s<0$, indicating that there is no pre-trend; the the parallel trend assumption \ref{eq:para} appears to be valid.
\footnote{Although insignificant, the estimates seem to decline slightly right before the treatment ($s=0$). This is an indication of anticipation for the upcoming inheritances. In robustness check \ref{subsec:antici}, I will use $s=-5$ to be the reference year to further examine the limited anticipation assumption. Additionally, I attempt to use $s=-1$ as the reference year, but there is significant pretend (results not shown).}
The estimates in the left part of Figure \ref{fig:combine2_basic_etpn_p3_bq3} decrease dramatically after $s>10$. This is because heirs in cohorts that can be tracked for more than 10 years after the transfer (cohorts 1996-2004) are relatively younger when they receive the transfer, and the effect of inheritances on the probability of self-employment is heterogeneous across age groups, as will be discussed in Section \ref{subsec:balance} and Section \ref{subsec:hetero_age}. 

Interestingly, the effect appears to persist for up to 10 years, although the estimation gets more imprecise when $s$ increases. In addition, there appears to be a tiny (but insignificant) drop in annual wage income right prior to the inheritances happen ($s=-2$ and $-1$). This can be due to (i) a portion of the heirs anticipate the inheritance and start to adjust their behaviors, (ii) parental death itself reduces the labor supply of heirs since transfers within 3 years following the death are also counted as inheritances in the baseline analysis, and (iii) heirs may have to take care of their ill parents. Although parental death can be a confounder of inheritances, the comparison of the two graphs in Figure \ref{fig:combine2_basic_wage_p3_bq3} indicates that it cannot be the only reason for the drop in annual wage income: given that all the inheritors have lost their parents within 3 years of $s=0$, I4-sized inheritances have a greater impact than I3-sized inheritances. In an extreme case, assuming all the drop in wages on receiving an I3 inheritance (-2.5\%) is due to parental death, the effect of an I4 inheritance on annual wage income is still around -1.8\% (4.3-2.5=1.8). This confirms the Carnegie effect. 

Figure \ref{fig:combine2_basic_etpn_p3_bq3} demonstrates that the likelihood of being self-employed rises by at least 1\% following the receipt of a bequest. Again, the sudden increase of the estimates depicted in the left graph in Figure \ref{fig:combine2_basic_etpn_p3_bq3}, when $s>10$ is the result of cohort heterogeneity, as will be discussed in Section \ref{subsec:balance}. It is worth noting that inheritances between $W^t$ and $2W^t$ (I3) seem to have a stronger positive effect on the probability of self-employment than those exceeding $2W^t$ (I4), especially for relatively young heirs ($s>10$). A possible explanation is that although inheritances ease the liquidity constraints faced by relatively young heirs, thereby increasing their likelihood of self-employment, too great a wealth transfer may curb entrepreneurship. This can be understood as the Carnegie effect on entrepreneurship. 

Compensated by the income from self-employment, the probability of which is increased by the receipt of an inheritance, the effect of inheritances on annual occupational income is
slightly more moderate (-2.2\% and -3.9\% within I3 and I4 groups, respectively) than the effect on annual wage income, as shown in Figure \ref{fig:combine2_basic_yrkinnt_p3_bq3}. This is particularly obvious for heirs who receive I3 inheritances: the effect is no longer significant 3 years after the transfers happen.

To conclude, receiving large inheritances from parents reduces annual wage and occupational income. The negative effect on earnings might persist for up to 10 years, and the effect is proportional to inheritance size. However, the magnitude of the negative effect (2.5-4.3\%) is roughly half of that in the previous studies on Nordic countries (7\%-8\%). Given that the death of a parent itself might be a confounder of receiving inheritances, the magnitude of the Carnegie effect may be smaller (but the effect still exists). On the other hand, large inheritances increase the probability of self-employment by at least 1\%, but an excess of inheritances appears to dampen this positive effect. This is another evidence confirming the existence of the Carnegie effect on entrepreneurship. The baseline estimates are strongly affected by cohort heterogeneity as $s$ increases (and thus fewer treatment cohorts are included in the regression). The cohort heterogeneity will be further examined in the subsequent section.


\section{Robustness check}
\label{sec:robust}

In this section, I probe the stability of my baseline results by modifying some of the key assumptions of my empirical strategy. I first avoid including gifts from other sources as much as possible and study only the effect of parental inheritances by further restricting their timing relative to the parent's death. As mentioned previously, the aggregated estimates are affected by the heterogeneous cohorts. Estimates aggregated from different cohorts may not be comparable. Therefore, I refine the aggregation scheme as suggested in the literature on staggered DiD method \citep{Callaway2020} to make sure that all aggregated estimates are based on the same cohorts. 

In Section \ref{subsec:birth}, I ease the birth-year constraint to make the sample more comparable with those in the literature. In Section \ref{subsec:near}, I further take the composition and size of the control group in regression into account. To test the limited anticipation assumption, I next adjust the reference year to be 5 years before inheritance. I allow individuals to receive small gifts or inheritances more than once and, separately, exclude self-employed heirs, who may be able to underreport their wage income to avoid tax in the robustness checks \ref{subsec:recur} and \ref{subsec:exclude}.

\subsection{Timing of inheritance relative to parental death}
\label{subsec:pdeath}
The exogeneity of the inheritances comes from the fact that parental death times are hard to anticipate accurately, and the Norwegian inheritance law requires that the majority of the deceased's property be transferred to their children. Nevertheless, limited by the data, I cannot observe the exact source of the gifts or inheritances directly. Therefore, I infer that an inheritance or gift received close to a parent's death is more likely to be a bequest and hence more exogenous and unpredictable. That is to say, the exogeneity of the transfer is dependent upon its relationship to parental death. If the transfers are not the result of parental death, they can be endogenous. For example, parents can lend money to their children who lost their jobs. The drop in wage income after receiving the transfer specified is thus inverse causality.

\begin{figure}[htbp]
    \centering
    \copyrightbox[b]{\includegraphics[width=1\textwidth]{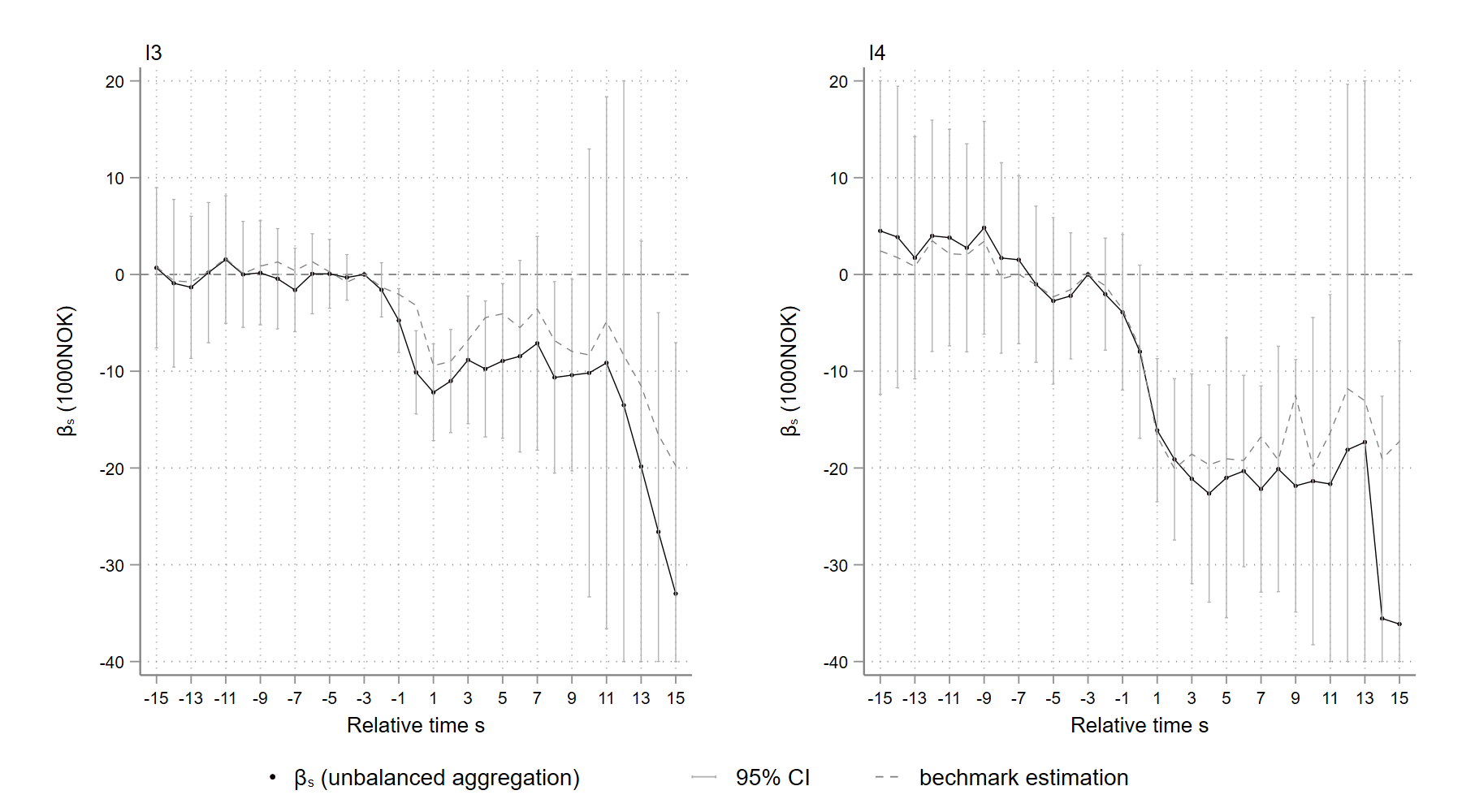}}{\scriptsize Note: $I_3 \in (W^t,2W^t]$, $I_4 \in (2W^t,\infty]$, $W^t$ is the mean annual wage in Norway in year $t$.}
    \caption{$\hat{ATT_s}$ on annual wage income: inheritances in the year of parental death}
    \label{fig:combine2_basic_wage_p3_bq0}
\end{figure}

In the benchmark analysis, inheritances or gifts realized within 3 years of a parent's death are treated as exogenous. This is a fairly broad time frame, and the gifts may, in part, come from resources other than advances of parental inheritances. I thus restrict inheritances to be transfers (i) within a year of parental death and, (ii) in the year of parental death (which excludes advances), separately. Figure \ref{fig:combine2_basic_wage_p3_bq0} depicts the effect of inheritances that happen in the parental death year on annual wage income. Because there are fewer observations, the estimates are less precise than those from the baseline. To increase the legibility of Figure \ref{fig:combine2_basic_wage_p3_bq0}, I trimmed the confidence intervals that are too long when $s>10$, which can be recognized easily from the figure.

It appears that the magnitude of the effect, in this case, is comparable to or somewhat greater than in the benchmark case. There may be a small number of endogenous gift recipients included in the benchmark regression, whose labor supply is not affected in the same manner as that of heirs. This could be attributable to the effect of parental death since this takes place in the same year as the inheritance. In any event, the estimates in Figure \ref{fig:combine2_basic_wage_p3_bq0} are very similar to the benchmark results. This implies that recipients who receive large gifts or inheritances in years close to parental death are comparable to those who receive gifts or inheritances in the year when their parents die. In other words, the gifts or inheritances defined in the baseline analysis are highly likely to be exogenous inheritances or advances on inheritances, rather than gifts given for other reasons.

The effect of inheritances on the other two outcomes, the probability of self-employment and annual occupational income, and the effect of inheritances received within a year around parental death are shown in Appendix Figures \ref{fig:combine2_basic_etpn_p3_bq0} - \ref{fig:combine2_basic_yrkinnt_p3_bq1}. The estimates in these figures are all similar to the baseline values.






\subsection{Balanced aggregation schemes}
\label{subsec:balance}
In the study period and under the 2-year limited anticipation assumption, the average treatment effect $\tau_s$ can be estimated by $\beta_s$ for all $s\in[-21,18]$ (only $s\in[-15,15]$ is shown in the baseline event plots). However, not all cohorts can be tracked for so many years. For example, in part A of Figure \ref{fig:heat_I3_attgt_ratio_p3_bq3_etpn} in Section \ref{sec:results}, the effect $\tau_{gs}$ are estimated by $\beta_{gs}$ for all the 19 cohorts when $s=-2$, as depicted by the grids next to the gray diagonal (reference time $s=-3$). However, when $s=18$, the effect can only be estimated for cohorts 1996, which is represented by the bright-yellow cell at the bottom right corner in Figure \ref{fig:heat_I3_attgt_ratio_p3_bq3_etpn}. As introduced in Section \ref{sec:data}, I restrict the individuals in my sample to those born between 1951 and 1975, which makes the age composition vary across cohorts. Consequently, when s>0, the greater $s$ is, the younger the heirs are on average when they receive inheritances in cohorts included in the regression. In the above example, $\beta_{-2}$ is the aggregate of $\beta_{g,-2}$, $g\in[1994,2014]$, which is estimated in heirs who are, on average, 48 years old in the inheritance year, whereas $\beta_{18} = \beta_{1994,18}$ and the average age of heirs in inheritance year in cohort 1994 is 36. The sudden drop in annual wage in the left graph in Figure \ref{fig:combine2_basic_wage_p3_bq3} and the dramatic increase in self-employment probability in Figure \ref{fig:combine2_basic_etpn_p3_bq3}, Section \ref{sec:results}, when $s>10$ is a result of the age composition of cohorts included in the regression. 

Formally, when aggregating the cohort-time specific treatment effect estimates $\beta_{gs}$ into time specific estimates $\beta_s$ based on the aggregation scheme \ref{eq:agg}, $\beta_{gs}$ estimated in all cohorts $g$ with the same relative event time $s$ ($s=g-t$) are aggregated into the same $\beta_s$. Denote the set of cohorts with the same relative event time $s$ as $G_s$. As $|s|$ increases, the size of set $G_s$ falls and the average age of inheritors in set $G_s$ gets younger. When the average treatment effect is heterogeneous among cohort $g$, varying composition of $G_s$ makes the aggregated $\beta_{s}$ incomparable between any two relative event time $s_1$ an $s_2$. In other words, the variance of the estimates $\beta_s$ is a consequence of both the dynamics of ATT $tau_s$ and the heterogeneity of ATT across cohorts. 

The challenge can be handled by choosing estimates $\beta_{gs}$ based on a fixed set of cohorts, $G^{\text{balance}}_{a,b}$, to aggregate, such that all the cohorts in set $G^{\text{balance}}_{a,b}$ can be observed $\forall s \in [a,b]$. This is called ``balanced aggregation scheme'' \citep{Callaway2020}. The balanced aggregation scheme can be written as equation \ref{eq:agg_balance},

\begin{equation}
\beta_s = \sum^{G^{\text{balance}}_{a,b}}_{g} \frac{N_{gs}}{N_s} \beta_{gs}
\label{eq:agg_balance} 
\end{equation}

Note that, with the exception of the composition of cohorts employed in aggregation, there is a trade-off between the length of the interval $[a,b]$, i.e. (balanced) horizon width, and estimation precision. The more expansive the horizon, the fewer cohorts can be observed within the entire interval $[a,b]$, and so the estimation is less precise. For example, in my study period between 1993 and 2017, if I set the balanced horizon width to be $[-10,10]$, under the 2-year anticipation assumption, $G^{\text{balance}}_{-10,10}=\{2003,2004\}$, while if the horizon is $[-5,10]$, cohorts 1998-2004 are in the balanced set $G^{\text{balance}}_{-5,10}$. In this robustness check section, I choose horizons $[-10,10]$, $[-5,10]$, $[-5,5]$ (cohorts 1998-2009 included), and $[-3,15]$ (cohorts 1996-2007 included) to refine the aggregation scheme and rule out the cohort heterogeneity effect (which will be further examined in Section \ref{subsec:hetero_age}) from the dynamics of estimated ATT. 


\begin{figure}[htbp]
  \centering
  \copyrightbox[b]{\includegraphics[width=1\textwidth]{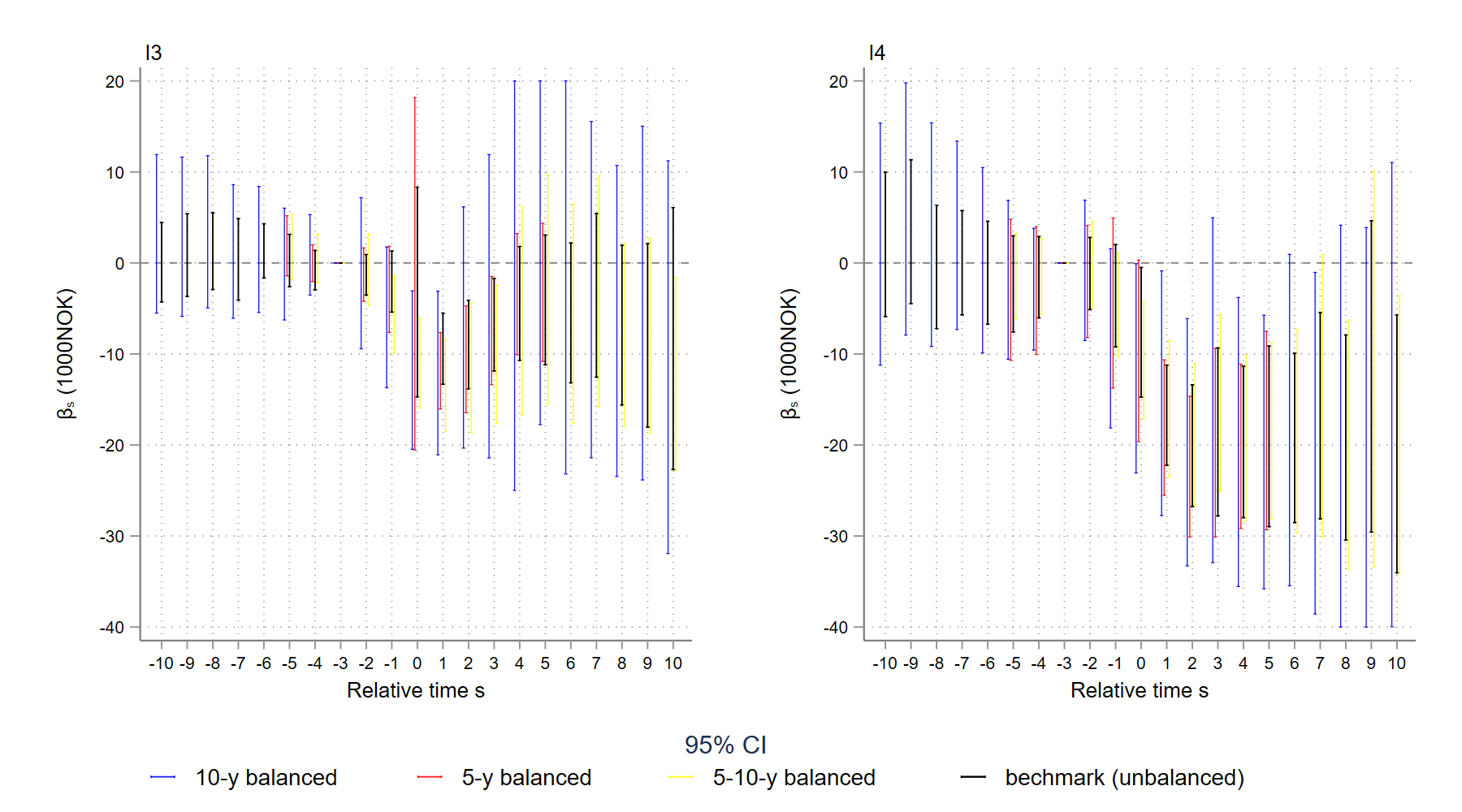}}{\scriptsize Note: $I_3 \in (W^t,2W^t]$, $I_4 \in (2W^t,\infty]$, $W^t$ is the mean annual wage in Norway in year $t$.}
  \caption{$\hat{ATT_s}$ on annual wage income with balanced aggregation horizons}
  \label{fig:combine2_balanced_wage_p3_bq3}
\end{figure}

The red, green and blue 95\% confidence interval bars in Figures \ref{fig:combine2_balanced_wage_p3_bq3} are based on three different balanced aggregation schemes $G^{\text{balance}}_{-5,5},G^{\text{balance}}_{-5,10}$ and $G^{\text{balance}}_{-10,10}$ separately with annual wage income as outcome. As discussed previously, confidence interval bars tend to be longer when the balanced horizon is wider because fewer cohorts are involved in the aggregation. The estimation is particularly imprecise with horizon $[-10,10]$ because only two cohorts are employed. The dynamics of the balanced aggregated estimates closely resemble the unbalanced aggregation results in the baseline analysis, so it is safe to say that, at least when $s<10$, the variation of the effect of inheritances on annual wage income in the baseline analysis is its own dynamics, instead of due to the heterogeneity of the effect across cohorts. The balanced aggregation results for effects on the probability of self-employment and annual occupational income are displayed in Appendix Figure \ref{fig:combine2_balanced_yrkinnt_p3_bq3} and \ref{fig:combine2_balanced_etpn_p3_bq3} and they are consistent with the baseline conclusion as well.

\begin{figure}[htbp]
  \centering
  \copyrightbox[b]{ \includegraphics[width=1\textwidth]{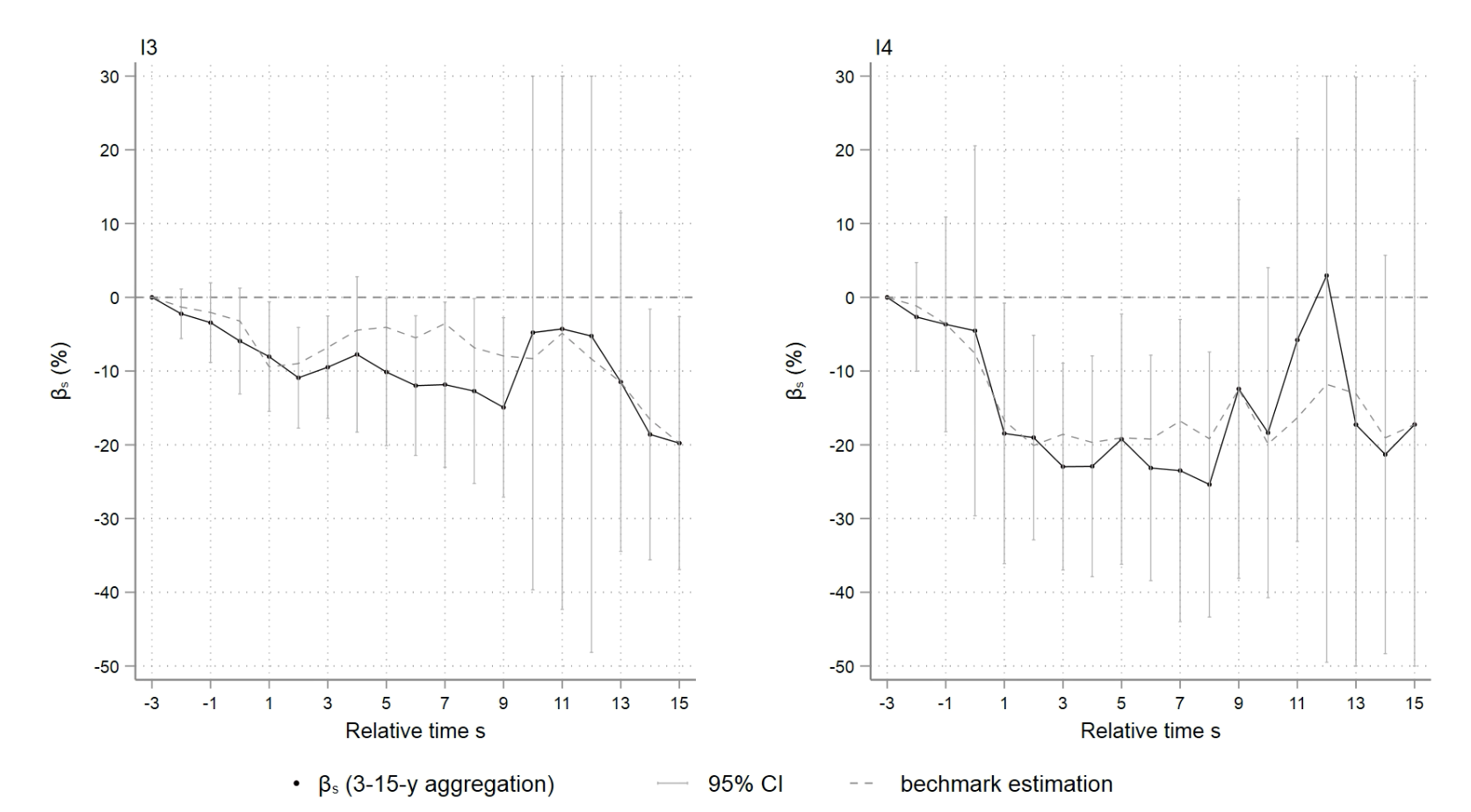}}{\scriptsize Note: $I_3 \in (W^t,2W^t]$, $I_4 \in (2W^t,\infty]$, $W^t$ is the mean annual wage in Norway in year $t$.}
  \caption{$\hat{ATT_s}$ on annual wage income: cohorts 1996-1999}
  \label{fig:combine2_basic_wage_p3_long_bq3}
\end{figure}
\begin{figure}[htbp]
  \centering
  \copyrightbox[b]{ \includegraphics[width=1\textwidth]{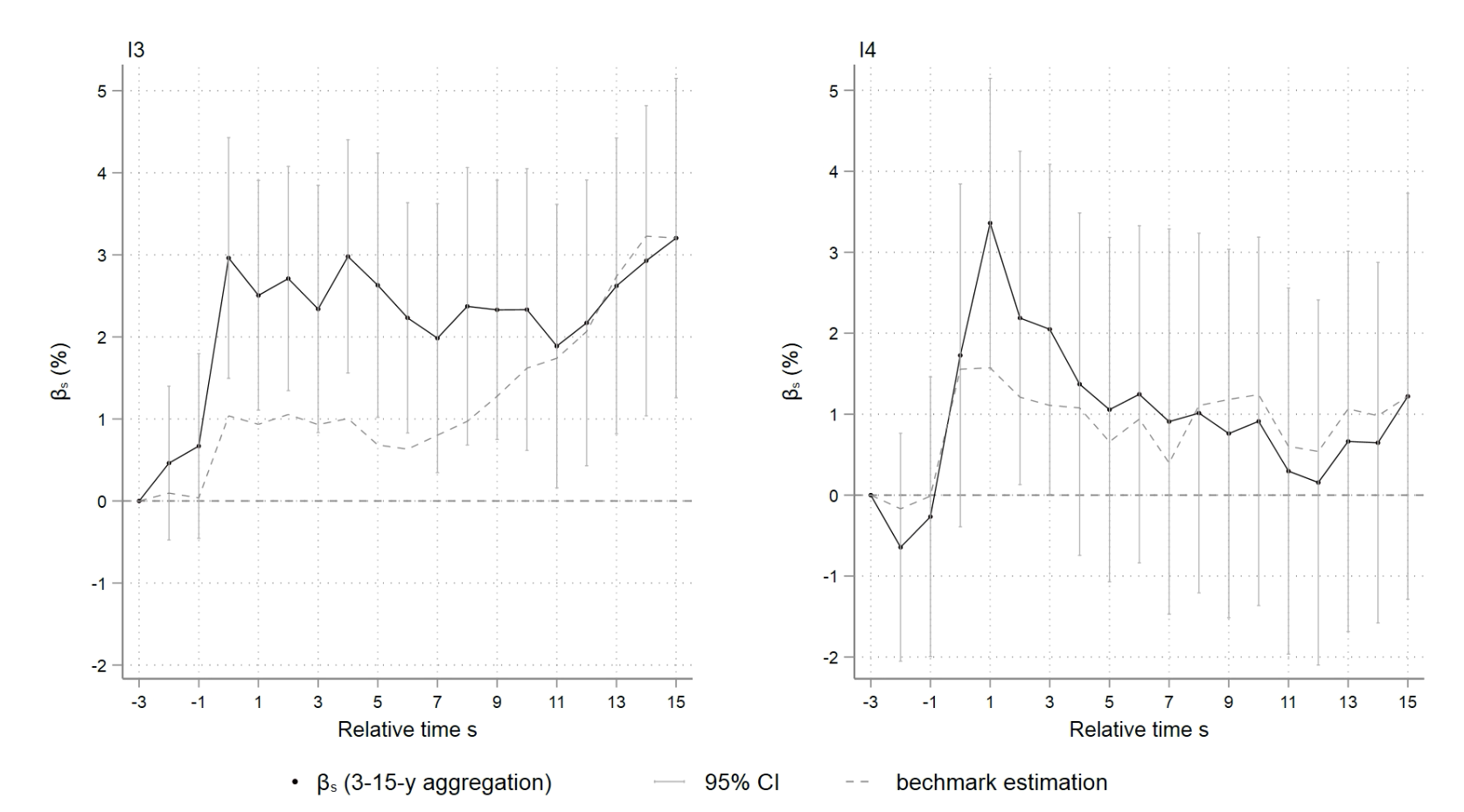}}{\scriptsize Note: $I_3 \in (W^t,2W^t]$, $I_4 \in (2W^t,\infty]$, $W^t$ is the mean annual wage in Norway in year $t$.}
  \caption{$\hat{ATT_s}$ on Pr(self-employment): cohorts 1996-1999}
  \label{fig:combine2_basic_etpn_p3_long_bq3}
\end{figure}

As noted, when $s>10$, the baseline estimates (Figures \ref{fig:combine2_basic_wage_p3_bq3}-\ref{fig:combine2_basic_etpn_p3_bq3}) in Section \ref{sec:results} are driven primarily by the age composition of the cohorts included in the aggregation. to track the dynamics of the effect over more than 10 years, I only include early cohorts, 1996-2004. Due to my sample's age restriction, people in these cohorts are relatively young in the year they receive their inheritances. The baseline results suggest that relatively young cohorts are more affected by inheritances. To exclude the effect of cohort heterogeneity and study the dynamics of the treatment effect on young heirs 
for more than 10 years, I adopt a balanced horizon $G^{\text{balance}}_{-3,15}$, which employs only cohorts 1996-1999, as depicted in Figures \ref{fig:combine2_basic_wage_p3_long_bq3} and \ref{fig:combine2_basic_etpn_p3_long_bq3}.
\footnote{The average age of the heirs who receive inheritances above national average annual income in cohorts 1996-1999 is 39. The annual wages of theses heirs in the year receiving I3- and I4-sized inheritances are 317 and 351 thousand NOK, separately.} 

As expected, the annual wage income and self-employment probability for relatively young heirs exhibit a greater response to I3-sized inheritances. The effect on the likelihood of self-employment is three times greater than in the baseline case. In contrast to the long-lasting impact of I3 inheritances, the right side of Figure \ref{fig:combine2_basic_etpn_p3_long_bq3} indicates that for inheritances more than twice the average national annual wage, relatively young heirs are more likely to be self-employed shortly after inheriting, but the effect is rather short-lived. This finding supports the hypothesis that substantial inheritances have a perverse effect on self-employment, easing heirs' liquidity constraints while simultaneously damping entrepreneurship in the long-run. The age heterogeneity of the effect is further examined in Section \ref{sec:hetero}.

\subsection{Birth year constraint}
\label{subsec:birth}

In the baseline analysis, I employ a 25-year long panel data set to trace the dynamics of the treatment effect. To ensure that all individuals are eligible to remain on the labor market participation for the entire study period, I limit the sample to those born between 1951 and 1975. Since the birth-year constraint unavoidably affects sample's age distribution, a source of cohort heterogeneity, the baseline estimates may be a direct result of the birth year constraint. I apply two eased birth-year constraints (at the expense of panel length) to determine whether the baseline results are sensitive to the birth-year constraints. One is applied in the literature, $[1944,1976]$ \citep{Boe2019}, and the other one, $[1948,1978]$, is a symmetrical extension of the baseline birth constraint by 6 years.

\begin{figure}[htbp]
  \centering
 \includegraphics[width=1\textwidth]{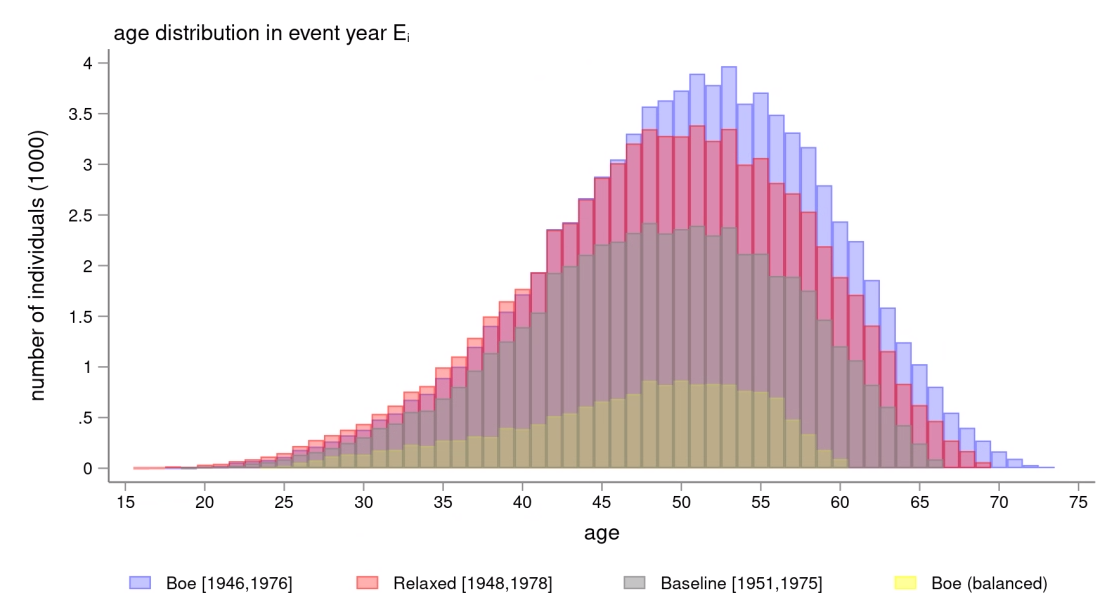}
  \caption{Age distribution of samples under different birth-year constraints}
  \label{fig:obs_birth_constraint_event_year_age}
\end{figure}

Figure \ref{fig:obs_birth_constraint_event_year_age} depicts the age distribution of the samples with varying birth year restrictions. The age distribution of individuals born between $[1944,1976]$ (light blue bars) is slightly more left-skewed, in comparison with the sample under baseline ($[1951,1975]$) and relaxed ($[1948,1978]$) birth-year constraints (light red and green bars, respectively). The (short) yellow bars in Figure \ref{fig:obs_birth_constraint_event_year_age} represent the age distribution of inheritors in the 2000-2004 cohorts based on the birth-year constraint $[1944,1976]$. These inheritors are the counterparts of the inheritors studied by \cite{Boe2019}.
\footnote{It is of particular interest to test whether the difference between my findings and those of the literature is a result of my birth-year constraint or my inheritance size criteria. In Appendix Table \ref{tab:match_boe}, I apply matching method on cohorts 2000-2004 under the birth-year constraint $[1944,1976]$ in my sample and obtain very similar estimates in comparison with \cite{Boe2019}, who studied the effect of inheritances on labor supply using Norwegian registry data.}

Appendix Table \ref{tab:des_boe} describes the aforementioned samples in greater detail. Appendix Figures \ref{fig:combine2_basic_wage_p3_44_76_bq3}-\ref{fig:combine2_basic_yrkinnt_p3_48_78_bq3} are event plots based on the samples restricted by birth years $[1944,1976]$ and $[1948,1978]$.
To provide a direct comparison with the literature, I further employ only the 2000-2004 cohorts and aggregate the cohort-year specific treatment effect using a balanced scheme $G^{\text{balance}}_{-6,6}$ in Appendix Figures \ref{fig:combine2_boe_wage_p3_44_76_bq3}-\ref{fig:combine2_boe_yrkinnt_p3_44_76_bq3}. These inheritor cohorts and the horizon ($[-6,6]$) are exactly the same as those employed by \cite{Boe2019}. The effect estimated in the preceding event plots is summarized in Table \ref{tab:summ}. 

According to Table \ref{tab:summ}, in samples with different birth-year constraints, the effect of inheritances varies, but does not diverge much from the baseline results. In addition, the effect magnitudes in Table \ref{tab:summ} are, for the most part, smaller than those found in the literature. The only estimates that are larger than the baseline are the balanced aggregated estimates with birth-year constraint $[1944,1976]$ (i.e., as per ``Boe (balanced)'', second row in Table \ref{tab:summ}). This is unsurprising since the average age of these cohorts is slightly younger than my baseline sample, as shown in Table \ref{tab:des_boe}, and the response to an inheritance of young heirs is greater than old heirs.


\begin{table}[htbp]
\centering
\def\sym#1{\ifmmode^{#1}\else\(^{#1}\)\fi}
\caption{Summary of the estimates under different birth year constraints}
\label{tab:summ}
\vspace{-5pt}

\begin{threeparttable}
\begin{tabular}{l*{8}{c}}
\toprule
Inheritances &      &\multicolumn{3}{c}{$I_3 \in (W^t,2W^t]$}&&\multicolumn{3}{c}{$I_4 \in (2W^t,\infty]$} \\
   				             \cmidrule{3-5}    \cmidrule{7-9}    
             &      & mean &$\hat{\beta}$& \%   &&    mean  &$\hat{\beta}$&   \%   \\
\midrule          
\multirow{2}{*}{baseline}  & wage  &  416 & -10  & -2.4  &&    470  &   -20  &  -4.3  \\
                           & ocpt  &  447 & -10  & -2.2  &&    513  &   -20  &  -3.9  \\
Boe    & wage  &  349 & -10  &-2.9   &&   378   &  -20   & -5.3  \\
 (balanced)                & ocpt  &  379 & -10  &-2.6   &&   414   &  -15   & -3.6  \\
Boe   & wage  &  387 & -5   &-1.3   &&   432   &  -10   & -2.3  \\
 (unbalanced)  			   & ocpt  &  416 & -10  & -2.4  &&    471  &   -10  &  -2.1  \\
\multirow{2}{*}{relaxed}   & wage  &  396 & -8   &-2.0   &&   443   &  -18   & -4.1  \\
                           & ocpt  &  425 & -10  & -2.4  &&    483  &   -18  &  -3.7  \\
\bottomrule
\end{tabular}
\begin{tablenotes}
      \small
      \item Note: (1) Baseline sample contains inheritors in cohorts $[1996,2014]$ and born between 1951 and 1975; sample ``Boe (balanced)'' are cohorts $[2000-2004]$, born in $[1944,1976]$; sample ``Boe (unbalanced)'' are cohorts $[1997-2010]$, born in $[1944,1976]$; sample ``relaxed'' are cohorts $[1999,2014]$, born in $[1948,1978]$ (2) ``wage'' and ``ocpt'' are average annual wage and occupational income in the year receiving inheritances, where occupational income = wage income + business income. (3) $W^t$ is the average annual wage in Norway in year $t$. (4) Income in 1,000 Norwegian kroner (NOK) at 2015 price.
    \end{tablenotes}
\end{threeparttable}
\end{table}


It is worth mentioning that the dynamics of the estimates are far less affected by cohort heterogeneity as $s$ increases than are the baseline estimates (See, for example, Appendix Figures \ref{fig:combine2_basic_etpn_p3_44_76_bq3} and \ref{fig:combine2_basic_etpn_p3_48_78_bq3}). This is because individuals in cohorts upon which the unbalanced aggregation is based when $s>10$ are, on average, older than those in the baseline estimation. For instance, under the birth-year constraint $[1944,1976]$, only individuals in cohort 1994 can be tracked for 14 years, and their average age in 1994 is 40. In contrast, the average age of the 1994 cohort in the baseline sample is 36. This age difference within the same cohort under different birth-year constraints also illuminates the source of the effect's heterogeneity. Individuals in different cohorts are exposed to inheritances across life stages while also being subject to varying macroeconomic environments. The boom and bust of business cycles may determine how individuals dispose of inheritances and the impact on their labor participation. For example, the probability of self-employment may be affected by business cycles and will be a source of cohort heterogeneity aside from age \citep{brunjes2013recession,congregado2012recession,meager1992does}. This will be a source of the cohort heterogeneity besides age composition. 
\footnote{This concern is mitigated in part by the fact that my study period is so long that it encompasses several business cycles, while the cohort heterogeneity appears unaffected.}
However, under different birth-year constraints, the effect estimated in the same cohort - hence with the same macroeconomic conditions - depends substantially on the age composition within the cohort. I, therefore, conclude that most of the effect heterogeneity can be traced to cohort age composition rather than simply macroeconomic conditions.

\subsection{Nearest-n-not-yet-treated cohorts as control}
\label{subsec:near}
\begin{figure}[htbp]
  \includegraphics[width=1\textwidth]{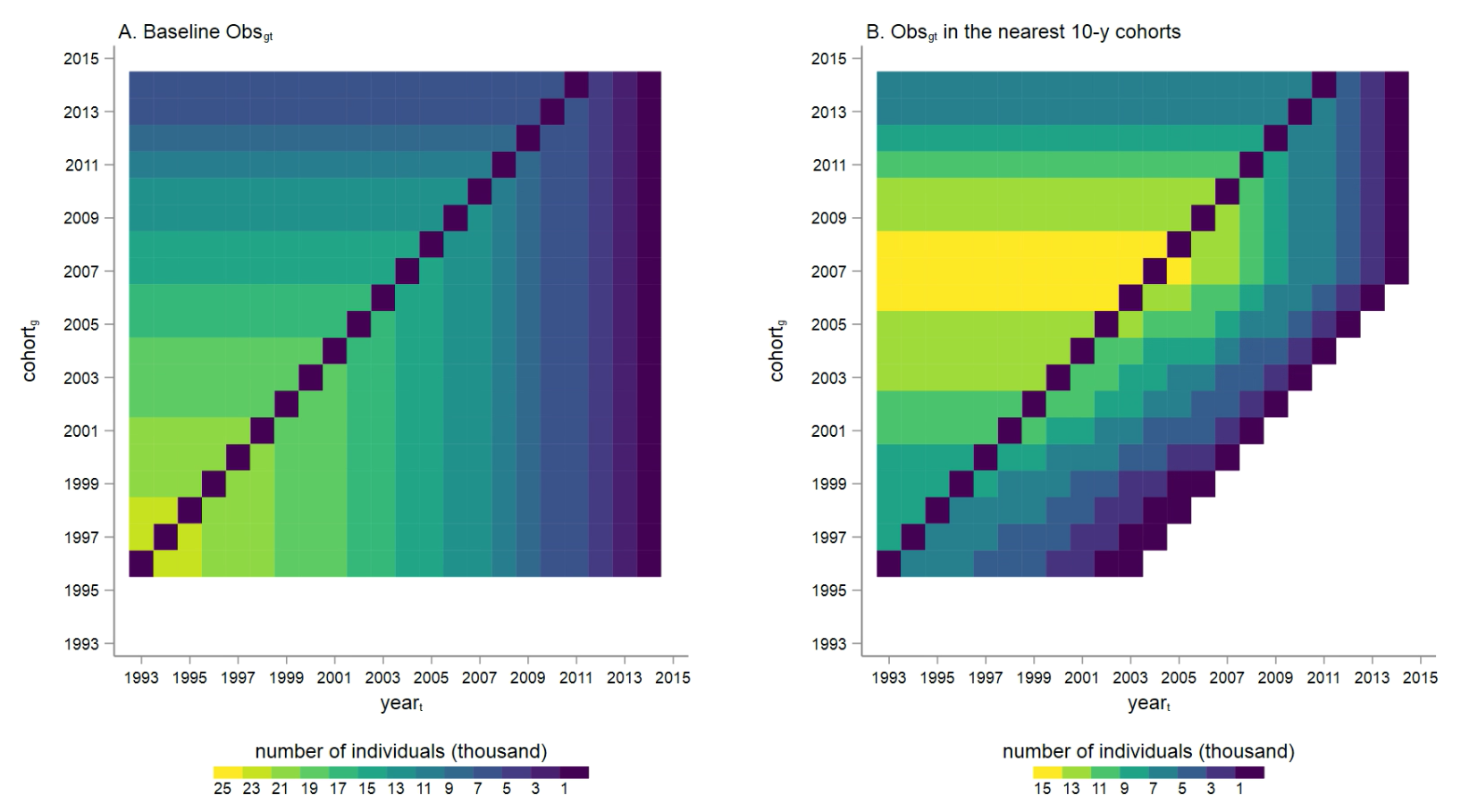}
  \caption{$\hat{ATT_s}$ using the nearest 10 cohorts as control group}
  \label{fig:heat_obs_ratio_bq3}
\end{figure}

The composition of the ``not-yet-treated'' cohorts employed as controls also varies by treatment cohort $g$ and calendar year $t$. As a result, these not-yet-treated cohorts may not constitute an appropriate counterfactual. This issue concerning the identification strategy is not discussed in the staggered DiD literature. The color intensity in the left portion of Figure \ref{fig:heat_obs_ratio_bq3} indicates the number of individuals in the not-yet-treated cohort given $g$ and $t$. Lighter colors correspond with a larger sample size. The dark cells near to the diagonal represent the omitted reference years, i.e. $t=g-3$. When $g$ and $t$ are close to 1996, the beginning of the study period, most of the other cohorts have not yet been treated, making the control group large. However, when the treatment cohort $g$ is close to $1996$, the control group includes cohorts both close to and far from $g$, such as cohort $2014$. In contrast, as $g$ increases, the size of the control (not-yet-treated) group shrinks, and the cohorts in the control group become closer to $g$.


If the not-yet-treated cohorts $g'$ have divergent time trends when $g' \gg g$, the parallel time trend assumption \ref{eq:para} is violated. Consequently, the estimates of $\tau_{gt}$ are biased when $g$ is close to 1996. To address this issue, I relax the conditional parallel trend assumption \ref{eq:para} by requiring that, at the very least, the conditional parallel trend holds when control group $g'$ is inside a small neighborhood of the treatment cohort $g$, i.e. $g' \in [g,g+n]$. Formally speaking,

\begin{equation}
E[Y^0_{g,t} - Y^0_{g,g-3} |X, G_{g} = 1] = E[Y^0_{g',t} - Y^0_{g',g-3} |X, D_{t} = 0, g'\in [g,g+n]]
\label{eq:para_nn} 
\end{equation}

The right side of Figure \ref{fig:heat_obs_ratio_bq3} depicts the size of the control group consisting only the nearest-$10$ not-yet-treated cohorts ($g' \in [g,g+10]$) for any given treated cohort $g$. The treatment in Figure \ref{fig:heat_obs_ratio_bq3} is inheritances in category I4 that are more than twice the average annual wage. The control groups are smaller than in the baseline case, in the left side of Figure \ref{fig:heat_obs_ratio_bq3}, because each control group contains at most 10 not-yet-treated cohorts. In addition, at every given relative event time $s$, the size of the control group for treatment groups, $g$, resemble each other more than they do in the baseline analysis because of the constraint that the control group falls within the interval $[g,g+10]$.
\footnote{There appears to be more observations in control groups for treatment cohorts 2006-2008. This is because in my sample, older people are more likely to receive large inheritances. Moreover, for treatment cohorts 2008-2014, there are fewer and fewer not-yet-treated cohorts in neighborhood $[g,g+10]$. As a result, the control groups are larger for treatment cohorts 2006-2008.}
However, for a given cohort $g$, the sample size of the nearest-$10$ not-yet-treated cohorts shrinks quickly when relative event time $s$ increases; this also occurs because the pool of not-yet-treated cohorts is smaller. When $s \ge 10$, there is no ``not-yet-treated'' cohort in neighborhood $[g,g+10]$; hence the heatmap is blank. There is thus a trade-off between horizon length and the size of the neighborhood $[g,g+10]$; that is, the degree to which the parallel trend assumption \ref{eq:para_nn} is relaxed.

\begin{figure}[htbp]
  \centering
  \copyrightbox[b]{ \includegraphics[width=1\textwidth]{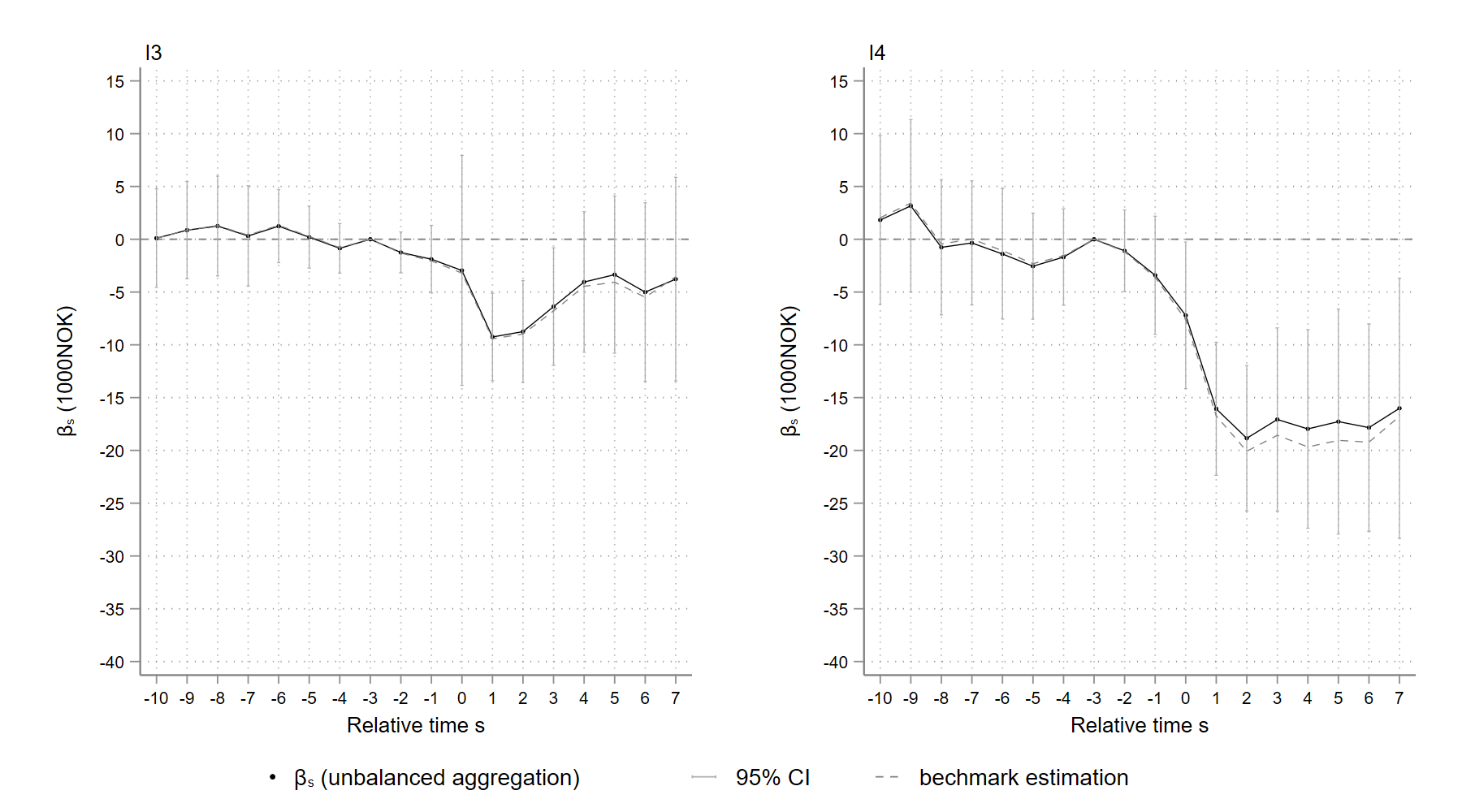}}{\scriptsize Note: $I_3 \in (W^t,2W^t]$, $I_4 \in (2W^t,\infty]$, $W^t$ is the mean annual wage in Norway in year $t$.}
  \caption{$\hat{ATT_s}$ using nearest-10 not-yet-treated cohorts as control}
  \label{fig:combine2_robust_wage_n10_p3_bq3}
\end{figure}


Figure \ref{fig:combine2_robust_wage_n10_p3_bq3} shows the DiD estimation results using annual wage income as the outcome variable and the nearest-$10$ not-yet-treated cohorts as controls. Under the local parallel trend assumption \ref{eq:para_nn}, the estimator should be less biased than the baseline regression - which implicitly employs the not-yet-treated cohorts $g' \in [g,g+\infty]$ as controls - if the time trend in the control group correlates with its composition. The horizon of the estimation using the nearest-$10$ not-yet-treated cohorts as controls is also shorter. The results in Figure \ref{fig:combine2_robust_wage_n10_p3_bq3} closely resemble those of the benchmark regression. This may be because most of the differences between control and treatment groups are captured by age profiles, which are controlled in the regression, and there are no unobservable characteristics omitted from the model. Appendix Figures \ref{fig:combine2_balanced_yrkinnt_p3_bq3} and \ref{fig:combine2_balanced_etpn_p3_bq3} depict the effects of inheritances on the probability of self-employment and annual occupational income with nearest-$10$ not-yet-treated cohorts as controls. They all support the baseline findings.


\subsection{Limited anticipation assumption}
\label{subsec:antici}

The validity of the estimates is contingent on the reference year selected. The potential outcome (earnings) should not be altered by the treatment (receiving an inheritance) in both treatment and control groups in and before the reference year. It is worth repeating that individuals arguably have a rough expectation of when and how much they can inherit. They, therefore, may have a labor supply smoothing plan in place long before receiving their inheritances. In this instance, the labor supply that has already been smoothed at first is the potential outcome, and the underlying assumption is that individuals in each cohort, on average, have the same smoothing behavior, until they receive their inheritances.

In the baseline event plot depicted in Figure \ref{fig:combine2_basic_wage_p3_bq3}, Section \ref{sec:results}, the treatment effect seems to turn negative immediately before the transfer ($s=-2$ and $-1$), indicating that a portion of recipients anticipate their inheritances and begin to adjust their labor supply prior to receiving the wealth transfer.
\footnote{The baseline event plots function as placebo tests by themselves since the estimates with $s<0$ are estimated separately, which can be interpreted as utilizing every period before the real event time (s=0) as ``pseudo-event time''. The long horizon in my study cast light on the pre-trend well in advance of the treatment.} 
In this section, I modify the reference year to be 5 years before the inheritance takes effect, assuming it is possible to anticipate receiving an inheritance no more than 4 years in advance. Under the 4-year anticipation assumption, the 2014-2017 cohorts are no longer valid ``not-yet-treated'' cohorts because they may be subject to treatment after 2017, which is not observable. Also, the 1993-1997 cohorts are not legitimate treatment groups because their corresponding reference years are before 1993 and are thus also unobservable.

\begin{figure}[htbp]
  \centering
  \copyrightbox[b]{ \includegraphics[width=1\textwidth]{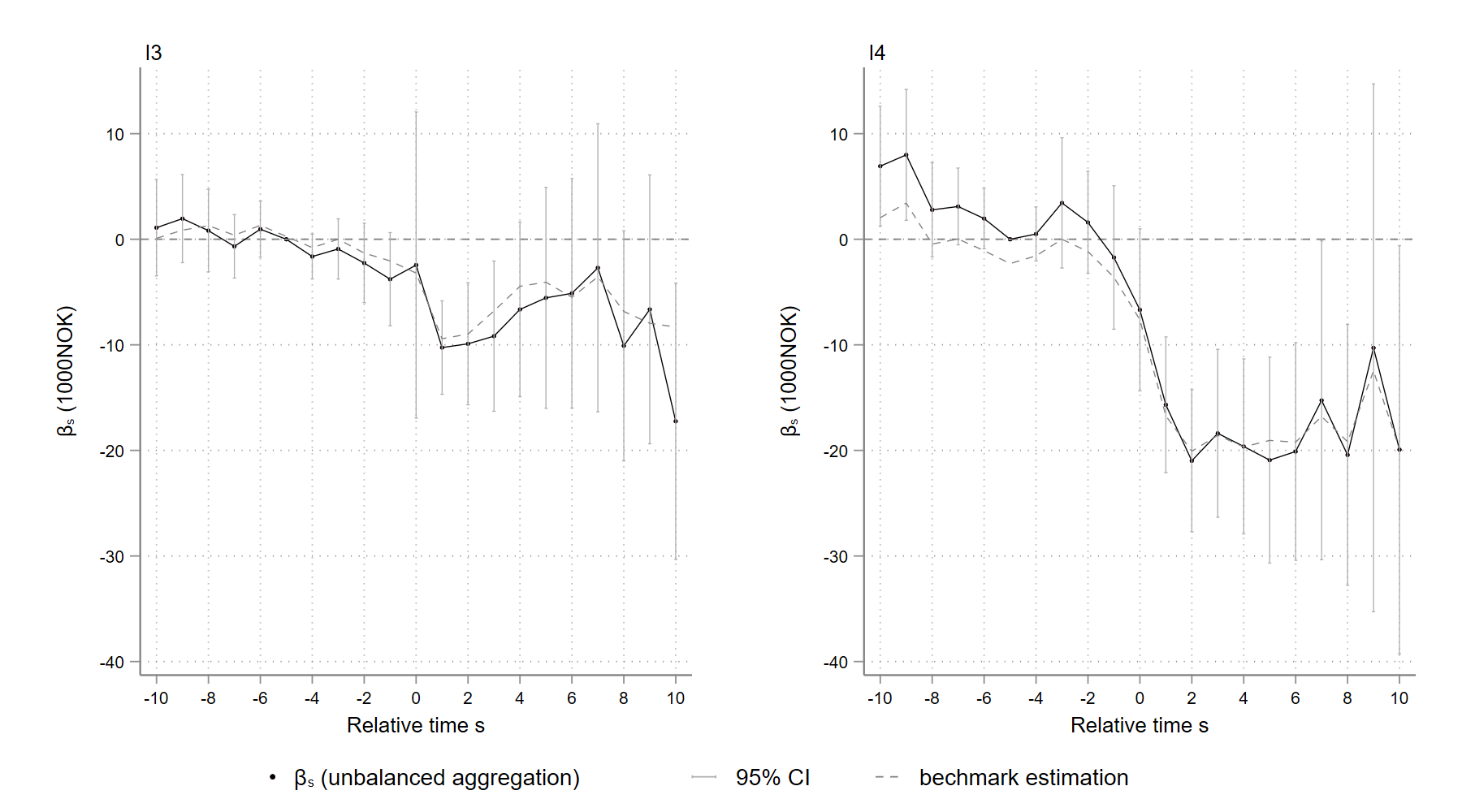}}{\scriptsize Note: $I_3 \in (W^t,2W^t]$, $I_4 \in (2W^t,\infty]$, $W^t$ is the mean annual wage in Norway in year $t$.}
  \caption{$\hat{ATT_s}$ on annual wage with $s=-5$ as reference}
  \label{fig:combine2_robust_wage_p5_bq3}
\end{figure}

Figure \ref{fig:combine2_robust_wage_p5_bq3} depicts the estimated effect of inheritances on annual wage income using $s=-5$ as the reference period. The results are very similar to the benchmark regression (with $s=-3$ as the base year). There is no significant pre-trend around $s=-5$, and the decrease in the treatment effect does not appear to have begun before $s=-2$; hence the baseline reference year, $s=-3$, is appropriate. With $s=-5$ as reference year, Appendix Figures \ref{fig:combine2_robust_etpn_p5_bq3} and \ref{fig:combine2_robust_yrkinnt_n10_p3_bq3} depict the effects of large inheritances on the probability of self-employment and annual occupational income, respectively. The baseline reference year, $s=-3$, is also valid in these plots.

\subsection{Multiple gifts}
\label{subsec:recur}

The baseline analysis is in respect of heirs who received a single gift or inheritance during the study period, thereby avoiding requiring extra assumptions for multiple treatments. However, 9\% of the population born between 1951 and 1975 received multiple gifts or inheritances over the study period. 
\footnote{99\% of the Norwegians born between 1951 and 1975 do not receive more than 4 gifts or inheritances during the study period (1993-2017). The covariance between the total amount (market value) and the number of gifts or inheritances is only 0.104.}
Approximately one-fifth of the recipients are excluded from the regression due to the prohibition on multiple gifts or inheritances. However, if the magnitude of the wealth shock effect is proportional to the wealth transfer, small gifts may not preclude identifying the effect of large inheritances on labor supply. Permitting individuals who inherit large bequests once to receive small gifts multiple times during the study period allows more observations to be included in the regression, making the estimation more precise.

\begin{figure}[htbp]
  \centering
  \copyrightbox[b]{ \includegraphics[width=1\textwidth]{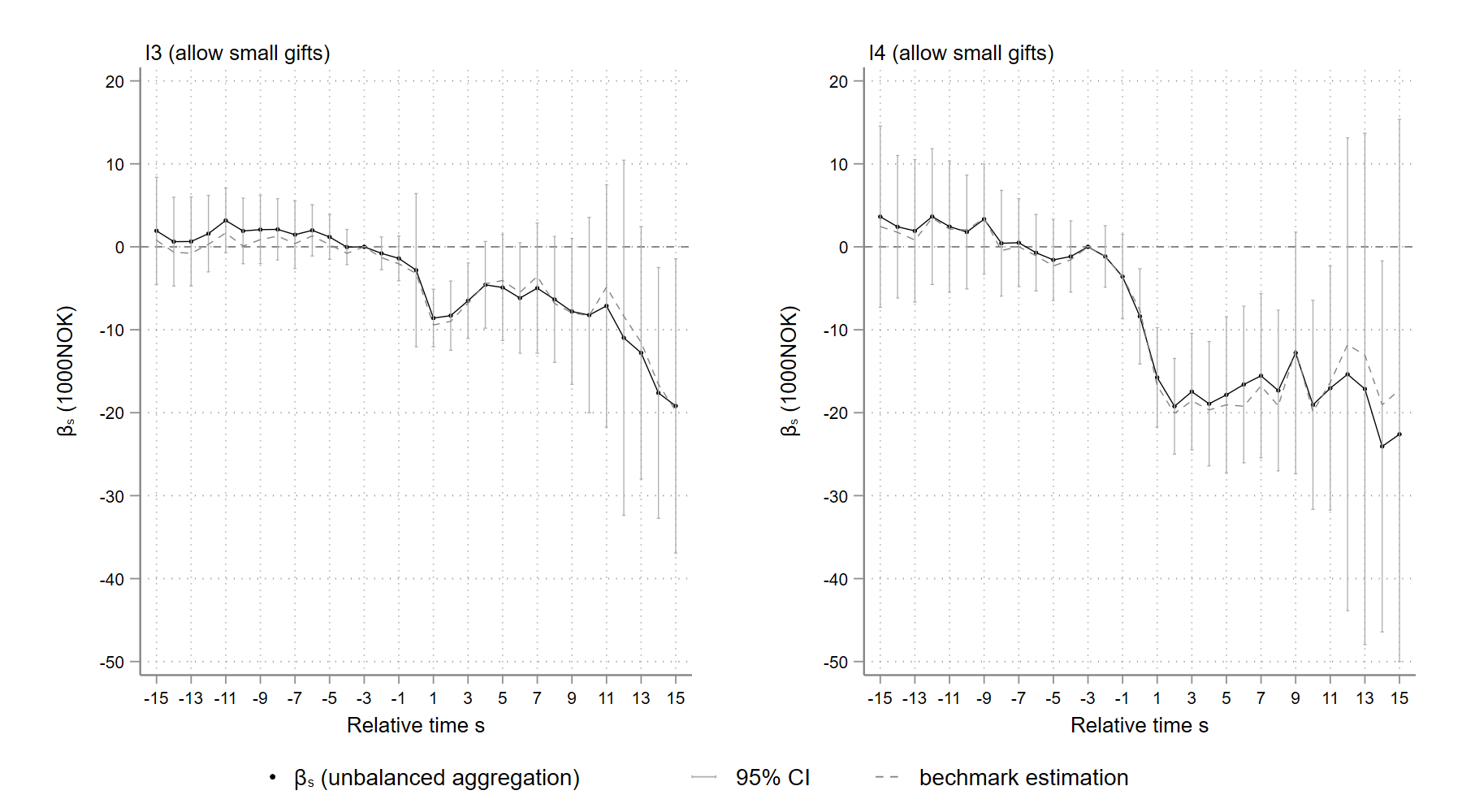}}{\scriptsize Note: $I_3 \in (W^t,2W^t]$, $I_4 \in (2W^t,\infty]$, $W^t$ is the mean annual wage in Norway. The notation $ig$ means small gifts are allowed in the sample.}
  \caption{$\hat{ATT_s}$ on annual wage allowing small gifts}
  \label{fig:combine2_ignore_wage_p3_bq3}
\end{figure}

In this section, I allow heirs of large inheritances I3 and I4 to receive gifts on multiple occasions, provided each gift is less than half of the average annual wage, i.e. $\tfrac{1}{2}W^t$, and the total amount of the gifts throughout the study period does not exceed $\tfrac{1}{2}W^t$. The eased restriction increases the sample size of recipients whose receive $I3$ and $I4$ inheritances by 23\% and 19\%, respectively. Figure \ref{fig:combine2_ignore_wage_p3_bq3} displays the estimation results allowing the multiple gifts with annual wage income as the outcome. Allowing multiple gifts or inheritances gives very similar results as that of the baseline analysis, suggesting that small gifts can be distinguished from large inheritances, and the former constitutes white noise in identifying the effect of inheritances on earnings.
The estimated effects on the probability of self-employment and annual occupational income are set out in Appendix Figures \ref{fig:combine2_ignore_etpn_p3_bq3} and \ref{fig:combine2_ignore_yrkinnt_p3_bq3}; these estimations are also comparable to the baseline findings.



\subsection{Excluding self-employment}
\label{subsec:exclude}

\begin{figure}[htbp]
  \centering
  \copyrightbox[b]{ \includegraphics[width=1\textwidth]{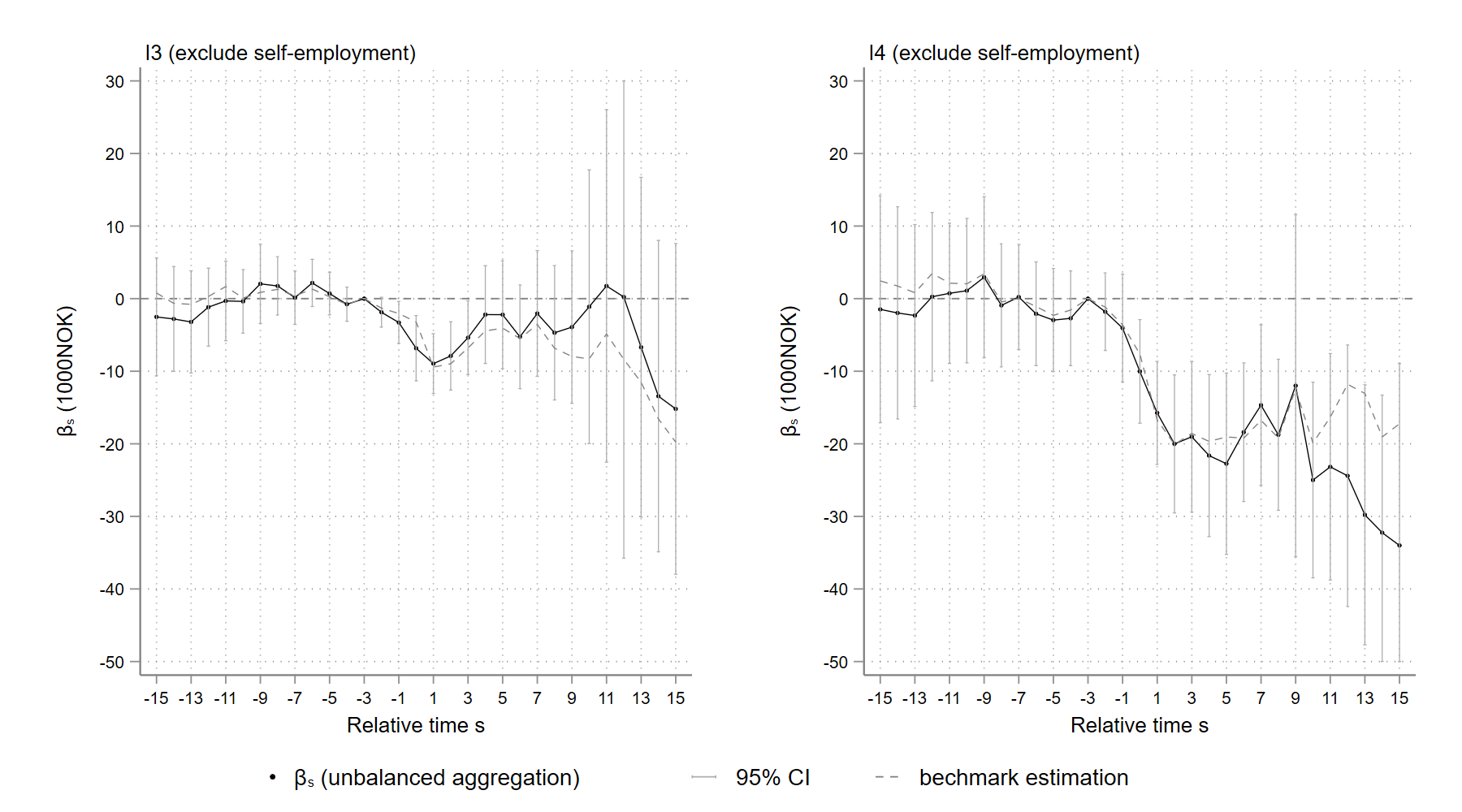}}{\scriptsize Note: $I_3 \in (W^t,2W^t]$, $I_4 \in (2W^t,\infty]$, $W^t$ is the mean annual wage in Norway. The notation $ig$ means small gifts are allowed in the sample.}
  \caption{$\hat{ATT_s}$ on annual wage excluding self-employment}
  \label{fig:combine2_basic_wage_p3_exclude_bq3}
\end{figure}

Individuals who are self-employed or have experience with self-employment may be able to manipulate the wage income they report to the tax authorities to evade tax, resulting in measurement errors in calculating their annual wages. If individuals begin self-employment after receiving an inheritance and intentionally under-report their yearly wage income, the magnitude of the negative effect on annual wage income will be overestimated. Figure \ref{fig:combine2_basic_wage_p3_exclude_bq3} depicts the estimated effect of the inheritance on the annual wage income of individuals with no business income during the entire 1993-2017 study period. The estimates are very similar to the baseline results when $s<10$. When $s>10$, however, there is a difference in the effect of inheritances for individuals who have never been self-employed. 

For inheritances I3 (between $W^t$ and $2W^t$ in size), when $s>10$, the magnitude of the estimates appears to be smaller than the baseline case, suggesting that the problem of wage under-reporting is likely to exist, especially in cohorts with a greater proportion of young heirs. According to the baseline event plot \ref{fig:combine2_basic_etpn_p3_bq3} (Section \ref{sec:results}) and the discussion on the age composition of the cohorts in the robustness check Section \ref{subsec:balance}, inheritances in category I3 have a greater impact on the probability of self-employment among heirs who receive inheritances when they are young, a group over-represented when $s>10$. This explains why eliminating self-employment affects the estimates more when $s>10$. In addition, since the proportion of young heirs who are self-employed is quite high, excluding them from the aggregation reduces the precision of the estimates.

In contrast to inheritance I3, it appears that inheritances exceeding $2W^t$ (I4) reduce the annual wage income of heirs who have never been self-employed comparatively more than in the baseline case when $s>10$. This can also be explained by the findings from the baseline event plot \ref{fig:combine2_basic_etpn_p3_bq3}: although substantial inheritances increase the probability of self-employment in general, they seem to inhibit the entrepreneurship of young heirs to some extent. If those who receive category I4 inheritances and start a business anyway are more determined and earn more than the average level, excluding them from the regression exaggerates the estimated the negative effect on earnings. Age group heterogeneity is discussed in greater detail in the following section.

\section{Heterogeneity}
\label{sec:hetero}

As suggested by the literature reviewed in Section \ref{sec:intro}, the influence of inheritances on labor supply can be heterogeneous by age and sex. In this section, I first investigate the heterogeneity of parental inheritance effects across sex and age groups. Furthermore, since heirs inheriting from their grandparents are often very young, they make it possible to study the effect of inheritances on very young recipients. In Section \ref{subsec:grand}, I examine the effect of inheritances on these very young heirs.


\subsection{Heterogeneity by sex}
\label{subsec:hetero_sex}

Inheritances may have different effect on the labor supply of men and women. Columns 2-3 and 6-7 in Table \ref{tab:hetero} describe the characteristics of male and female heirs in the years of receiving inheritances I3 and I4- sized inheritances, respectively. Although the average inheritance amount is fairly balanced between the sexes, men have greater income and wealth on average. The probability of male heirs being self-employed is also substantially greater than that for female heirs.

\begin{table}[htbp] \centering
\caption{Sample mean in event year $E_i$ by sex and age groups}
\label{tab:hetero}
\vspace{-4pt}
\begin{threeparttable}

\begin{tabular}{l*{10}{c}}
\toprule
inheritance size:  &\multicolumn{4}{c}{ $I_3 \in (W^t,2W^t]$} &&\multicolumn{4}{c}{$I_4 \in (2W^t,\infty]$} \\
            \cmidrule{2-5} \cmidrule{7-10}   
groups:    &\multicolumn{1}{c}{female}&\multicolumn{1}{c}{male}&\multicolumn{1}{c}{young}&\multicolumn{1}{c}{old}&&\multicolumn{1}{c}{female}&\multicolumn{1}{c}{male}&\multicolumn{1}{c}{young}&\multicolumn{1}{c}{old}\\
\midrule                                      
inheritance year&       2008&   2007&2004 &2011  &&   2009 &  2008 &    2005     &2012        \\
inheritance amount&       547&     546&495  &610   &&   1,688 &  1,694 &  1,496      &1,895        \\
wage income&             319&     527&393  &444   &&     352 &    593 &    436      &506        \\
occupational income&      334&     575&422  &477   &&     374 &    658 &    470      &557        \\
Pr(self-employment)& 0.09&    0.17&0.13 &0.12  &&    0.11 &   0.18 &   0.15      &0.14        \\
gross wealth&             802&   1,500&821  &1,503 &&   1,734 &  3,030 &  1,628      &3,142        \\
-debt rate &              0.50&    0.68&0.80 &0.48  &&    0.28 &   0.39 &   0.50	     &0.27     \\
age         &              48&      48&42   &55    &&      49 &     49 &    42       &56        \\
education   &             4.2&     4.4&4.1  &4.3   &&     4.6 &    4.8 &    4.7      &4.7        \\
\midrule
no. of individuals & 15,854     &  13,929  & 16,351  & 13,432  &&  12,212   &  11,712    & 12,222     &     11,702   \\
\bottomrule

\end{tabular}

\begin{tablenotes}
      \small
      \item Note: (1) Income and wealth are deflated with CPI to $1,000$ NOK at 2015 price. (2) $W^t$ is the mean annual wage in Norway in year $t$. (3) young/old heirs inherit before/after 50 years old. (4) occupational income is the summation of wage income and business income; self-employment is defined as having non-zero business income in a certain year; $edu$ is an ordered 0-8 categorical variable as defined by SSB: \href{https://www.ssb.no/klass/klassifikasjoner/36/}{https://www.ssb.no/klass/klassifikasjoner/36/}, e.g. edu=4 for upper secondary education. 
    \end{tablenotes}
\end{threeparttable}
\end{table}
 
\begin{figure}[htbp]
    \centering
    \copyrightbox[b]{\includegraphics[width=1\textwidth]{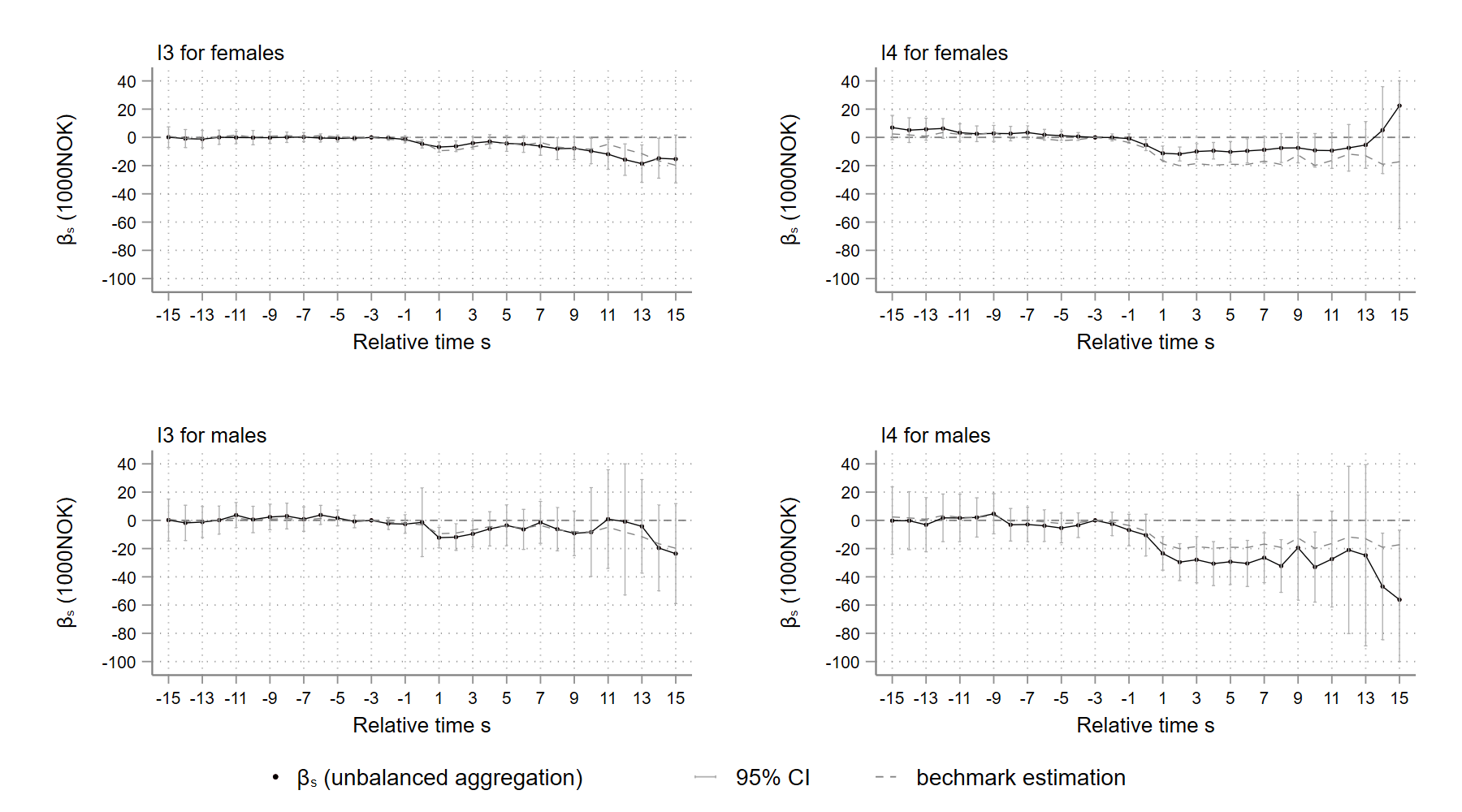}}{\scriptsize Note: $I_3 \in (W^t,2W^t]$, $I_4 \in (2W^t,\infty]$, $W^t$ is the mean annual wage in Norway in year $t$.}
    \caption{$\hat{ATT_s}$ on annual wage income by sex group}
    \label{fig:combine2_basic_wage_p3_bq3_sex}
\end{figure}
\begin{figure}[htbp]
    \centering
    \copyrightbox[b]{\includegraphics[width=1\textwidth]{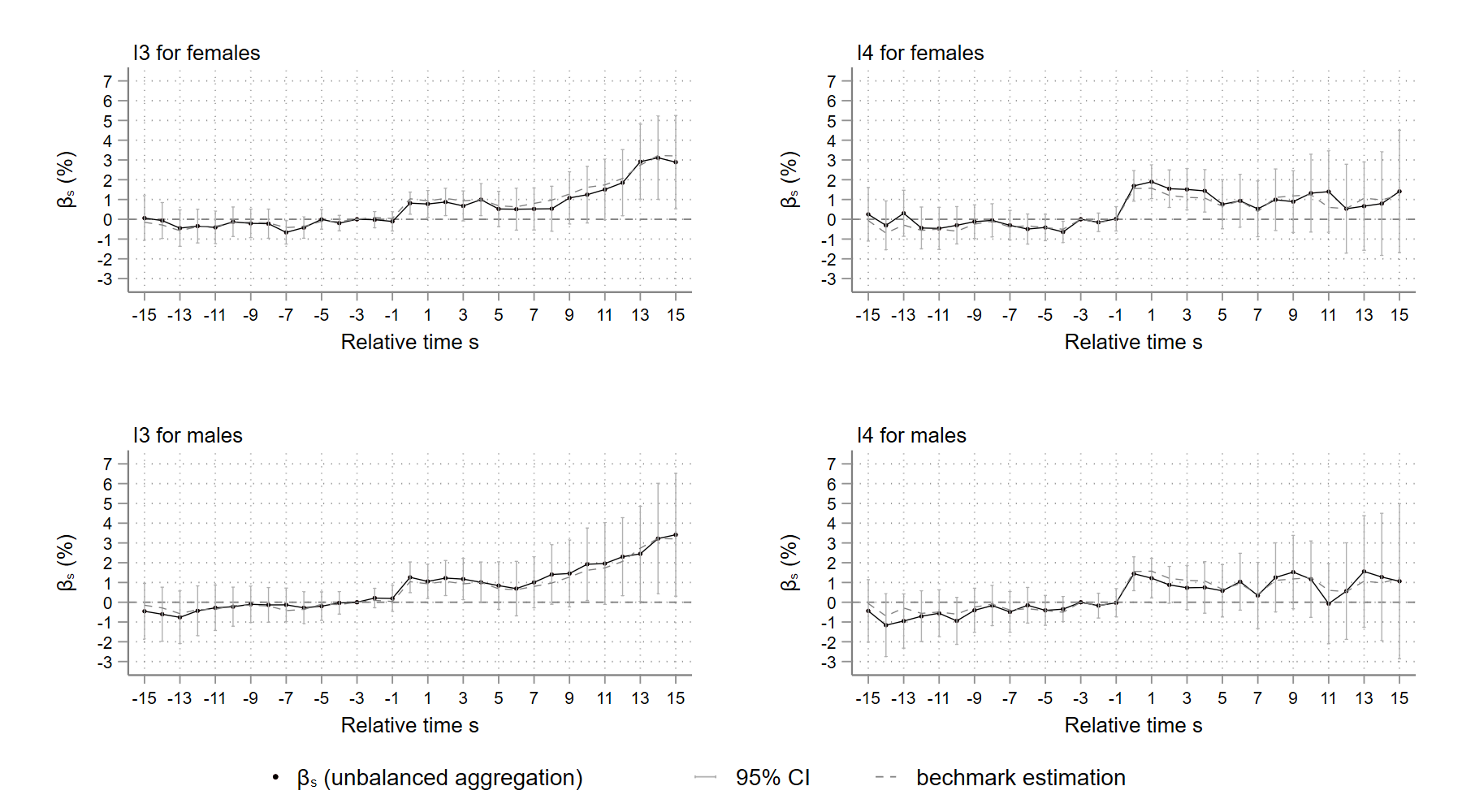}}{\scriptsize Note: $I_3 \in (W^t,2W^t]$, $I_4 \in (2W^t,\infty]$, $W^t$ is the mean annual wage in Norway in year $t$.}
    \caption{$\hat{ATT_s}$ on Pr(self-employment) by sex group}
    \label{fig:combine2_basic_etpn_p3_bq3_sex}
\end{figure}
\begin{figure}[htbp]
    \centering
    \copyrightbox[b]{\includegraphics[width=1\textwidth]{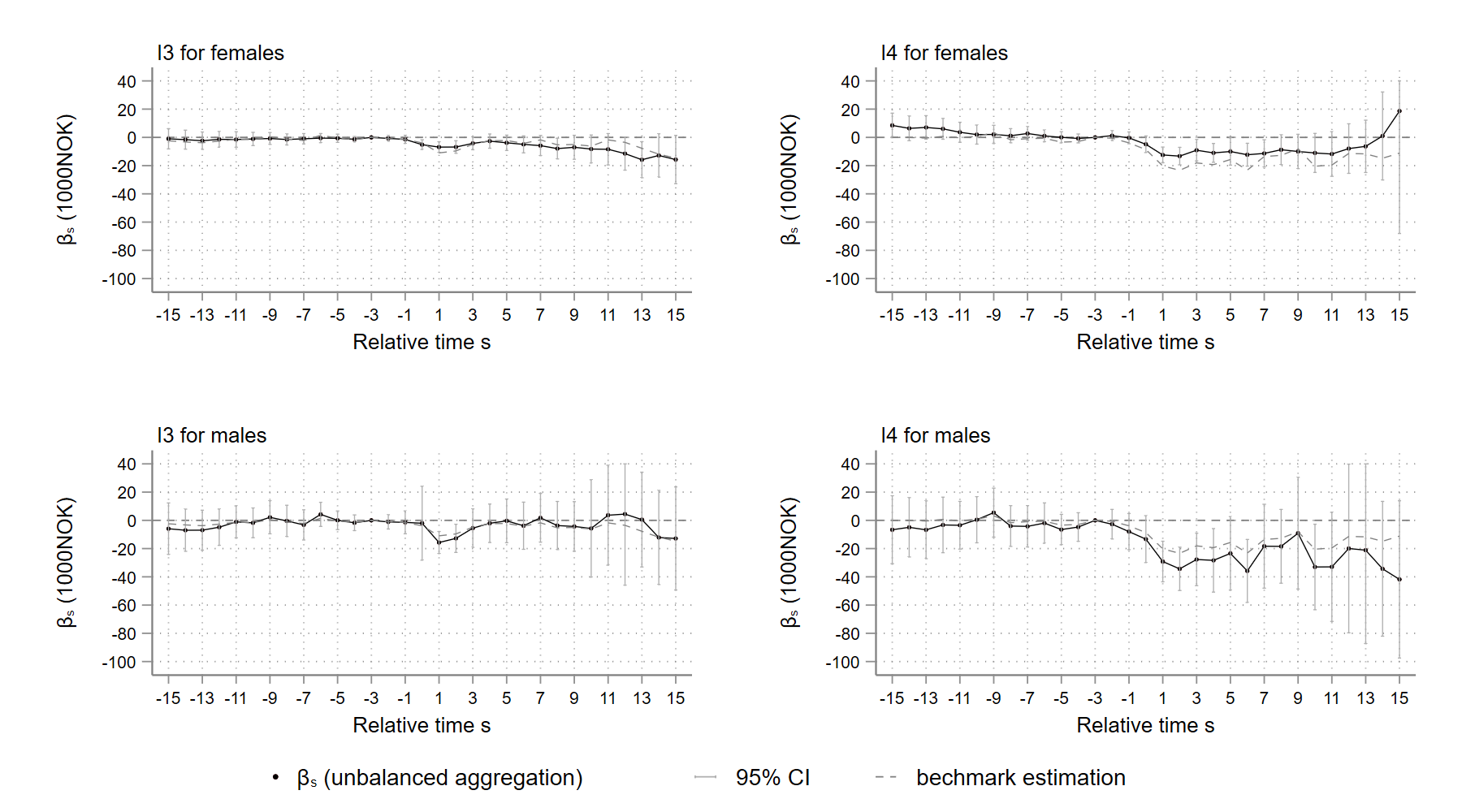}}{\scriptsize Note: $I_3 \in (W^t,2W^t]$, $I_4 \in (2W^t,\infty]$, $W^t$ is the mean annual wage in Norway in year $t$.}
    \caption{$\hat{ATT_s}$ on annual occupational income by sex group}
    \label{fig:combine2_basic_yrkinnt_p3_bq3_sex}
\end{figure}

Figures \ref{fig:combine2_basic_wage_p3_bq3_sex} -\ref{fig:combine2_basic_yrkinnt_p3_bq3_sex} compare the effects of large inheritances I3 and I4 on men and women inheritors' annual wage income, probability of self-employment, and annual occupational income. According to the first column of Figure \ref{fig:combine2_basic_wage_p3_bq3_sex}, the absolute effect of inheritances on male heirs (-12,000 NOK) is greater than that on female heirs (-8,000 NOK) immediately following the treatment. However, because female heirs, on average, earn less than male heirs, the effect of the inheritance relative to the average annual wage income within each group is roughly equivalent (-2\%). When the inheritance size exceeds $2W^t$, as shown in the second column of Figure \ref{fig:combine2_basic_wage_p3_bq3_sex}, male heirs' wages drop (-5\%) roughly twice as much as female heirs' wage (-2.8\%). Interestingly, the depiction in Figure \ref{fig:combine2_basic_etpn_p3_bq3_sex} demonstrates that the effect of inheritances on the likelihood of self-employment is similar for male and female heirs, regardless of the size of inheritances. 

The findings reveal that the greater yearly wage loss of male heirs who inherit more than $2W^t$, seen in Figures \ref{fig:combine2_basic_wage_p3_bq3_sex} is the result of their decreasing their labor supply relatively more than do female heirs, rather than being be result of their higher likelihood of self-employment. As a result, the annual occupational income of men receiving I4 inheritances also drops (-4.5\%) more than that of female heirs (-2.7\%). These findings contradict those in the existing literature \citep{doorley2016labour}.



\subsection{Heterogeneity by age groups}
\label{subsec:hetero_age}

As observed in the baseline results (Figures \ref{fig:combine2_basic_wage_p3_bq3}-\ref{fig:combine2_basic_yrkinnt_p3_bq3} in Section \ref{sec:results}) and in robustness check Section \ref{subsec:balance}, heirs who receive inheritances at a reasonably young age appear to have a relatively more pronounced response than their older counterparts. To analyze age heterogeneity, I define young and old heirs as those who receive inheritances before and after 50 years old. Each age group accounts for around half of the sample.
\footnote{I also attempt to categorize the sample into additional age groups, but there are insufficient observations to provide precise estimates.}
Table \ref{tab:hetero} also describes the characteristics of the heirs by age group in the years of receiving I3- and I4- sized inheritances. Young heirs are, on average, 13 years younger than old heirs. Young heirs earn and amass less wealth than their older counterparts in the year receiving inheritance. The debt rate and probability of being self-employed are higher for young than for old heirs. 
\footnote{Note that although the estimation is by age groups, the unbalanced aggregation results are still affected by age composition heterogeneity across cohorts. As $s$ increases, the inheritors in the treatment cohorts gets younger on average.}

\begin{figure}[htbp]
    \centering
    \copyrightbox[b]{\includegraphics[width=1\textwidth]{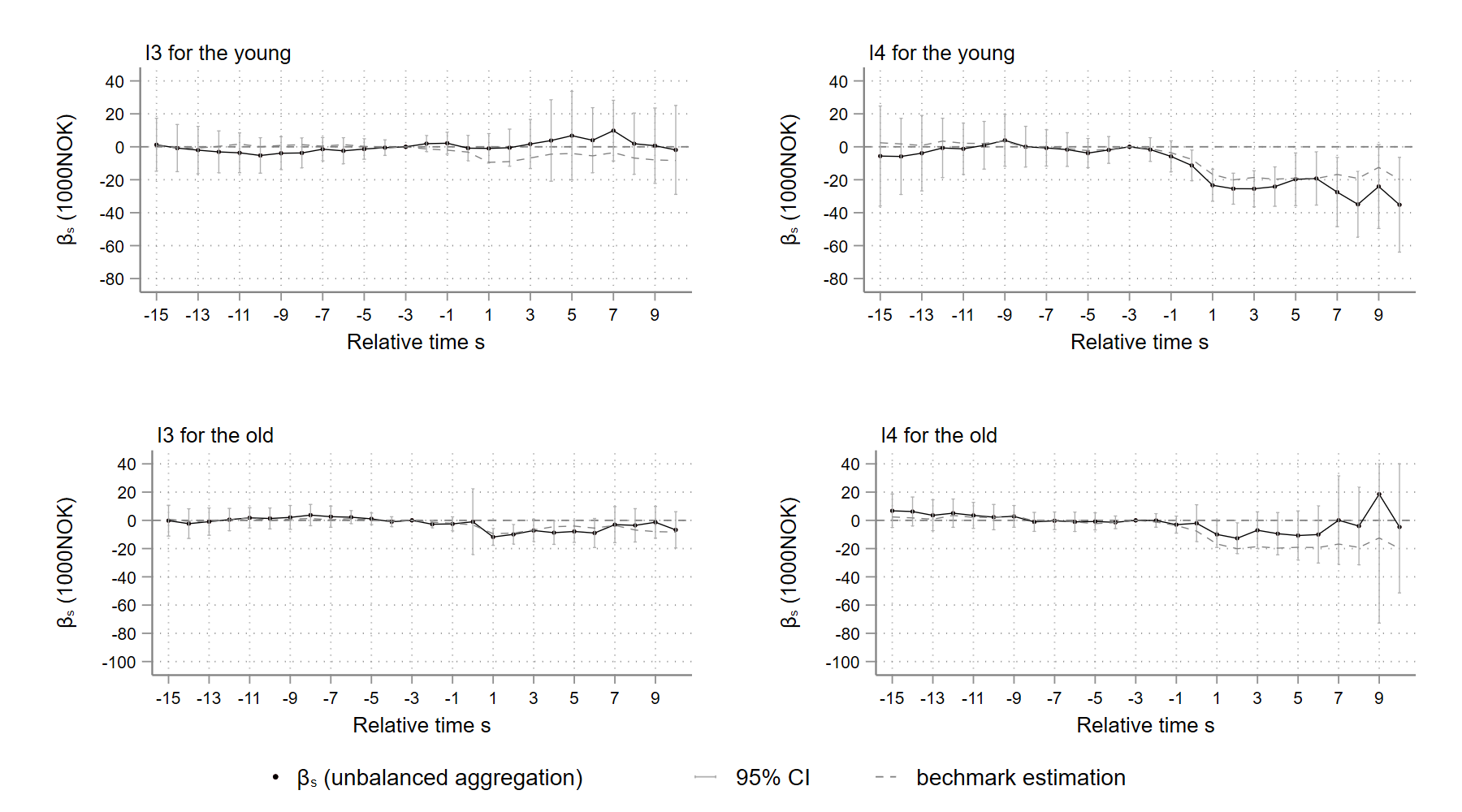}}{\scriptsize Note: $I_3 \in (W^t,2W^t]$, $I_4 \in (2W^t,\infty]$, $W^t$ is the mean annual wage in Norway in year $t$.}
    \caption{$\hat{ATT_s}$ on annual wage income by age group}
    \label{fig:combine2_basic_wage_p3_bq3_age}
\end{figure}

\begin{figure}[htbp]
    \centering
    \copyrightbox[b]{\includegraphics[width=1\textwidth]{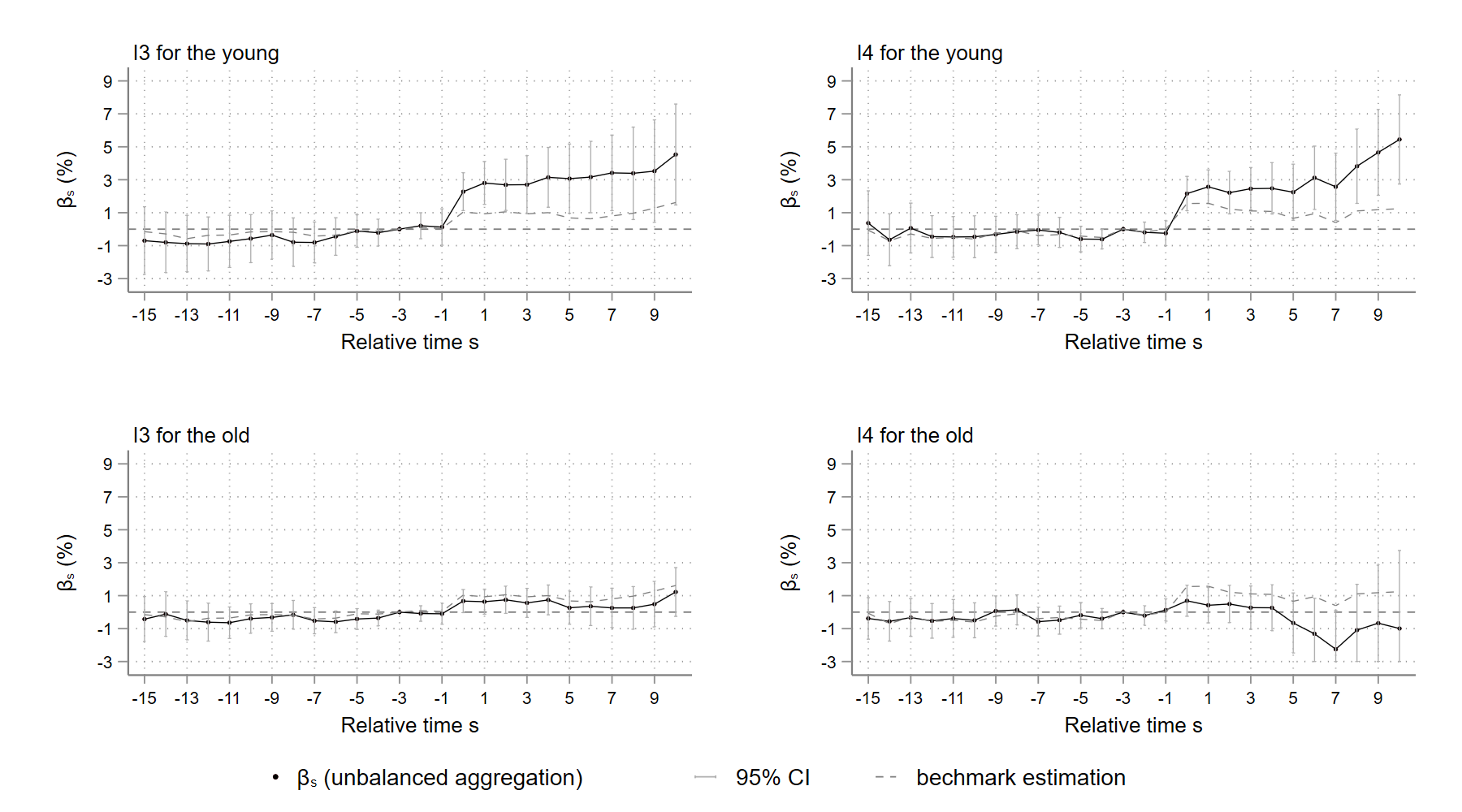}}{\scriptsize Note: $I_3 \in (W^t,2W^t]$, $I_4 \in (2W^t,\infty]$, $W^t$ is the mean annual wage in Norway in year $t$.}
    \caption{$\hat{ATT_s}$ on Pr(self-employment) by age group}
    \label{fig:combine2_basic_etpn_p3_bq3_age}
\end{figure}

\begin{figure}[htbp]
    \centering
    \copyrightbox[b]{\includegraphics[width=1\textwidth]{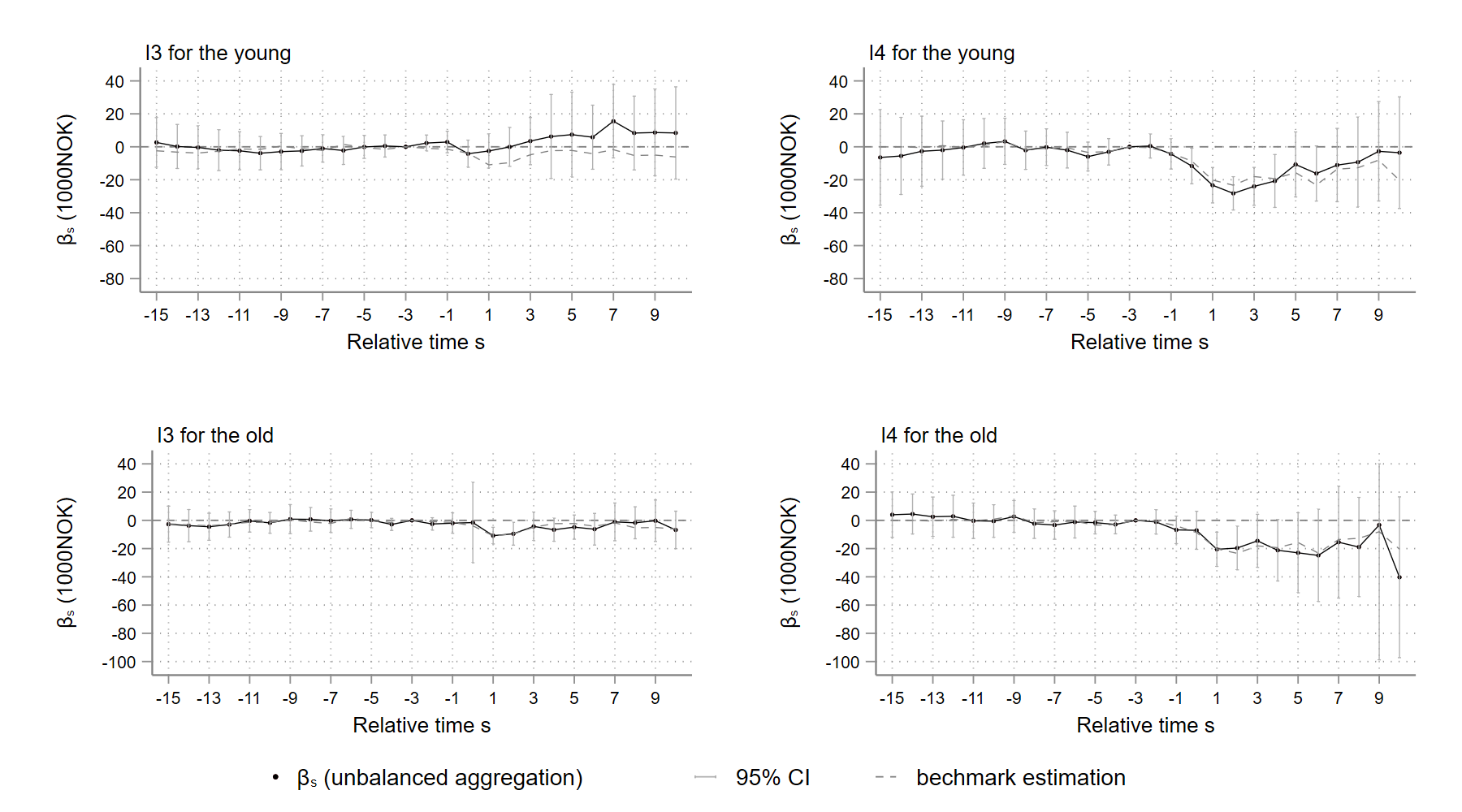}}{\scriptsize Note: $I_3 \in (W^t,2W^t]$, $I_4 \in (2W^t,\infty]$, $W^t$ is the mean annual wage in Norway in year $t$.}
    \caption{$\hat{ATT_s}$ on annual occupational income by age group}
    \label{fig:combine2_basic_yrkinnt_p3_bq3_age}
\end{figure}

The event plots by age group are depicted in Figures \ref{fig:combine2_basic_wage_p3_bq3_age} -\ref{fig:combine2_basic_yrkinnt_p3_bq3_age}. Because the age composition varies with cohorts and there are far fewer elderly heirs in the early cohorts (e.g. cohort 1996), the number of observations becomes extremely small as $s$ grows. I, therefore, trim the horizon to be $s \in [-10,10]$. According to Column 1 of Figure \ref{fig:combine2_basic_wage_p3_bq3_age}, inheritances of size I3 have no effect on the annual wage income of the young heirs, while the estimates (both absolute and relative effects) for old heirs are similar to the baseline case. However, when inheritances are greater than $2W^t$ in the second column in Figure \ref{fig:combine2_basic_wage_p3_bq3_age}, the decrease in the annual wage income of young heirs (-5.7\%) is much greater than that for old heirs (-2\%).


The results in Figure \ref{fig:combine2_basic_etpn_p3_bq3_age} partially explain this phenomenon. The probability of self-employment among young heirs increases by 3\%-5\% after obtaining I3-sized inheritances; whereas the probability of self-employment among older heirs is unaffected or even decreases. In other words, young heirs are more likely to quit their current jobs and become self-employed after receiving an inheritance, thereby reducing their annual wage income; however, large inheritances have minimal effect on the labor supply of older heirs who are close to retirement. As a result, the annual occupational income of young heirs decreases less than that of old heirs after receiving an I3 inheritance, as demonstrated in Column 1 of Figure \ref{fig:combine2_basic_yrkinnt_p3_bq3_age}. Interestingly, although the annual occupational income of young heirs drops significantly after receiving an I4 inheritance, as seen in Column 2 of Figure \ref{fig:combine2_basic_yrkinnt_p3_bq3_age}, the effect diminishes gradually; by contrast, the negative effect on old heirs persists for several years. These findings are in line with \cite{Boe2019} and \cite{Brown2010}, who find the negative effect of inheritances is greater in magnitude for individuals close to retirement, as mentioned in Section \ref{sec:intro}.

\subsection{Bequests from grandparents}
\label{subsec:grand}

The average age of the ``young'' heirs defined in the preceding section is 42 years old, which may not be sufficiently young to change careers after inheriting. In contrast, those who inherit from their grandparents may be very young. To further examine the effect of inheritances on young heirs, I focus in this section on those who inherit from their grandparents. Similar to the baseline analysis, I define inheritances from grandparents to be gifts or inheritances received within 3 years of the grandparents' death, provided the heirs' parents do not pass away during that time.

\begin{figure}[htbp]
    \centering
    \includegraphics[width=1\textwidth]{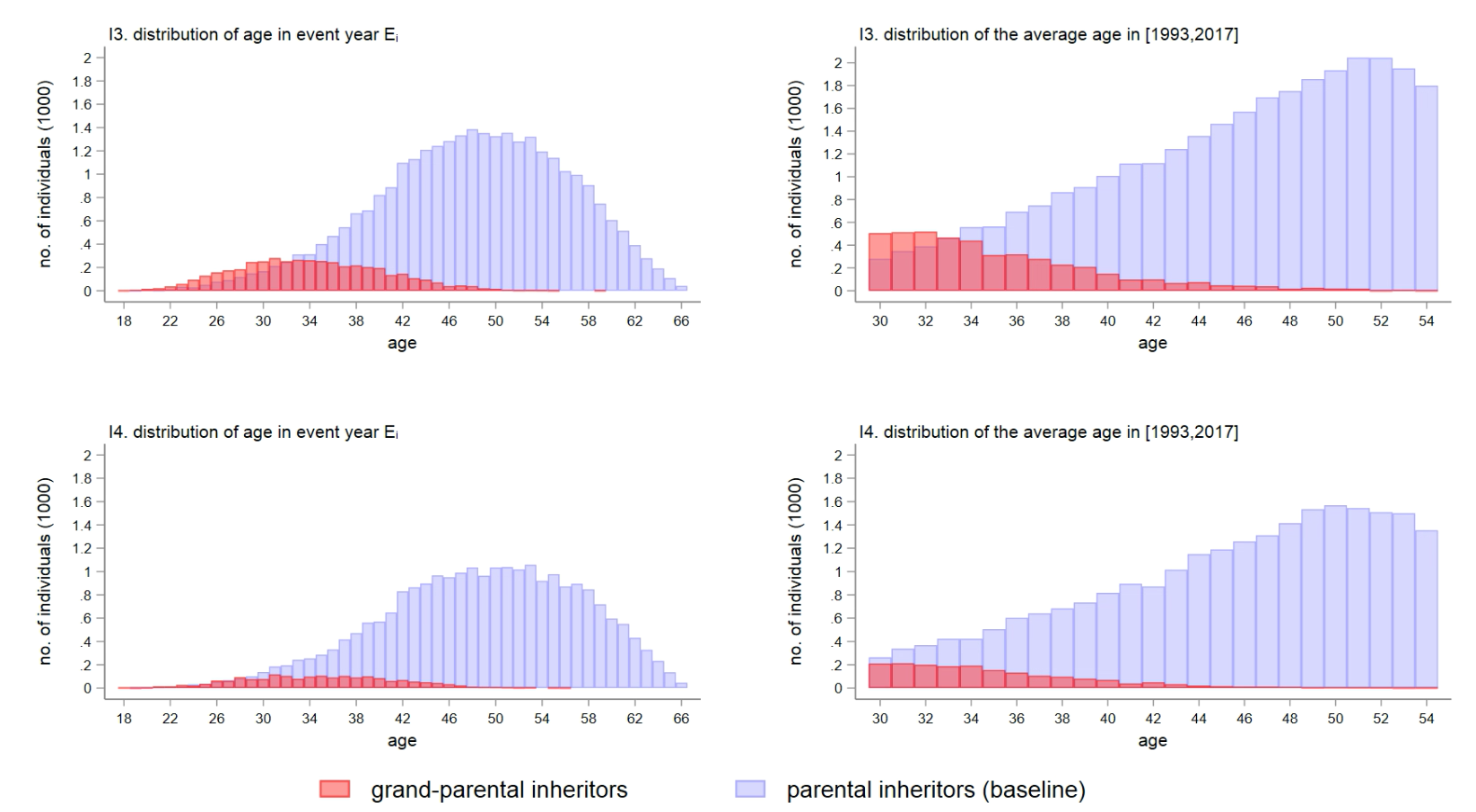}
    \caption{Age distribution of parental and grand-parental inheritors}
    \label{fig:obs_age_g_p}
\end{figure}

Figure \ref{fig:obs_age_g_p} contrasts the age distribution of heirs who receive inheritances from their grandparents (light-red bars) with that of heirs who inherit from their parents (baseline sample, light-blue bars). As expected, heirs inheriting from grandparents are younger (around 35 years old) than the baseline sample as a whole (47 years old).Since these heirs are typically quite young when they inherit, they may adjust not just their labor supply but also their educational level (e.g. dropping out from school to become self-employed).

The sample of heirs inheriting from their grandparents is also much smaller because:
(i) only 44\% of the recipients' grandparents' information is traceable in my sample.
\footnote{Only 22\% of the grandparents' information is traceable in the registry data for the entire Norwegian population born between 1951 and 1975, therefore my sample over-represent the individuals with traceable grandparent information.} 
(ii) Since the economic boom in Norway started in the 1970s, large bequests from grandparents are less common compared to those from parents. Furthermore, since the inheritance law places children ahead of grandchildren in the intestate order of succession, inheritances from grandparents are only permissible in certain circumstances. For example, before the inheritance tax was abolished in 2014, the deceased might pass their wealth directly to their grandchildren to ensure it is only taxed once. It is also conceivable that the deceased's children and grandchildren will jointly inherit businesses from the deceased. In any event, the sample of those receiving bequests from grand-parents is subject to different selections from the baseline sample, and the estimates may lack external validity.

\begin{table}[htbp] \centering
\caption{Sample mean in event year $E_i$: heirs of grandparents' bequests}
\label{des_event_gbq}
\vspace{-4pt}
\begin{threeparttable}
\begin{tabular}{l*{5}{c}}
\toprule
Interval &all&I1&I2&I3&I4 \\
\midrule
\multicolumn{5}{l}{\textbf{A.Heirs' information}} \\

inheritance year&       2003&     2003&     2003&     2004&     2005\\
inheritance amount&      275&       86&      239&      476&    1,577\\
labor income&            347&      332&      359&      373&      366\\
occupational income&     366&      350&      375&      394&      403\\
Pr(self-employment)&       0.13&     0.11&     0.12&     0.16&     0.27\\
capital income&           21&       13&       17&       32&       89\\
gross wealth&            653&      459&      607&      852&    2,046\\
-debt rate &            1.21&	  1.54&	    1.30&  	  1.07&	   0.59 \\
age         &             34&       33&       34&       34&       35\\
male        &           0.51&     0.49&     0.50&     0.53&     0.56\\
education   &            4.7&      4.6&      4.7&      4.8&      4.8\\
\midrule
\multicolumn{5}{l}{\textbf{B.Parental information}} \\
Pr(self-employment)$_f$&   0.36  & 0.35  & 0.35  &  0.37  &  0.44   \\
Pr(self-employment)$_m$&   0.20  & 0.20  & 0.19  &  0.20  &  0.26   \\
gross wealth$_f$       &   1,425 & 1,067 & 1,414 &  1,884 &  3,445   \\
gross wealth$_m$       &   598   &  467  & 624   &  750   &  1,252  \\
age$_f$       &   60   &  60  & 61   &  61   &  62  \\
age$_m$       &   58   &  57  & 58   &  59   &  60  \\
\midrule
no. of individuals  & 30,405& 16,059&7,962&4,518 &1,866 \\

\bottomrule
\end{tabular}
\begin{tablenotes}
      \small
      \item Note: (1) Income and wealth are deflated with CPI to $1,000$ NOK at 2015 price. (2) $W^t$ is the mean annual wage in Norway in year $t$. (3) occupational income is the summation of wage income and business income; self-employment is defined as having non-zero business income in a certain year; $edu$ is an ordered 0-8 categorical variable as defined by SSB: \href{https://www.ssb.no/klass/klassifikasjoner/36/}{https://www.ssb.no/klass/klassifikasjoner/36/}, e.g. edu=4 for upper secondary education. (4) Subscripts $f$ and $m$ denote heirs' fathers and mothers separately.
    \end{tablenotes}
\end{threeparttable}
\end{table}


Panel A of Table \ref{des_event_gbq} sets out heirs inheriting from their grandparents by the size of the inheritance. Heirs to I3- and I4-sized inheritances are much more likely to be self-employed and have a greater level of education than those in the baseline sample. A possible explanation is that these heirs inherit businesses (or an entrepreneurial family culture) from their grandparents. This argument is backed by descriptions of heirs' parents in panel B of Table \ref{des_event_gbq}. In the years when heirs receive inheritances from their grandparents, more than 35\% of the heirs' fathers and 19\% of their mothers have business income; these levels are much higher than for individuals in the baseline sample. It appears that these households have a culture of self-employment.

\begin{figure}[htbp]
    \centering
    \copyrightbox[b]{\includegraphics[width=1\textwidth]{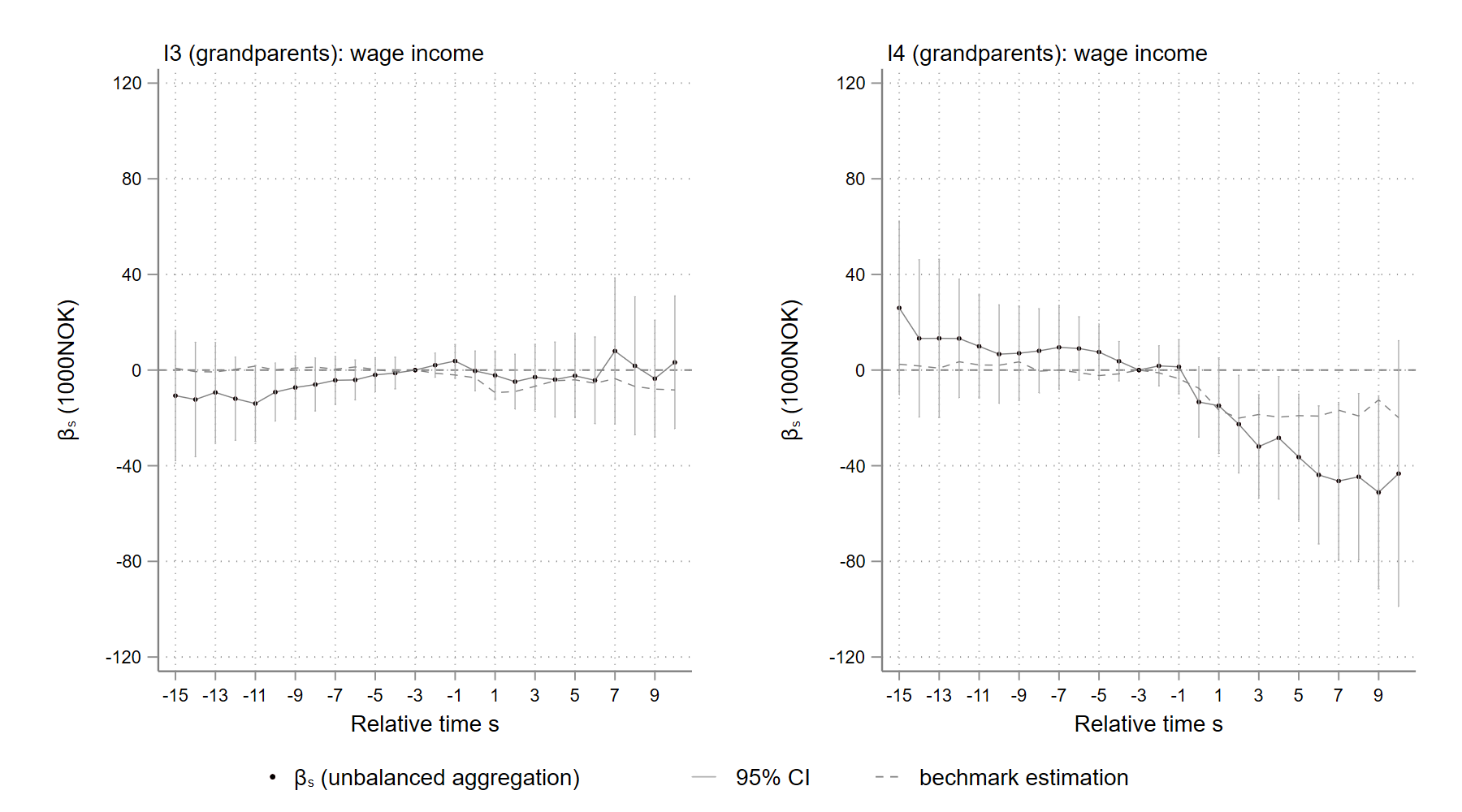}}{\scriptsize Note: $I_3 \in (W^t,2W^t]$, $I_4 \in (2W^t,\infty]$, $W^t$ is the mean annual wage in Norway in year $t$.}
    \caption{$\hat{ATT_s}$ on annual wage income: inheritances from grandparents}
    \label{fig:combine2_basic_wage_p3_gbq3}
\end{figure}

\begin{figure}[htbp]
    \centering
    \copyrightbox[b]{\includegraphics[width=1\textwidth]{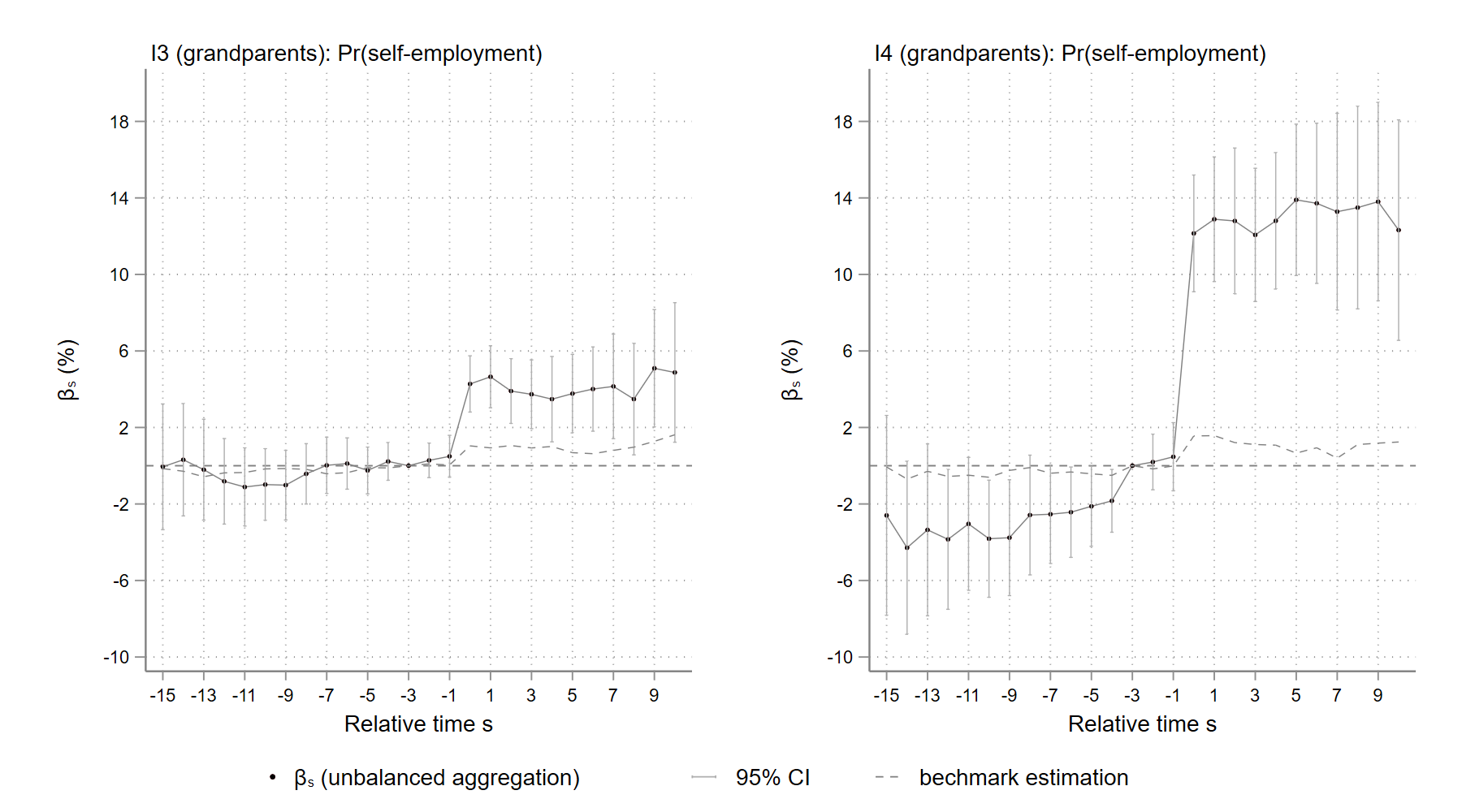}}{\scriptsize Note: $I_3 \in (W^t,2W^t]$, $I_4 \in (2W^t,\infty]$, $W^t$ is the mean annual wage in Norway in year $t$.}
    \caption{$\hat{ATT_s}$ on Pr(self-employment): inheritances from grandparents}
    \label{fig:combine2_basic_etpn_p3_gbq3}
\end{figure}

\begin{figure}[htbp]
    \centering
    \copyrightbox[b]{\includegraphics[width=1\textwidth]{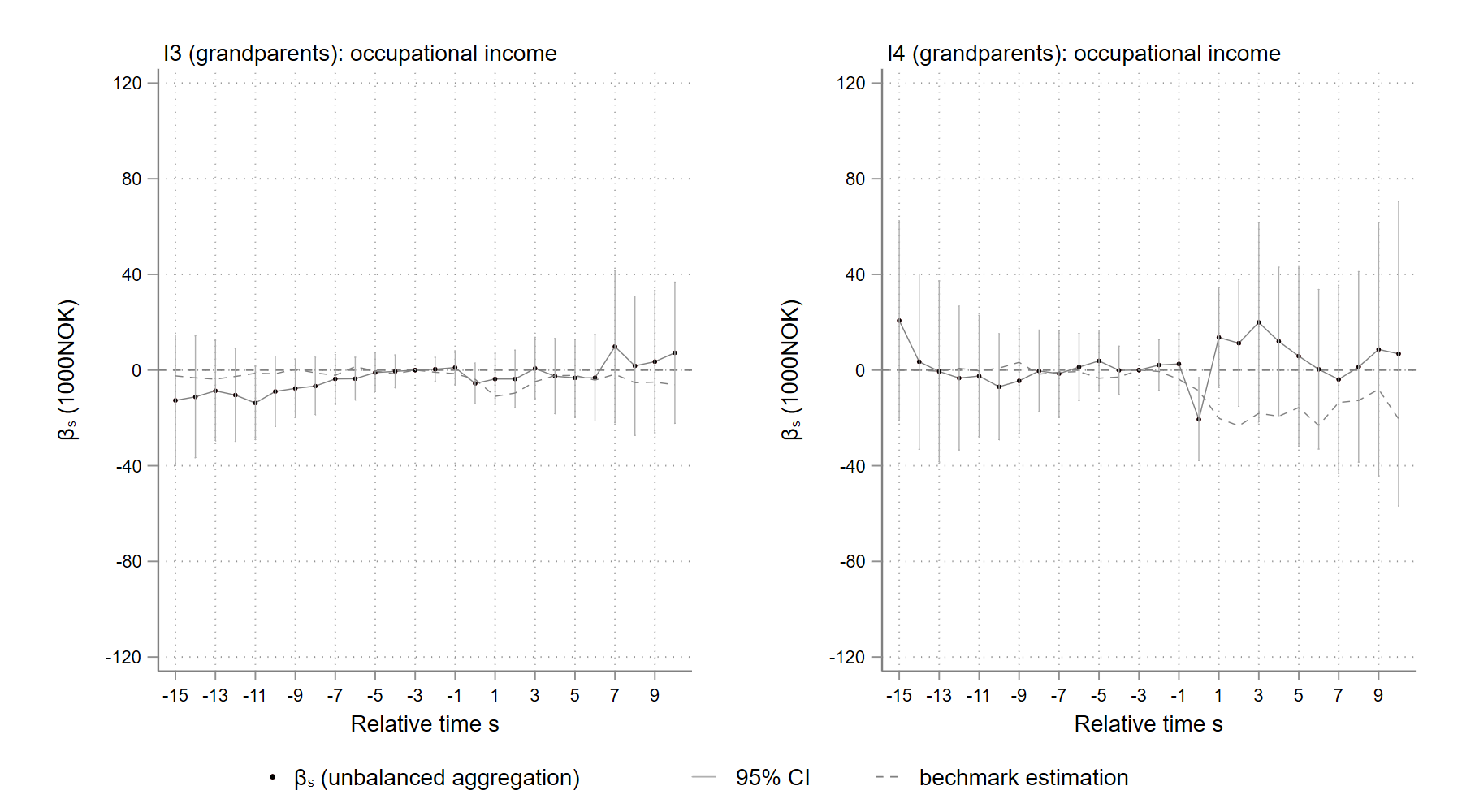}}{\scriptsize Note: $I_3 \in (W^t,2W^t]$, $I_4 \in (2W^t,\infty]$, $W^t$ is the mean annual wage in Norway in year $t$.}
    \caption{$\hat{ATT_s}$ on annual occupation income: inheritances from grandparents}
    \label{fig:combine2_basic_yrkinnt_p3_gbq3}
\end{figure}

Figures \ref{fig:combine2_basic_wage_p3_gbq3}-\ref{fig:combine2_basic_yrkinnt_p3_gbq3} illustrate the effect of I3- and I4- sized inheritances from grandparents on (very young) heirs' earnings and their probability of being self-employed. The pattern in these plots are very slimier to for the (relatively) young heirs in Figures \ref{fig:combine2_basic_wage_p3_bq3_age}-\ref{fig:combine2_basic_yrkinnt_p3_bq3_age} in Section \ref{subsec:hetero_age}, but more dramatic: receiving an I3-sized inheritance from grandparents has little effect on heirs' annual wage and occupational income, but increases the probability of being self-employed by around 3\%. In contrast, an inheritance from grandparents greater than $2W^t$ (I4) cut the young heirs' annual wage income by 11\% and increases their probability of being self-employed by 14\%, albeit with a slight pre-trend. These effects are approximately three times greater than the baseline estimates. In addition, the effect of I4-sized inheritances from grandparents on annual occupational income declines immediately after inheritances are transferred.


These findings suggest that the extent of heterogeneity across age groups is positively correlated with the age difference between the groups. The tendency appears to be more apparent the younger the heirs are. Another reason for the dramatically large effect on the probability of self-employment is that grandchildren may be less affected by the death of their grandparents than their parents. The grandchildren may not need to spend much time caring for their ailing grandparents. The death of the grandparents is thus unlikely to be a strong confounder of inheritances in comparison with parental death.

\begin{figure}[htbp]
    \centering
    \copyrightbox[b]{\includegraphics[width=1\textwidth]{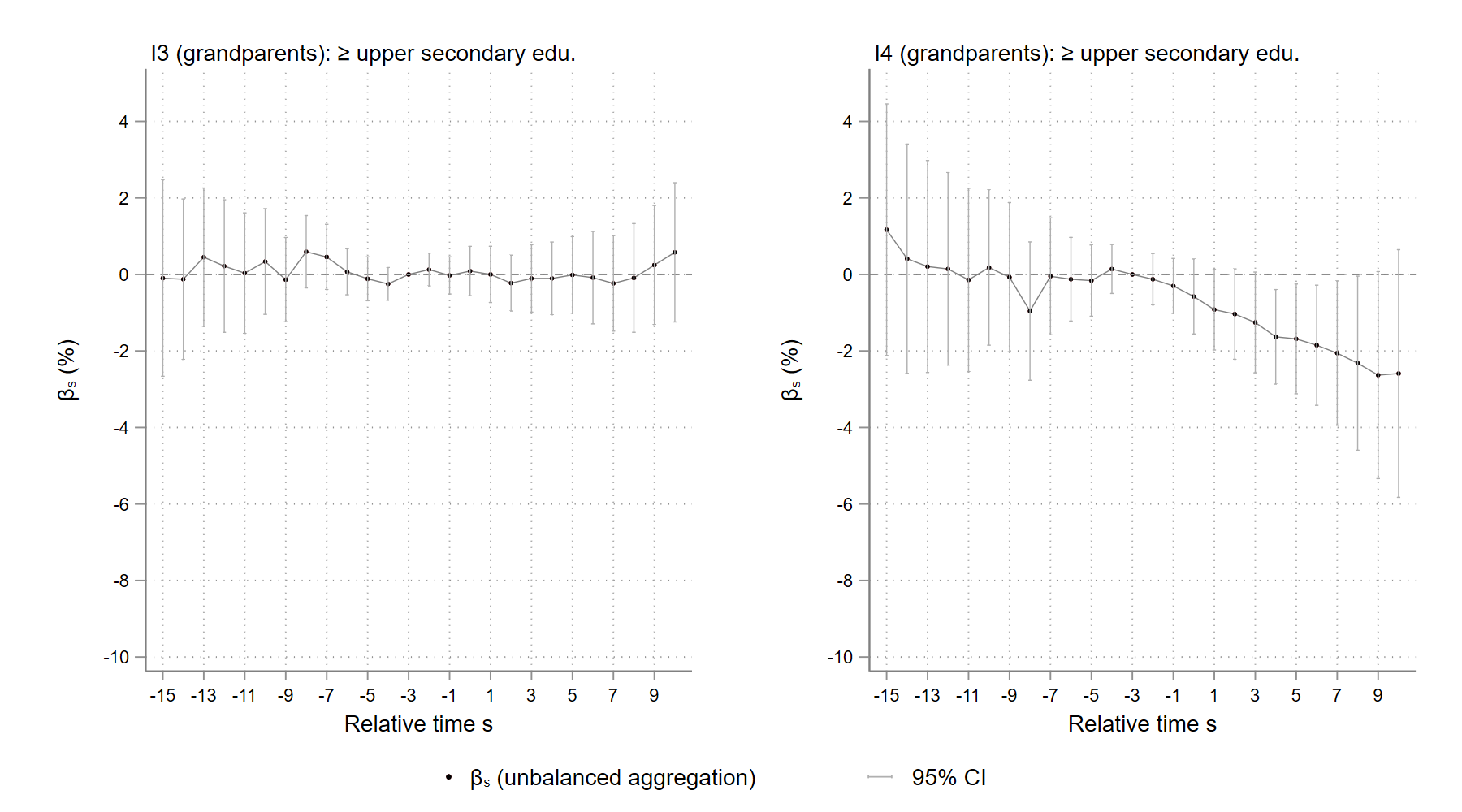}}{\scriptsize Note: $I_3 \in (W^t,2W^t]$, $I_4 \in (2W^t,\infty]$, $W^t$ is the mean annual wage in Norway in year $t$.}
    \caption{$\hat{ATT_s}$ on education: inheritances from grandparents}
    \label{fig:combine2_basic_eduav_p3_gbq3}
\end{figure}

As mentioned previously, heirs inheriting from their grandparents are generally young enough to change their educational level after inheriting. Figure \ref{fig:combine2_basic_eduav_p3_gbq3} depicts the effect of grandparents' inheritances on the probability of inheritors attaining more than an upper secondary education. Inheritances of size I3 seem to have no effect on the level of education of heirs, but when an inheritance exceeds twice the national annual wage, the probability of heirs having a post-secondary degree drops by around 2\%. This finding is consistent with the increased probability of self-employment. Very young heirs may quit school and begin working for themselves after receiving such huge inheritances.

\clearpage
\section{Conclusion}
\label{sec:coclude}

Inheritances have complex effects on the labor supply of heirs. On the one hand, I find empirical evidence of the so-called Carnegie effect. On average, large inheritances from parents reduce inheritors' annual wage income by 2\% to 4.3\%. The magnitude of the adverse effect is proportionate to the size of the bequest, demonstrating that parental death is not the primary cause of the income decline. The impact of inheritances on annual wage income is roughly half of that suggested by existing studies employing Nordic administrative data. On the other hand, inheritances alleviate the liquidity constraints of heirs and can encourage entrepreneurship. I find that inheritances exceeding the average national wage income increase the heir's probability of being self-employed by at least 1\%, which can somewhat compensate for their loss of wage income. Interestingly, inheritances that are more than twice the average national annual wage may, to some extent, dampen entrepreneurship, having a more moderate positive effect on the probability of self-employment.

The discrepancy between my findings and those in the literature is mainly attributable to methodology, rather than sample selection -my sample yields equivalent results when the same methodology is applied (in Appendix Section \ref{asubsec:boe}). The staggered DiD method allows me to avoid the potential selection on unobservable problem and the pitfall of the conventional TWFE model. The 25-year long panel data provides a more convincing pre-trend test and makes it possible to study the dynamics of inheritances' effect on income for more than 20 years, twice as long as previous studies. I find that the negative effect of inheritances on annual wage income can persist for as many as 10 years, which had not been previously identified. The staggered DiD approach also provides a tool for disentangling the cohort heterogeneity and treatment effect dynamics. Estimates based on unbalanced aggregation reveal that younger heirs are more responsive than older heirs, while estimates based on balanced aggregation illustrate how the treatment impact varies over time.

The effects of inheritance on labor supply differ by gender and age. Male inheritors are more likely to reduce their supply of labor than female heirs, whereas the effect on the likelihood of self-employments seems to be gender-neutral. The annual wage income of young heirs also declines more than that of older heirs in the case of inheritances that are more than twice Norway's average annual wage income. But this is partially attributable to the fact that young heirs are more likely to become self-employed after receiving substantial inheritances, and their occupational income is thus not as impacted; the younger the heirs are when they receive their inheritance, the more apparent this becomes. For very young heirs, those who, on average, receive large bequests from grandparents in their early 30s, the effect of inheritances can be so dramatic that their annual wage income decreases by 17\%, and the probability that they will become self-employed rises by 14\%. Large bequests from grandparents also lower the likelihood of young heirs attaining a post-secondary education by 2\%. 

As is the case in the existing literature \citep{Boe2019}, the limitation of my study is that I cannot observe the exact source of the gifts or inheritances directly from the data. Rather, I infer that these are inheritances or advances on inheritances based on the timing of transfers in relation to parental death. This may result in measurement errors in the treatment. The possibility that gifts from living relatives are endogenous undermines my causal argument. Further investigation of the specific source of inheritances is required. 

Parental death is also a potential confounder for inheritances. Intuitively, taking care of sick relatives and grieving the loss of a family member may reduce the labor supply. However, inheritances usually happen simultaneously with parental death, making it difficult to disentangle these two effects. As a result, the magnitude of the Carnegie effect may be overestimated. In accordance with Norwegian inheritance law, as discussed in Section \ref{sec:data}, the majority of inheritances take effect on the death of the last surviving parent. Future research could examine the impact on the labor supply of the loss of a parent where the other parent is still alive.

I evaluate the impact of receiving inheritances of various sizes and conclude that larger inheritances have a stronger impact on annual wage income. It would be interesting for future studies to analyze the effect of inheritances as a continuous function of the value of inheritances; this will allow us to identify the marginal effect of inheritances on labor supply, which is more analogous to the research objects in this field.

\clearpage
\bibliographystyle{aer}
\bibliography{carnegie}

\appendix \clearpage
\section*{Appendix}
\label{sec:append}

\setcounter{section}{0}
\setcounter{subsection}{0}
\setcounter{figure}{0}
\setcounter{table}{0}
\setcounter{equation}{0}
\renewcommand{\thetable}{A\arabic{table}}
\renewcommand{\thefigure}{A\arabic{figure}}
\renewcommand{\thesubsection}{A\arabic{subsection}}

\renewcommand*{\theHsection}{\thesection}
\renewcommand*{\theHsubsection}{\thesubsection}
\renewcommand*{\theHtable}{\thetable}
\renewcommand*{\theHfigure}{\thefigure}

\subsection{Replication of \cite{Boe2019}}
\label{asubsec:boe}

This section replicates the benchmark findings of \cite{Boe2019}. Similar to the literature, I match individuals who received inheritances only once between 2000 and 2004 with non-recipients using nearest-one propensity score matching method. Non-recipients are those who did not receive any inheritances/gifts during 1994-2010, i.e. the 6 years prior to 2000 and the 6 years following 2004. Three years before receiving inheritances is the year of matching. Identical to my baseline analysis, I define inheritances as inheritances/gifts in the tax data within 3 years of parental death. I restrict the sample to individuals who were born between 1944 and 1976 so that they can participate in the labor market over the entire study period $[1994,2010]$. The age distribution of matched large-sum inheritors in Appendix Figure \ref{fig:obs_boe_event_year_age} resembles Figure 1 on page 737 in the study of \cite{Boe2019}. 

\begin{figure}[htbp]
    \centering
    \copyrightbox[b]{\includegraphics[width=0.925\textwidth]{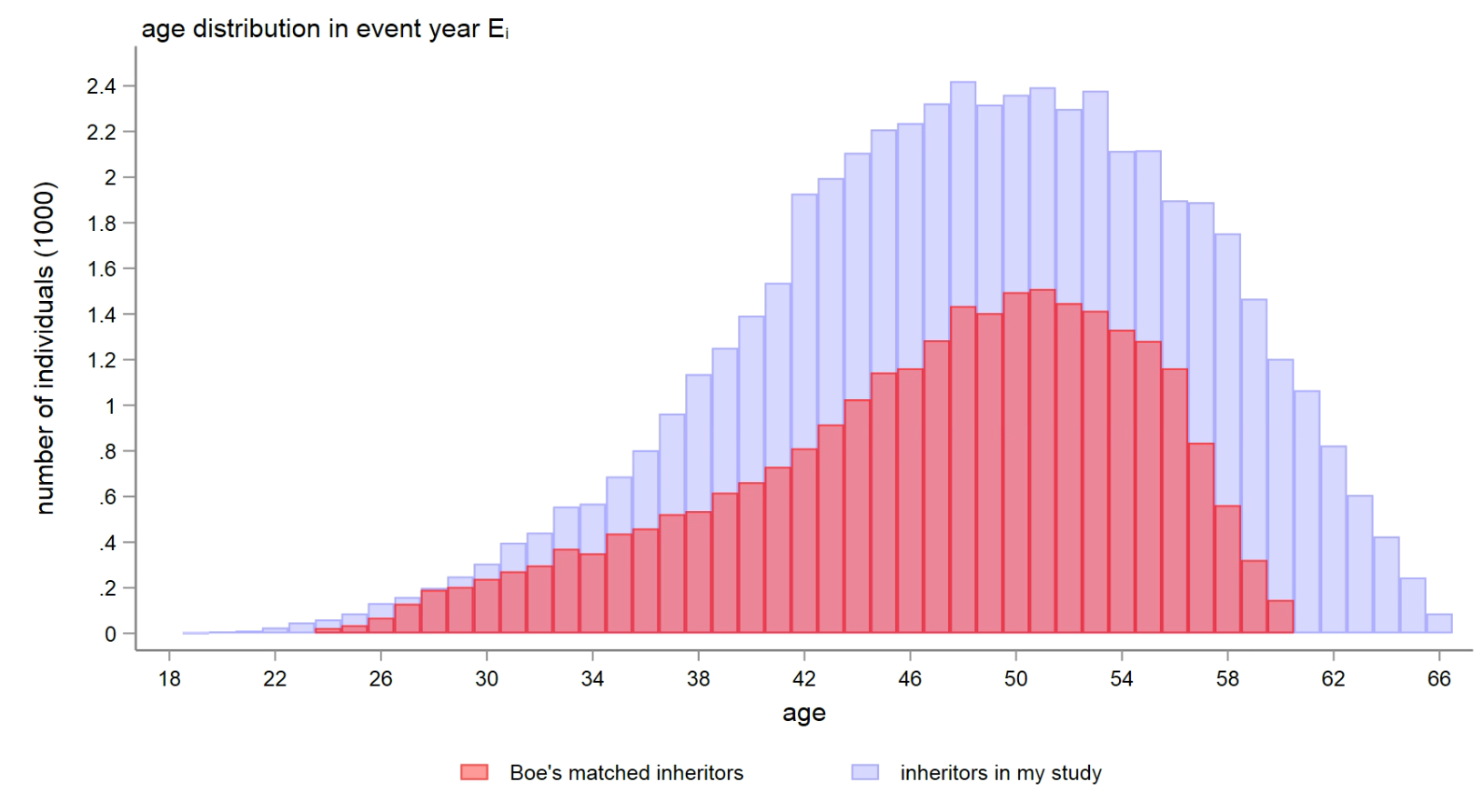}}{\scriptsize Note: ``inheritors in my study'' denotes individuals who inherit $I_3 \in (W^t,2W^t]$ or $I_4 \in (2W^t,\infty]$, $W^t$ is the mean annual wage in Norway in year $t$. }
    \caption{Age distribution of matched large-sum inheritors and I3, I4 inheritors in my study (replication of Boe et al. (2019) Figure 1, pp737.)}
    \label{fig:obs_boe_event_year_age}
\end{figure}

\begin{table}[htbp]\centering
\begin{threeparttable}

\def\sym#1{\ifmmode^{#1}\else\(^{#1}\)\fi}
\caption{Effect of inheritance on wage income (replication of Bø et al. table 3 pp741.)\label{tab:match_boe} \vspace{-5pt}}

\begin{tabular}{l*{6}{c}}
\toprule
                     &\multicolumn{2}{c}{All inheritors}                    &&                \multicolumn{2}{c}{Large-sums}  \\
                           \cmidrule{2-3}                                                        \cmidrule{5-6} 
\multicolumn{1}{c}{Year t}&\multicolumn{1}{c}{Est.}&\multicolumn{1}{c}{SE}  &&   \multicolumn{1}{c}{Est.}&\multicolumn{1}{c}{SE} \\
\midrule
\multicolumn{6}{l}{A. all non-inheritors to be matched} \\
\multicolumn{1}{c}{-6} &0.0618\sym{**}	& 0.0206 && -0.0050      & 0.0328  \\
\multicolumn{1}{c}{-5} &0.0484\sym{*}	& 0.0205 && 0.0079       & 0.0327   \\
\multicolumn{1}{c}{-4} &0.0130	    & 0.0205 && -0.0161      & 0.0329   \\
\multicolumn{1}{c}{-3} &-0.0044	    & 0.0203 && 0.0003       & 0.0332   \\
\multicolumn{1}{c}{-2} &0.0168	    & 0.0207 && -0.0206      & 0.0336   \\
\multicolumn{1}{c}{-1} &0.0172	    & 0.0212 && -0.0148      & 0.0345  \\
\multicolumn{1}{c}{0 } &-0.0010	    & 0.0218 && -0.0619	     & 0.0356  \\
\multicolumn{1}{c}{+1} &0.0007	    & 0.0225 && -0.0951\sym{**} & 0.0367  \\
\multicolumn{1}{c}{+2} &0.0142	    & 0.0230 && -0.0664      & 0.0374  \\
\multicolumn{1}{c}{+3} &0.0128	    & 0.0234 && -0.0779\sym{*}	 & 0.0381  \\
\multicolumn{1}{c}{+4} &0.0275	    & 0.0239 && -0.0680	     & 0.0386  \\
\multicolumn{1}{c}{+5} &0.0272	    & 0.0244 && -0.0700	     & 0.0396  \\
\multicolumn{1}{c}{+6} &0.0157	    & 0.0250 && -0.1070\sym{**} & 0.0406  \\
Matches  &\multicolumn{2}{c}{73,108} && \multicolumn{2}{c}{26,845} \\

\midrule
\multicolumn{6}{l}{B. non-inheritors with parental death to be matched} \\
\multicolumn{1}{c}{-6} &-0.0040	&  0.0203 &&   0.0003      & 0.0329  \\
\multicolumn{1}{c}{-5} &-0.0123	&  0.0202 &&  -0.0061      & 0.0326   \\
\multicolumn{1}{c}{-4} &-0.0329	&  0.0203 &&  -0.0325      & 0.0328   \\
\multicolumn{1}{c}{-3} &-0.0380	&  0.0202 &&  -0.0222      & 0.0330   \\
\multicolumn{1}{c}{-2} &-0.0050	&  0.0205 &&  -0.0385      & 0.0335   \\
\multicolumn{1}{c}{-1} &-0.0116	&  0.0210 &&  -0.0522      & 0.0342  \\
\multicolumn{1}{c}{0 } &-0.0316	&  0.0216 &&  -0.0935\sym{**} & 0.0353  \\
\multicolumn{1}{c}{+1} &-0.0345	&  0.8223 &&  -0.0997\sym{**} & 0.0366  \\
\multicolumn{1}{c}{+2} &-0.0093	&  0.8229 &&  -0.0555      & 0.0374  \\
\multicolumn{1}{c}{+3} &-0.0109	&  0.0233 &&  -0.0641	   & 0.0380  \\
\multicolumn{1}{c}{+4} &-0.0155	&  0.0237 &&  -0.0446	   & 0.0387  \\
\multicolumn{1}{c}{+5} &-0.0115	&  0.0241 &&  -0.0703	   & 0.0395  \\
\multicolumn{1}{c}{+6} &-0.0076	&  0.0249 &&  -0.0840\sym{*} & 0.0407  \\
Matches  &\multicolumn{2}{c}{73,378} && \multicolumn{2}{c}{26,843} \\

\bottomrule

\end{tabular}
\begin{tablenotes}
      \small
      \item Note: (1) Each row is the wage difference in year $t \in [-6,6]$, where $t=0$ is the year of inheritance. (2) Large-sums are those who inherited more than 30,000 NOK (deflated by CPI in 2015 prices) (3) The dependent variable is inverse hyperbolic sine (IHS) transformed annual wage in year t, $ihs(wage_t)$. (4) *** $p < 0.001$. ** $p<0.01$. * $p<0.05$. (5) Matches are the number of matched pairs using nearest-one propensity score mathcing method.
     \end{tablenotes}
\end{threeparttable}

\end{table}

I then compare the annual wage income of matched inheritors and non-inheritors for the time interval $s\in [-6,6]$, where year $s=0$ is the year of receipt. Appendix Table \ref{tab:match_boe} is a replication of table 3 on page 741 in the research of \cite{Boe2019}. In panel A of Appendix Table \ref{tab:match_boe}, all non-inheritors are in the ``pool of non-treated'' to be matched with inheritors; while in panel B, only non-inheritors whose parents died during the same period as inheritors matched are in the non-treated pool. The results in panels A and B resemble the literature. This indicates that the difference between these results and those obtained by using the staggered DiD method to the same inheritor sample in Section \ref{subsec:birth} is due to the methodology, not the sample selection. It appears that matching method tend to over estimate the effect of inheritances on annual wage income.

It is important to note that the sample of inheritors in Table \ref{tab:match_boe} is around half that of the literature. This may be because I restrict inheritances to occur close to parental death, but in the literature, researchers can observe whether a transfer is an inheritance or not (but they do not know the source of inheritances, either). There may be inheritances and advances covered by the literature that were not included from my sample.

\clearpage
\subsection{Staggered DiD vs. TWFE}
\label{asubsec:twfe}

When treatment is staggered, the conventional dynamic TWFE model confronts three challenges. Firstly, like the age–period–cohort conundrum, indicators for cohort
, calendar year and relative event time (calendar year minus event year) are perfectly co-linear in a dynamic TWFE model. At least one additional period must be omitted to break the co-linearity (i.e. at least two periods omitted in total), which imposes stronger assumptions on the omitted periods. Secondly, the specification of a dynamic TWFE model includes both the not-yet-treated cohorts and the already treated cohorts in the regression as the control group. The comparison between a treatment group and the already treated cohorts is called a ``forbidden comparison'' in the literature. Lastly, the conventional dynamic TWFE model assumes a homogeneous treatment effect among cohorts, which may not hold in this study since different cohorts have difference age compositions. When the dynamic treatment effect is heterogeneous among cohorts, the aforementioned ``forbidden comparison'' makes the conventional TWFE estimators biased \citep{Borusyak2021}.
\footnote{Denote the average treatment effect in cohort $g$ and year $t$ as $\tau_{gt}$. As illustrated by \cite{Borusyak2021}, \cite{GoodmanBacon2021} and \cite{Sun2020}, under the parallel trend assumption, conventional TWFE model estimates the weighted summation of $\tau_{gt}$. When the treatment effect is heterogeneous among cohorts, the ``forbidden comparison'' 
generates inexplicable(e.g. negative) weight to $\tau_{gt}$.}. When the TWFE estimators are biased, the pre-trend test based on the estimators is also invalid \citep{Sun2020}

\begin{figure}[htbp]
    \centering
    \copyrightbox[b]{\includegraphics[width=1\textwidth]{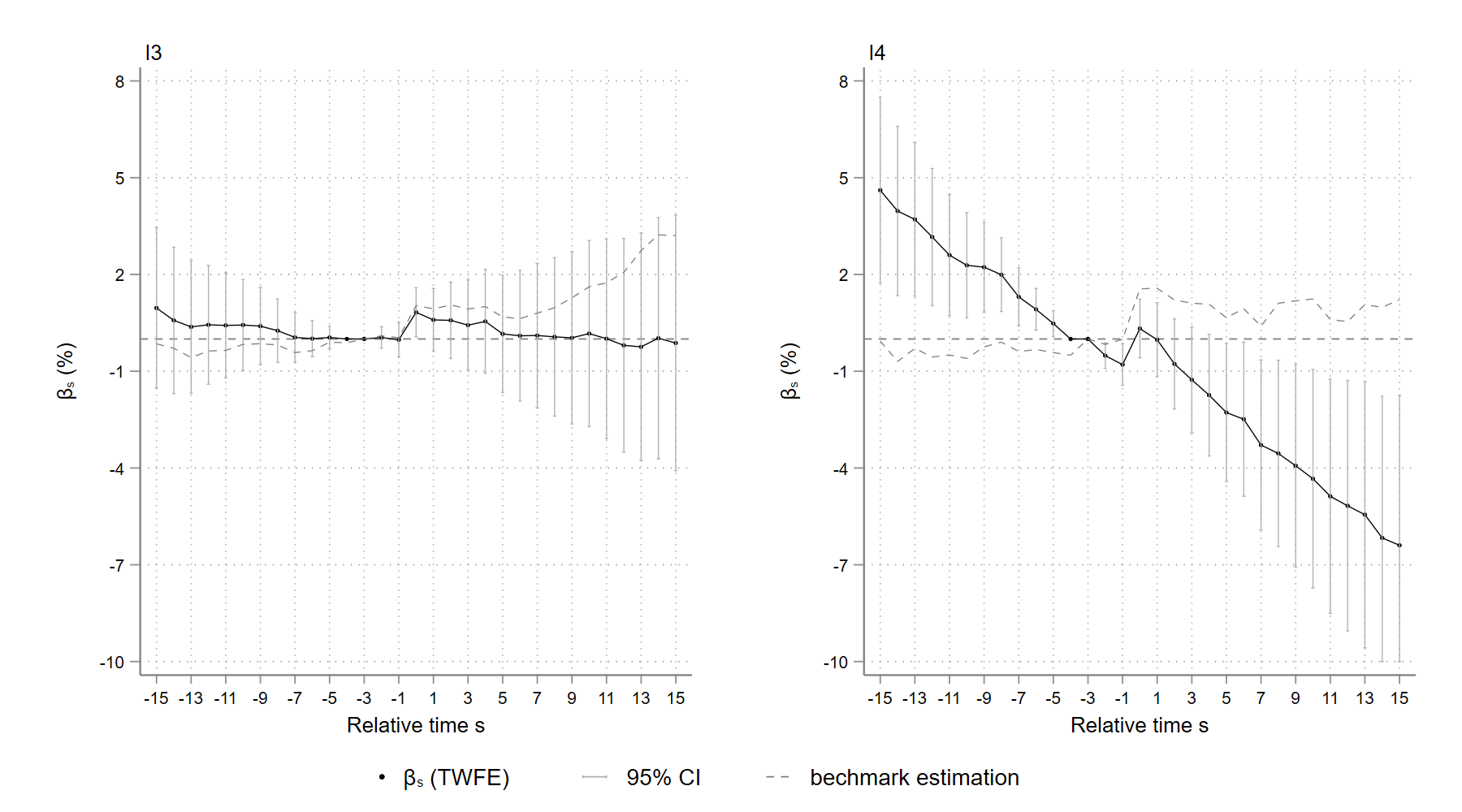}}{\scriptsize Note: $I_4 \in (2W^t,\infty]$, $W^t$ is the mean annual wage in Norway in year $t$.}
    \caption{$\tau_{s}$ of size $I4$ on Pr(self-employment) estimated by TWFE model}
    \label{fig:combine2_twfe_etpn_p3_bq3}
\end{figure}

Figure \ref{fig:combine2_twfe_etpn_p3_bq3} is an example of estimating the effect of inheritances using traditional TWFE model with year $s=-3$ and $-4$ as reference periods. Due to the aforementioned problems, TWFE estimates are invalid and curve in the right part of Figure \ref{fig:combine2_twfe_etpn_p3_bq3} shows a very clear pre-trend.

In contrast, the staggered DiD method uses only not-yet-treated cohorts as controls,

\begin{equation}
\begin{split}
\beta_{gt} &= E[Y^1_{g,t} - Y^0_{g,g-3} |age_t, G_{g} = 1] - E[Y^0_{g',t} - Y^0_{g',g-3} |age_t, D_{t} = 0] 
\end{split}
    \label{eq:did1} 
\end{equation}

Where $Y^1$ and $Y^0$ are the potential earnings if one receives inheritances or not. $g'$ represents the not-yet-treated cohorts. $t=g-3$ is 3 years ``before the treatment'' used as the reference period.

Under the conditional parallel trend assumption \ref{eq:para}, $\beta_{gt}$ estimates the cohort-year specific average treatment effect $\tau_{gt}$ (ATT):

\begin{equation}
\begin{split}
\beta_{gt} &= E[Y^1_{g,t} - Y^0_{g,g-3} |age_t, G_{g} = 1] - E[Y^0_{g',t} - Y^0_{g',g-3} |age_t, D_{t} = 0] \\
&=E[Y^1_{g,t} - Y^0_{g,g-3} |age_t, G_{g} = 1] - E[Y^0_{g,t} - Y^0_{g,g-3} |age_t, G_{g} = 1] \\
&=E[Y^1_{g,t} - Y^0_{g,t}| age_t, G_{g} = 1] \\
&=\tau_{gt}
\end{split}
    \label{eq:did_att} 
\end{equation}

The aggregation weights in scheme \ref{eq:agg} are all positive, which avoids the negative weights in conventional TWFE model \citep{GoodmanBacon2021,Callaway2020}.

\clearpage

\subsection{heatmaps of $\beta_{gs}$ in benchmark analysis}
\label{asubsec:heat}

\begin{figure}[htbp]
    \centering
    \copyrightbox[b]{\includegraphics[width=0.92\textwidth]{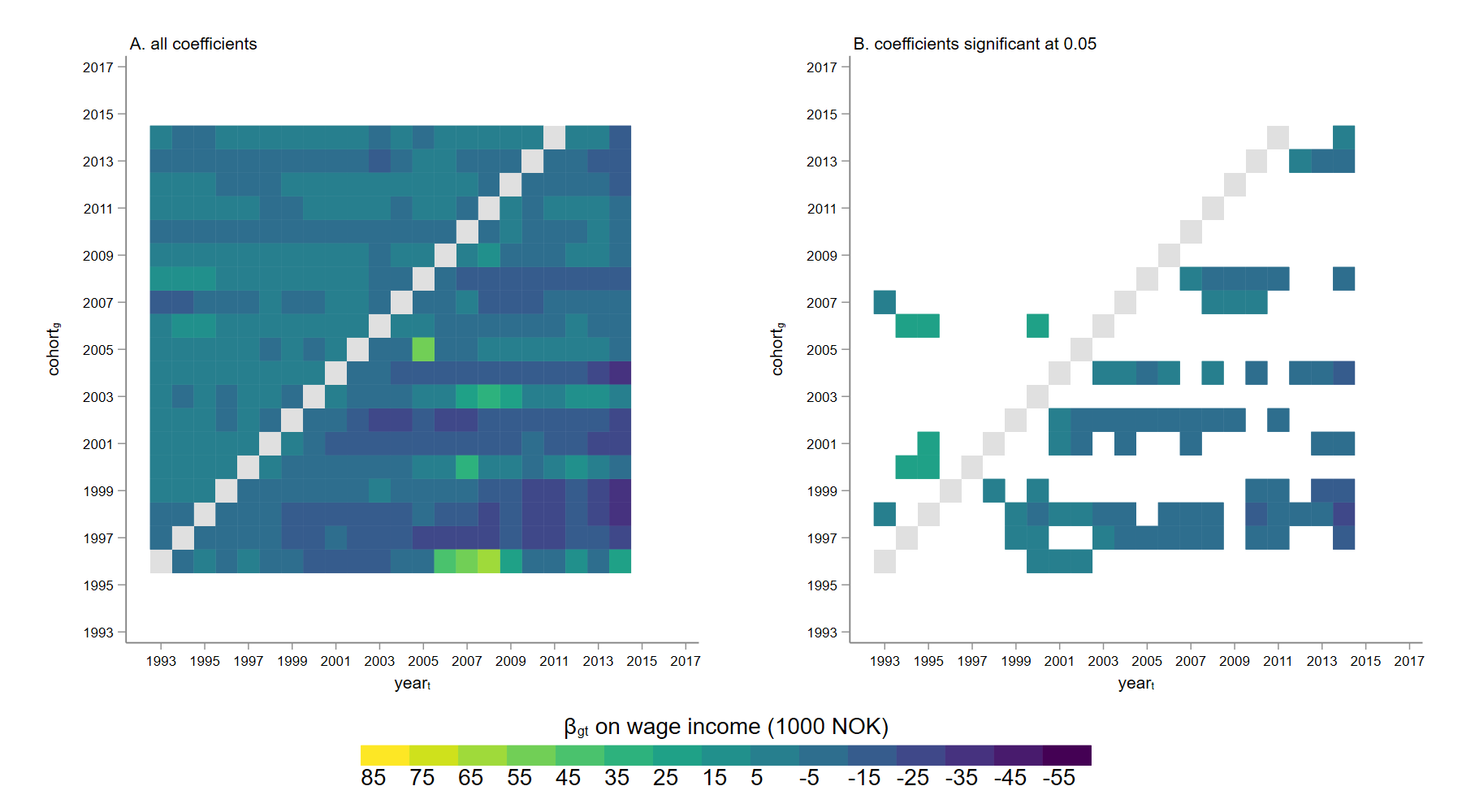}}{\scriptsize Note: $I_3 \in (W^t,2W^t]$, $W^t$ is the mean annual wage in Norway in year $t$.}
    \caption{$\hat{ATT_{gs}}$ of size $I3$ inheritances on annual wage}
    \label{fig:heat_I3_attgt_ratio_p3_bq3_wage}
\end{figure}
\begin{figure}[htbp]
    \centering
    \copyrightbox[b]{\includegraphics[width=0.92\textwidth]{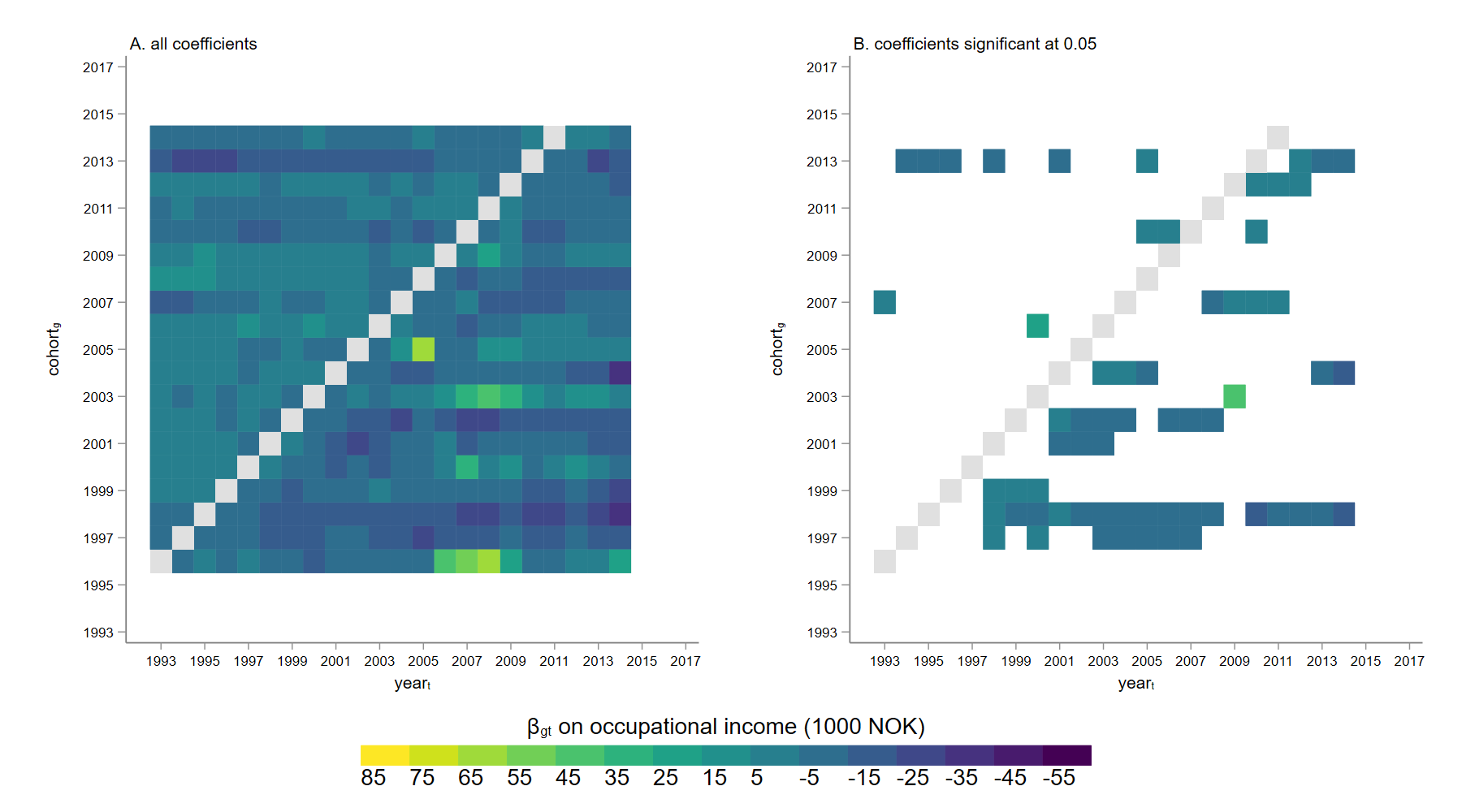}}{\scriptsize Note: $I_3 \in (W^t,2W^t]$, $W^t$ is the mean annual wage in Norway in year $t$.}
    \caption{$\hat{ATT_{gs}}$ of size $I3$ inheritances on annual occupational income}
    \label{fig:heat_I3_attgt_ratio_p3_bq3_yrkinnt}
\end{figure}
\begin{figure}[htbp]
    \centering
    \copyrightbox[b]{\includegraphics[width=0.925\textwidth]{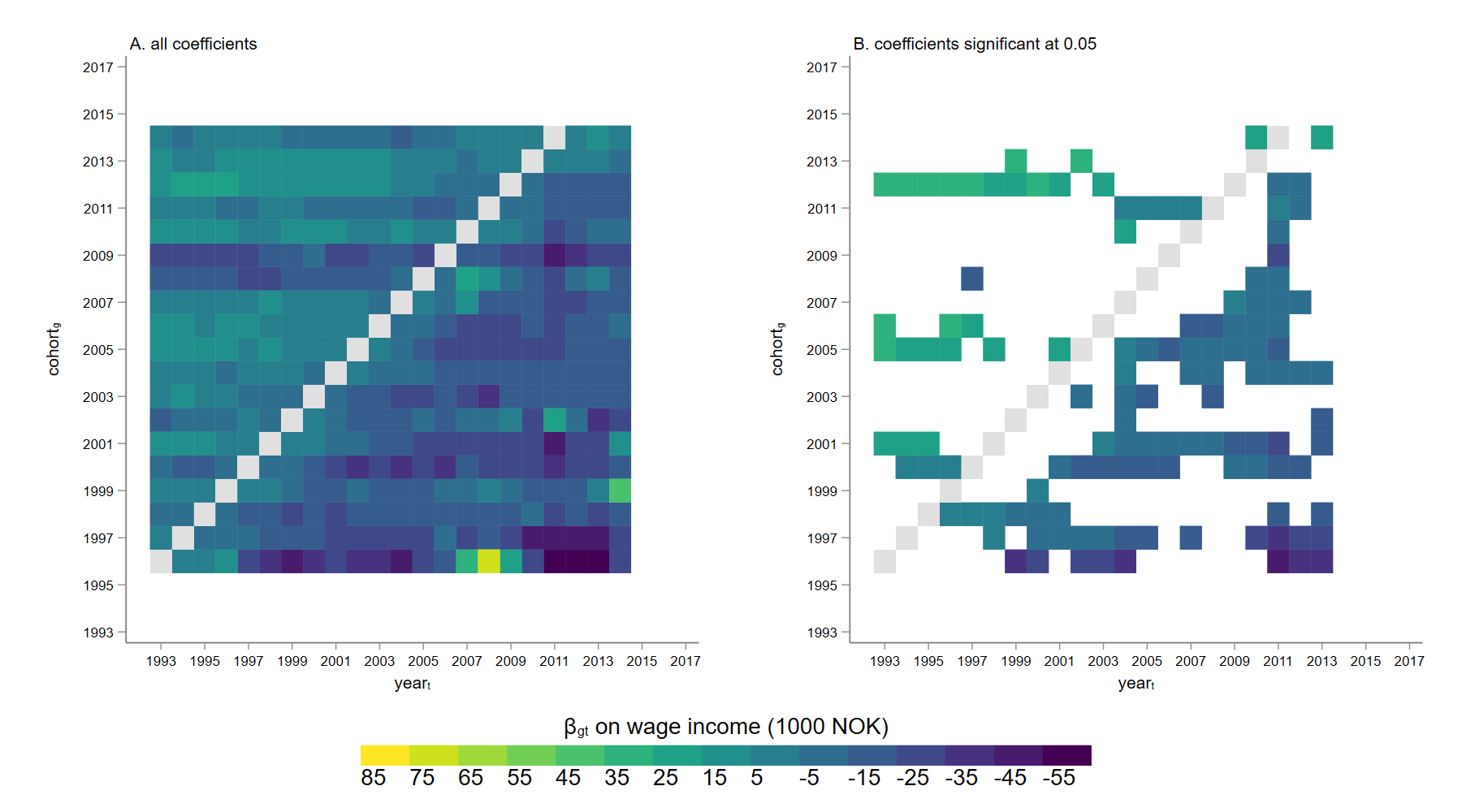}}{\scriptsize Note: $I_4 \in (2W^t,\infty]$, $W^t$ is the mean annual wage in Norway in year $t$.}
    \caption{$\hat{ATT_{gs}}$ of size $I4$ inheritances on annual wage}
    \label{fig:heat_I4_attgt_ratio_p3_bq3_wage}
\end{figure}
\begin{figure}[htbp]
    \centering
    \copyrightbox[b]{\includegraphics[width=0.925\textwidth]{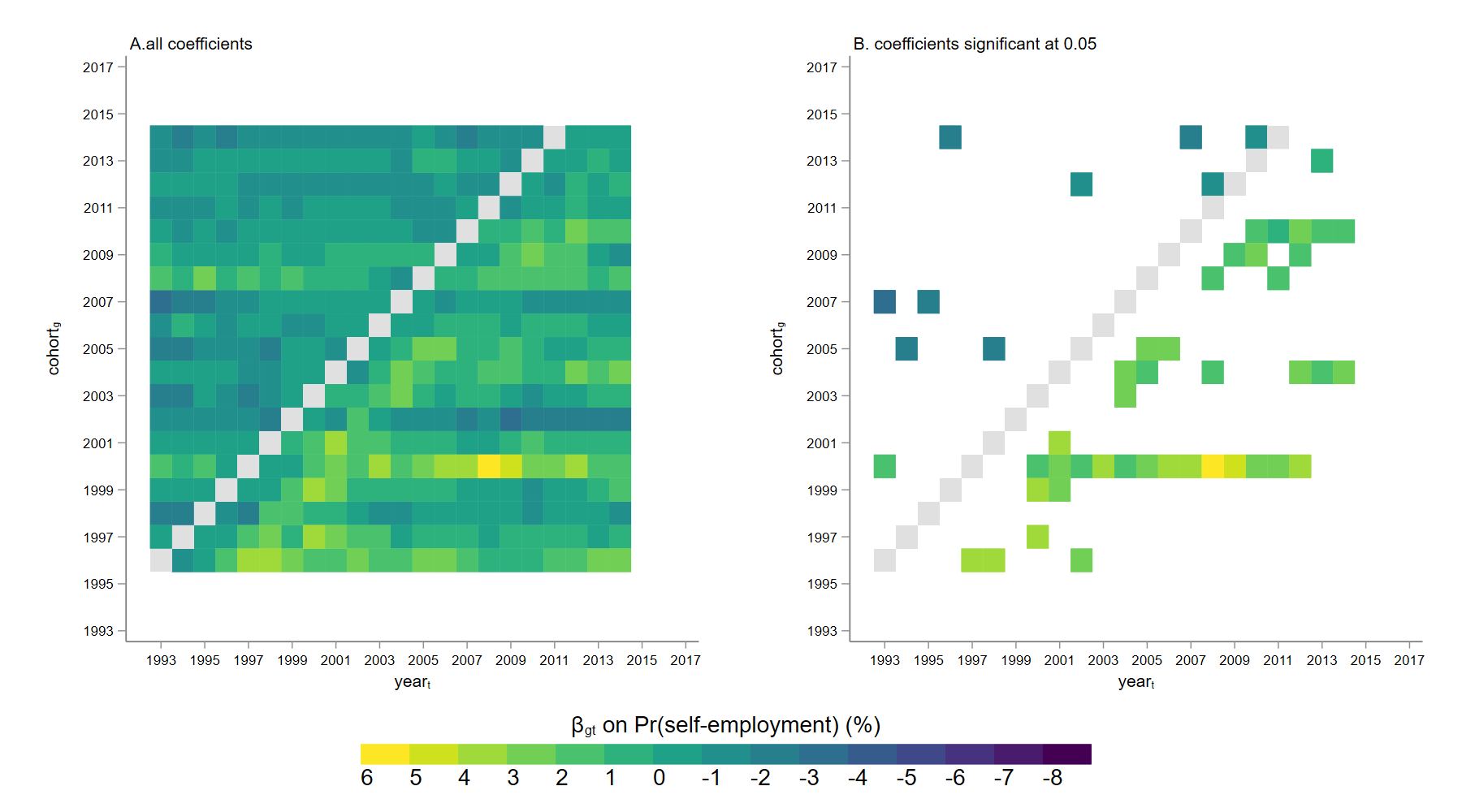}}{\scriptsize Note: $I_4 \in (W^t,2W^t]$, $W^t$ is the mean annual wage in Norway in year $t$.}
    \caption{$\hat{ATT_{gs}}$ of size $I4$ inheritances on Pr(self-employment)}
    \label{fig:heat_I4_attgt_ratio_p3_bq3_etpn}
\end{figure}
\begin{figure}[htbp]
    \centering
    \copyrightbox[b]{\includegraphics[width=0.925\textwidth]{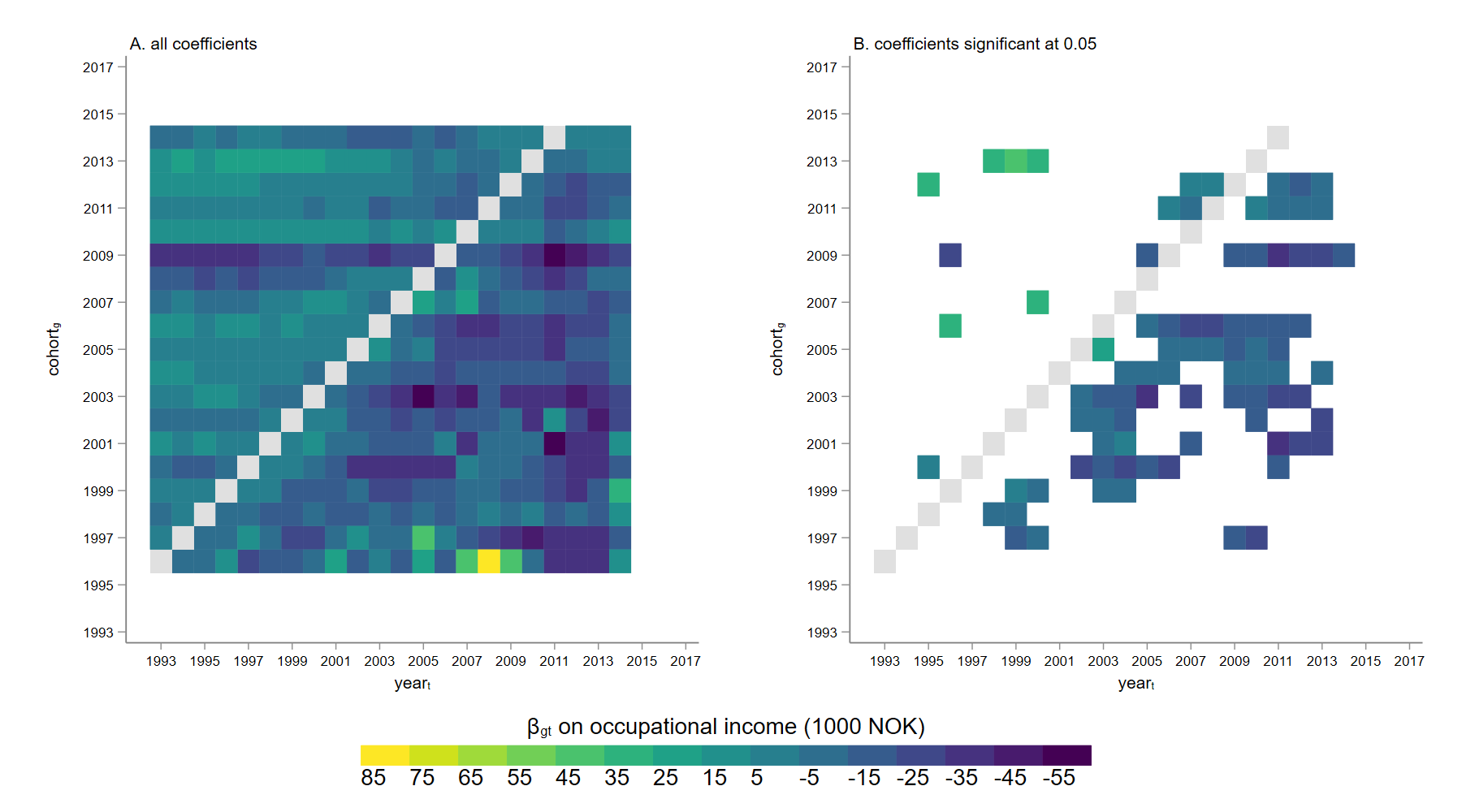}}{\scriptsize Note: $I_4 \in (2W^t,\infty]$, $W^t$ is the mean annual wage in Norway in year $t$.}
    \caption{$\hat{ATT_{gs}}$ of size $I4$ inheritances on annual occupational income}
    \label{fig:heat_I4_attgt_ratio_p3_bq3_yrkinnt}
\end{figure}

\clearpage
\subsection{Baseline estimates for the effects of small inheritances I1 and I2}
\label{asubsec:small}

\begin{figure}[htbp]
    \centering
    \copyrightbox[b]{\includegraphics[width=0.92\textwidth]{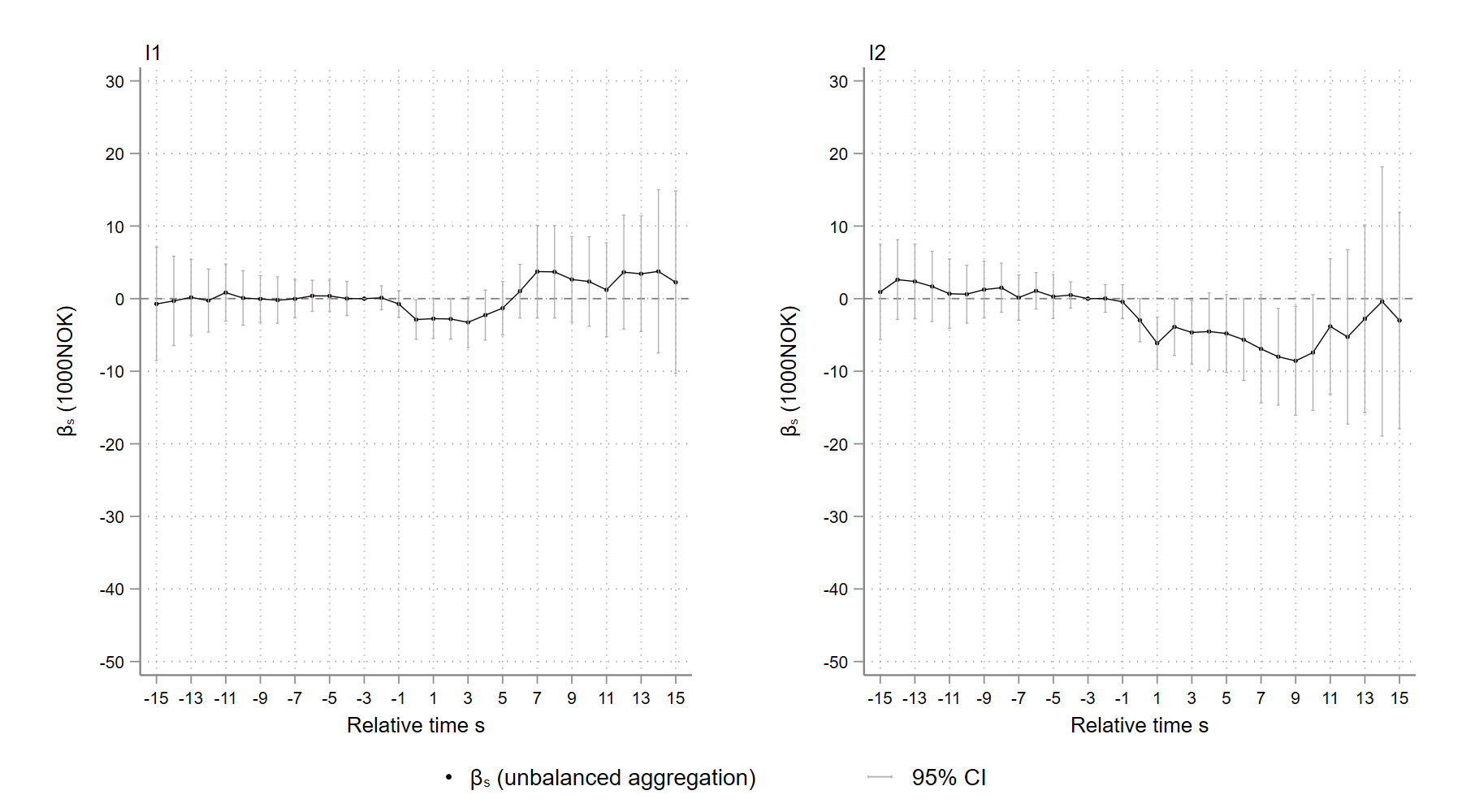}}{\scriptsize Note: $I_1 \in (0,\tfrac{1}{2}W^t], I_2 \in (\tfrac{1}{2}W^t,W^t]$, $W^t$ is the mean annual wage in Norway in year $t$.}
    \caption{$\hat{ATT_s}$ on annual wage income}
    \label{fig:combine2_basic_I1I2_wage_p3_bq3}
\end{figure}
\begin{figure}[htbp]
    \centering
    \copyrightbox[b]{\includegraphics[width=0.92\textwidth]{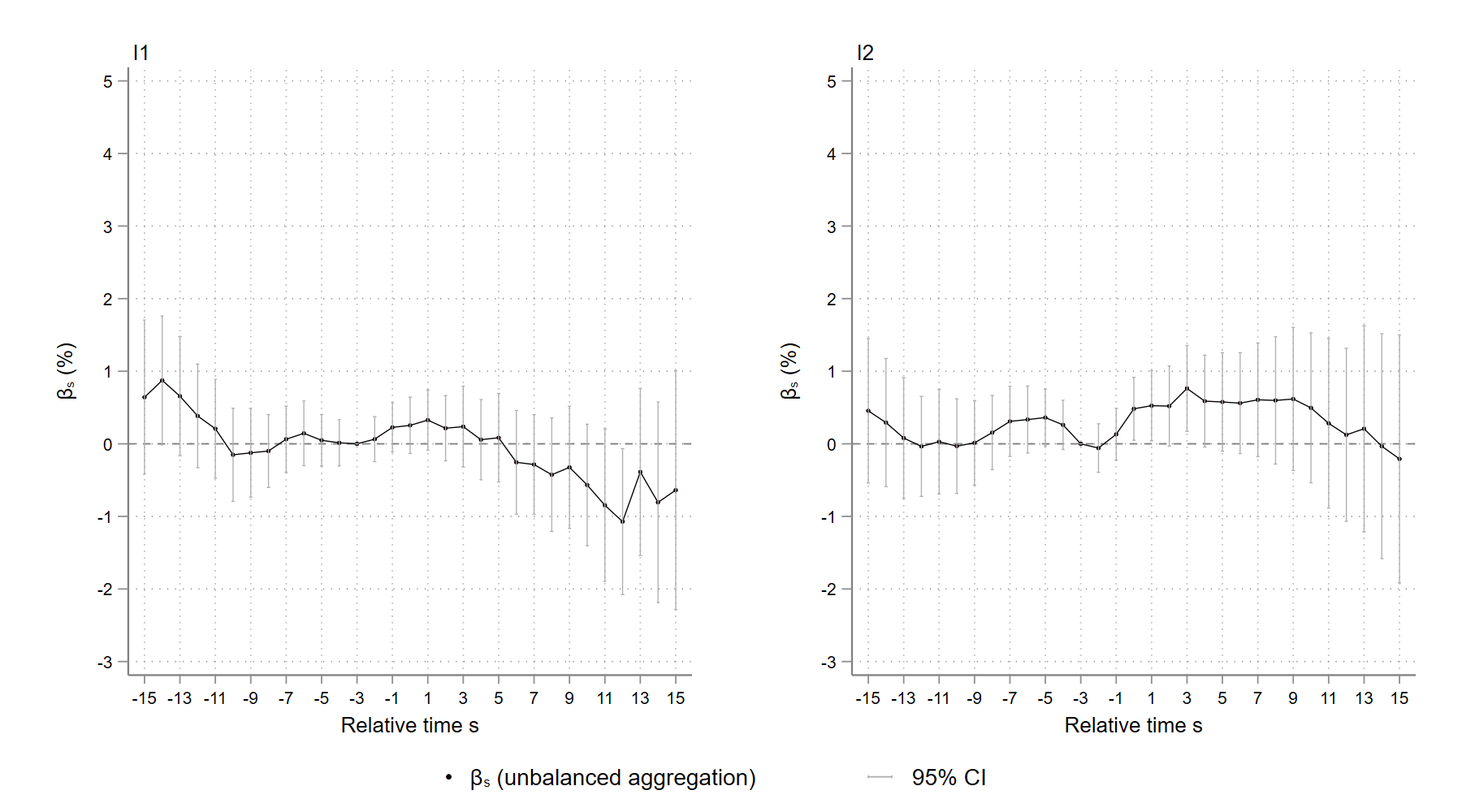}}{\scriptsize Note: $I_1 \in (0,\tfrac{1}{2}W^t], I_2 \in (\tfrac{1}{2}W^t,W^t]$, $W^t$ is the mean annual wage in Norway in year $t$.}
    \caption{$\hat{ATT_s}$ on Pr(self-employment)}
    \label{fig:combine2_basic_I1I2_etpn_p3_bq3}
\end{figure}
\begin{figure}[htbp]
    \centering
    \copyrightbox[b]{\includegraphics[width=0.925\textwidth]{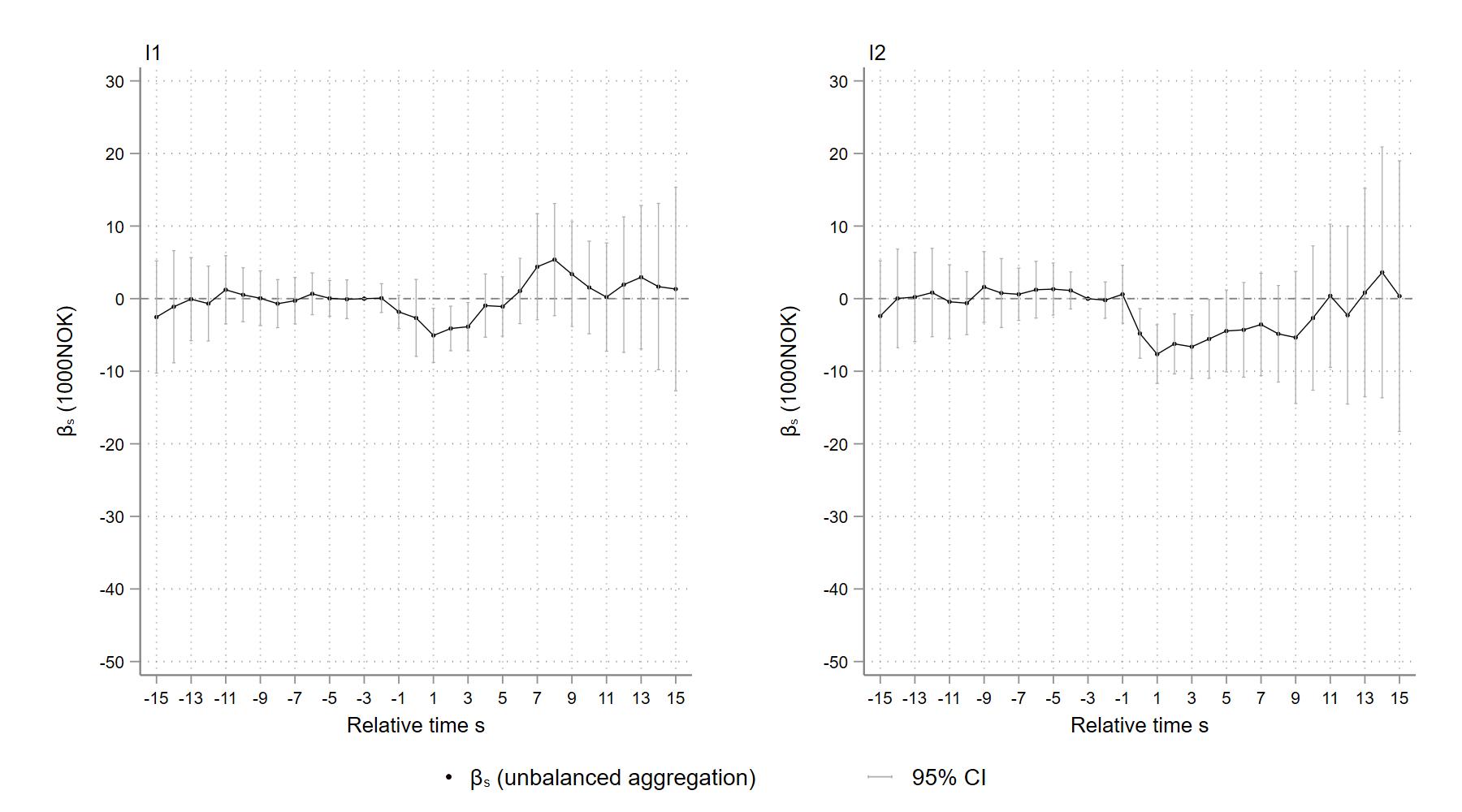}}{\scriptsize Note: $I_1 \in (0,\tfrac{1}{2}W^t], I_2 \in (\tfrac{1}{2}W^t,W^t]$, $W^t$ is the mean annual wage in Norway in year $t$.}
    \caption{$\hat{ATT_s}$ on annual occupational income}
    \label{fig:combine2_basic_I1I2_yrkinnt_p3_bq3}
\end{figure}

\clearpage
\subsection{Appendix for robustness checks}
\label{asubsec:robust}

\subsubsection{Parental death timing}
\label{asubsubsec:pd}

\begin{figure}[htbp]
    \centering
    \copyrightbox[b]{\includegraphics[width=0.925\textwidth]{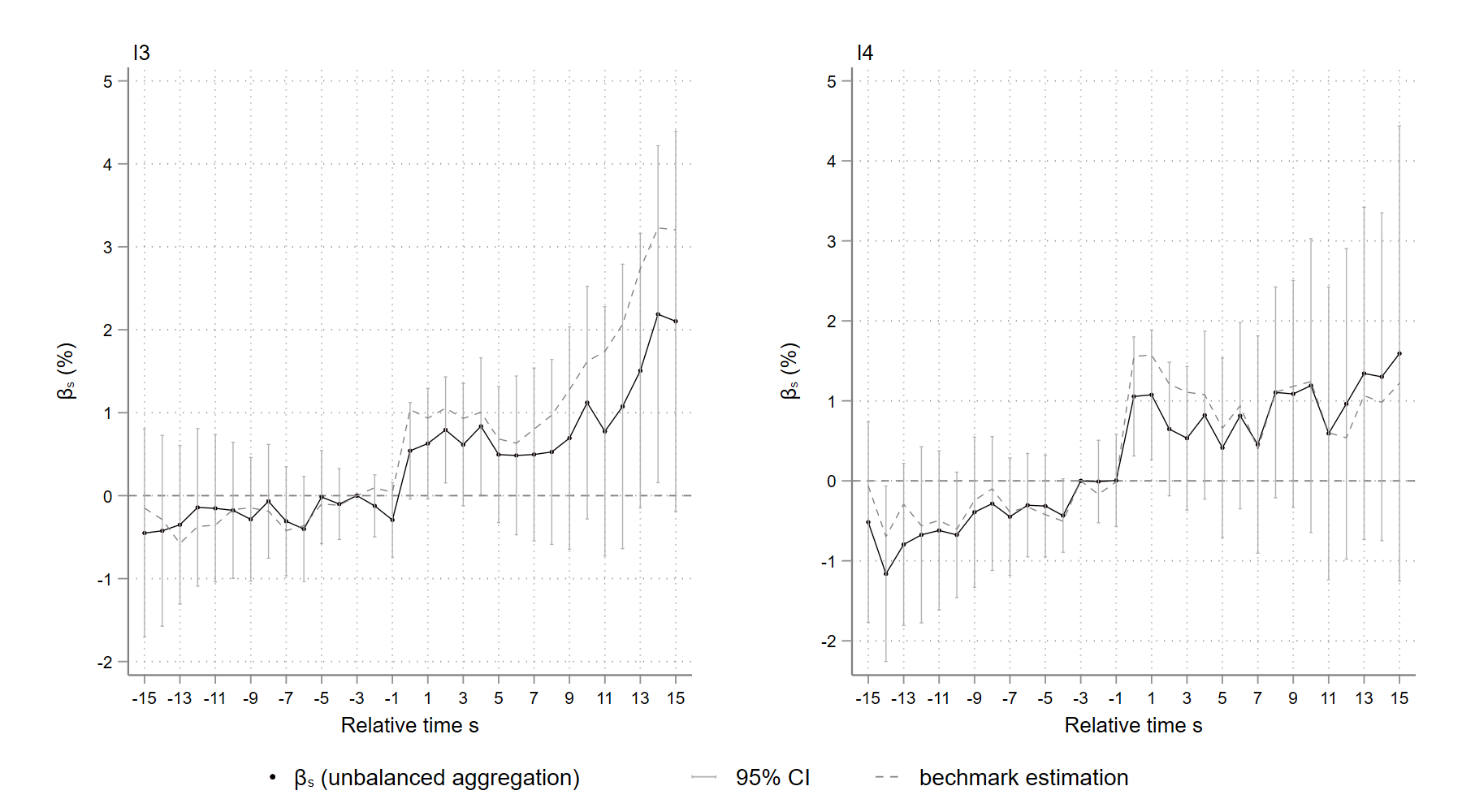}}{\scriptsize Note: $I_3 \in (W^t,2W^t], I_4 \in (2W^t,\infty]$, $W^t$ is the mean annual wage in Norway in year $t$.}
    \caption{$ATT_s$ on Pr(self-employment): inheritances in the year of parental death}
    \label{fig:combine2_basic_etpn_p3_bq0}
\end{figure}
\begin{figure}[htbp]
    \centering
    \copyrightbox[b]{\includegraphics[width=0.925\textwidth]{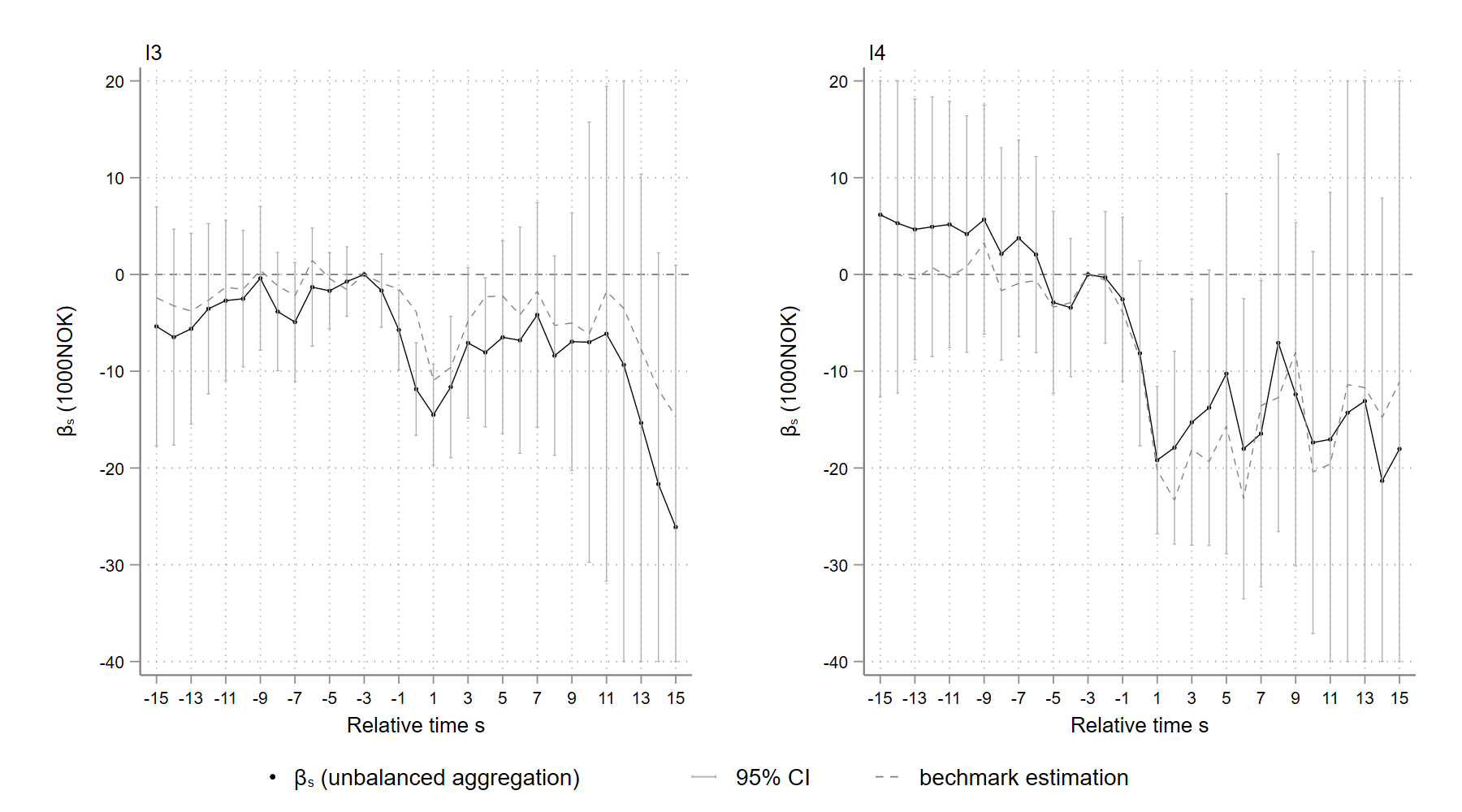}}{\scriptsize Note: $I_3 \in (W^t,2W^t], I_4 \in (2W^t,\infty]$, $W^t$ is the mean annual wage in Norway in year $t$.}
    \caption{$ATT_s$ on annual occupational income: inheritances in the year of parental death}
    \label{fig:combine2_basic_yrkinnt_p3_bq0}
\end{figure}
\begin{figure}[htbp]
    \centering
    \copyrightbox[b]{\includegraphics[width=0.925\textwidth]{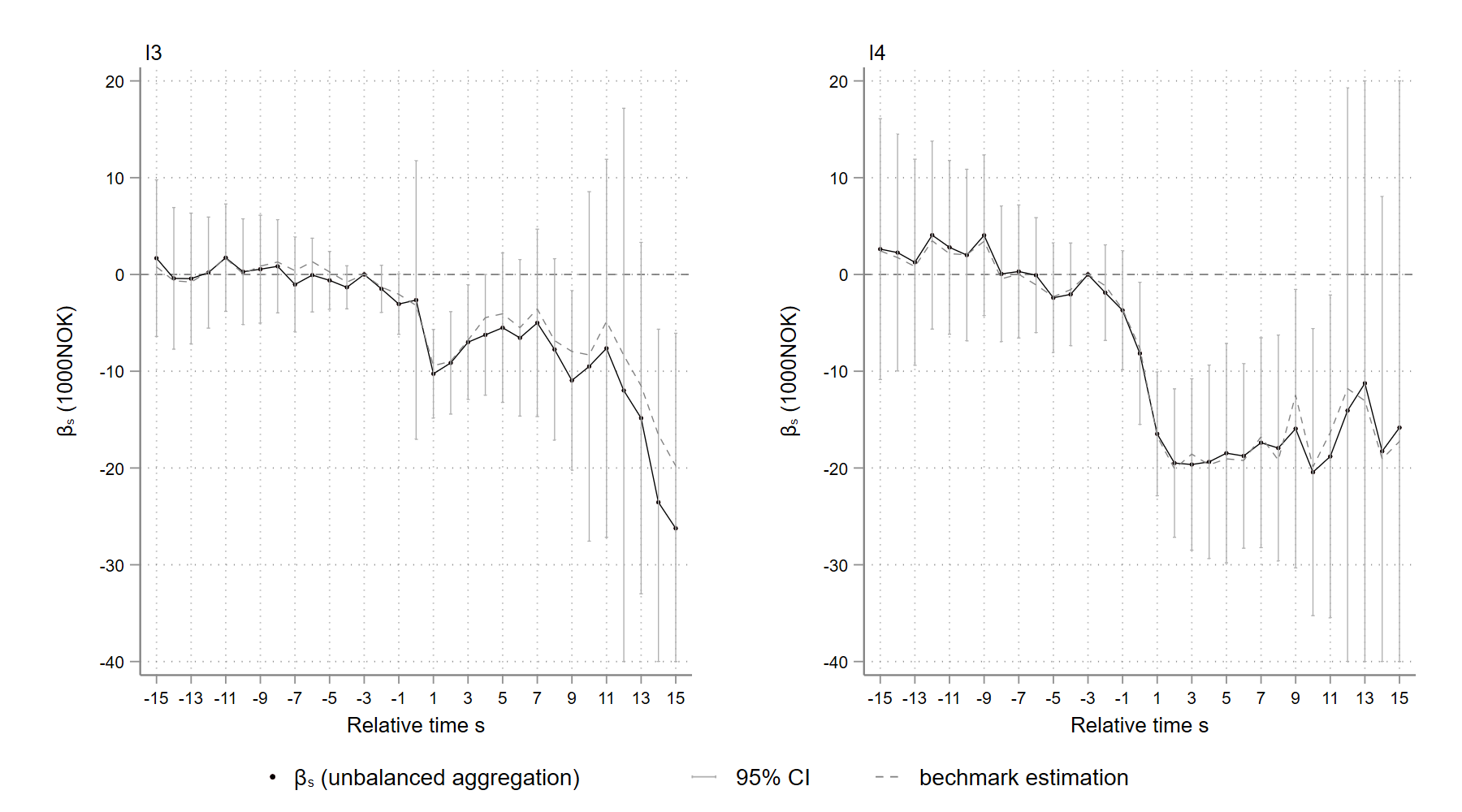}}{\scriptsize Note: $I_3 \in (W^t,2W^t], I_4 \in (2W^t,\infty]$, $W^t$ is the mean annual wage in Norway in year $t$.}
    \caption{$ATT_s$ on annual wage income: inheritances 1-year around parental death}
    \label{fig:combine2_basic_wage_p3_bq1}
\end{figure}
\begin{figure}[htbp]
    \centering
    \copyrightbox[b]{\includegraphics[width=0.925\textwidth]{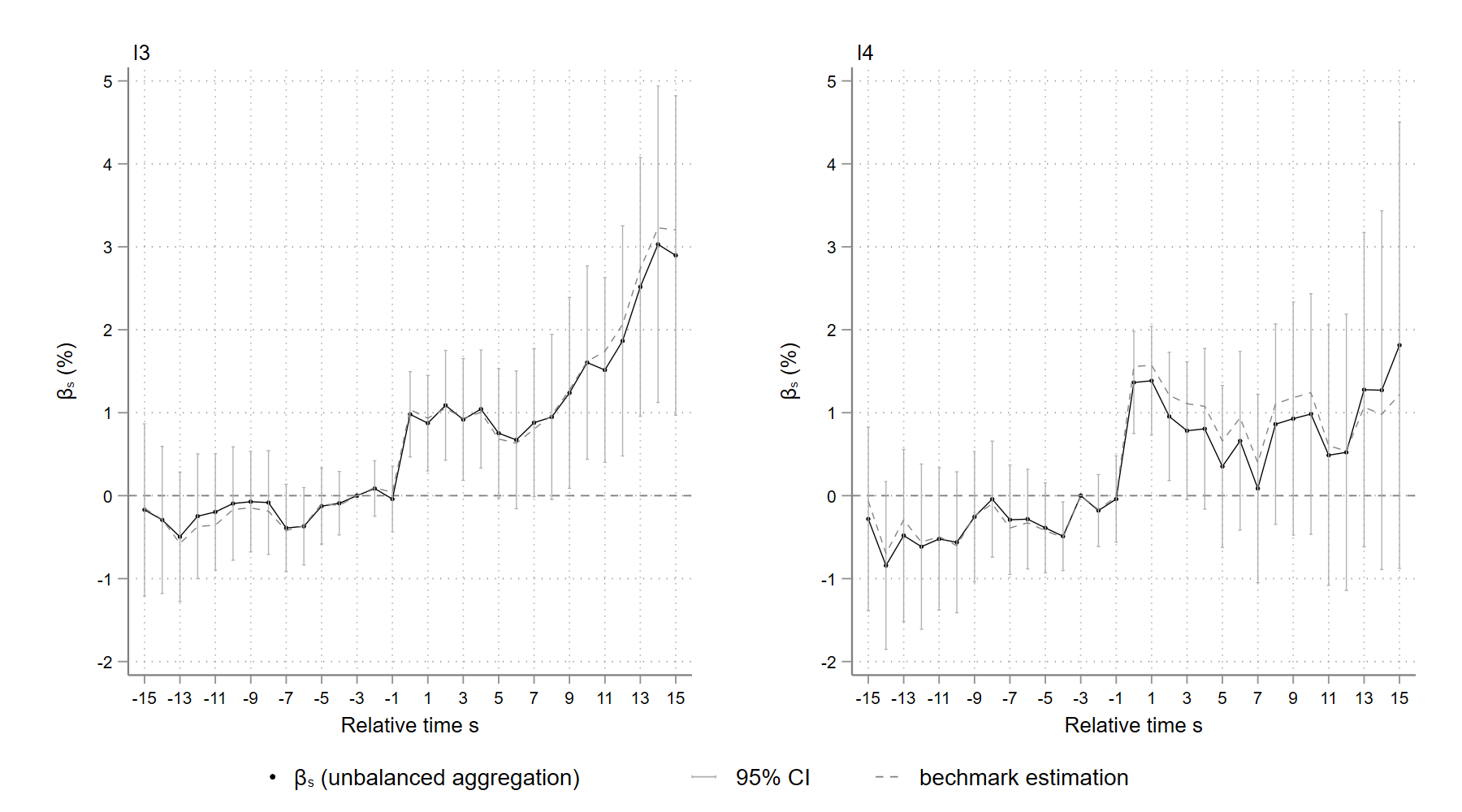}}{\scriptsize Note: $I_3 \in (W^t,2W^t], I_4 \in (2W^t,\infty]$, $W^t$ is the mean annual wage in Norway in year $t$.}
    \caption{$ATT_s$ on Pr(self-employment): inheritances 1-year around parental death}
    \label{fig:combine2_basic_etpn_p3_bq1}
\end{figure}
\begin{figure}[htbp]
    \centering
    \copyrightbox[b]{\includegraphics[width=0.925\textwidth]{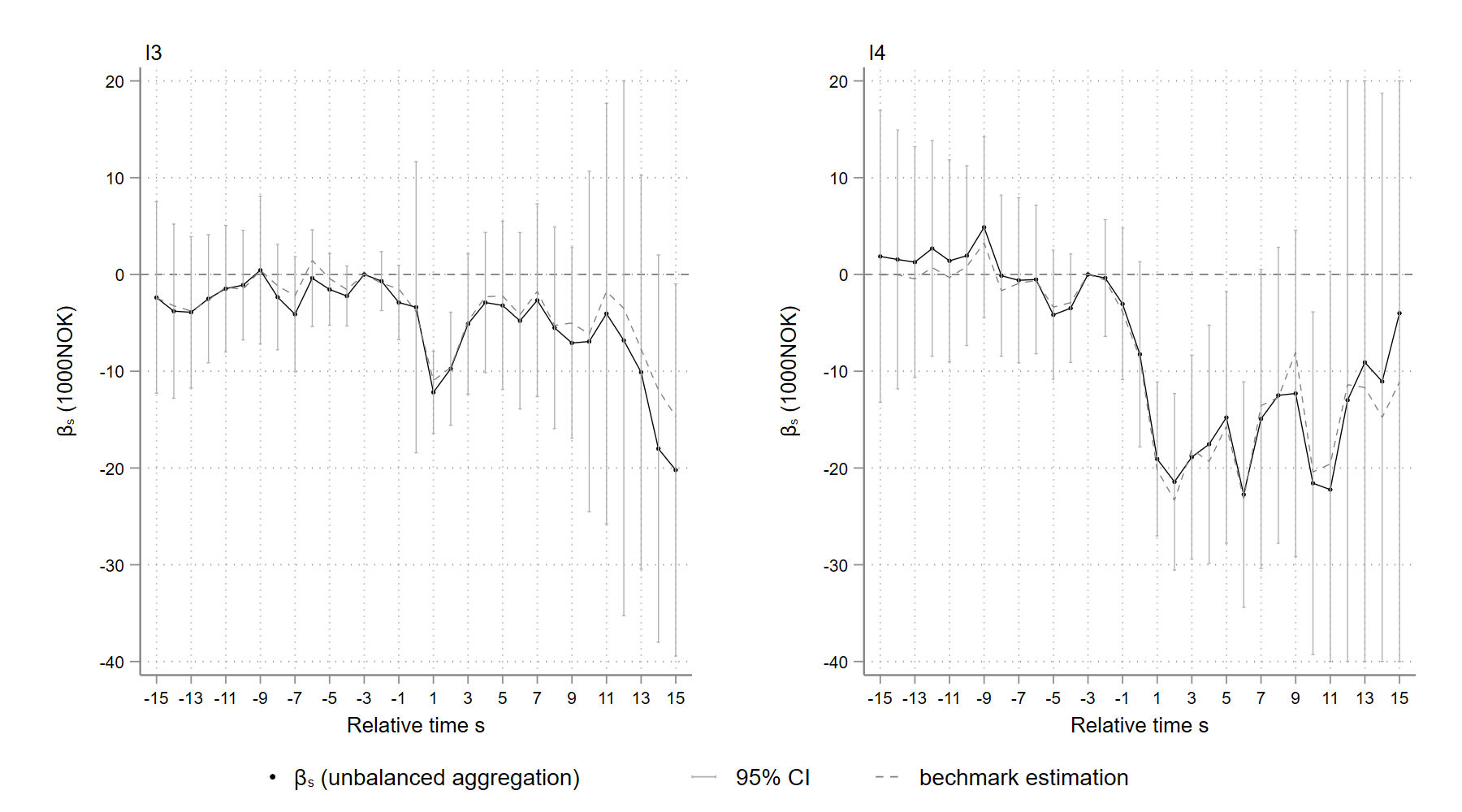}}{\scriptsize Note: $I_3 \in (W^t,2W^t], I_4 \in (2W^t,\infty]$, $W^t$ is the mean annual wage in Norway in year $t$.}
    \caption{$ATT_s$ on annual occupational income: inheritances 1-year around parental death}
    \label{fig:combine2_basic_yrkinnt_p3_bq1}
\end{figure}

\clearpage

\subsubsection{Balanced aggregation}
\label{asubsubsec:balance}

\begin{figure}[htbp]
  \centering
  \copyrightbox[b]{\includegraphics[width=0.925\textwidth]{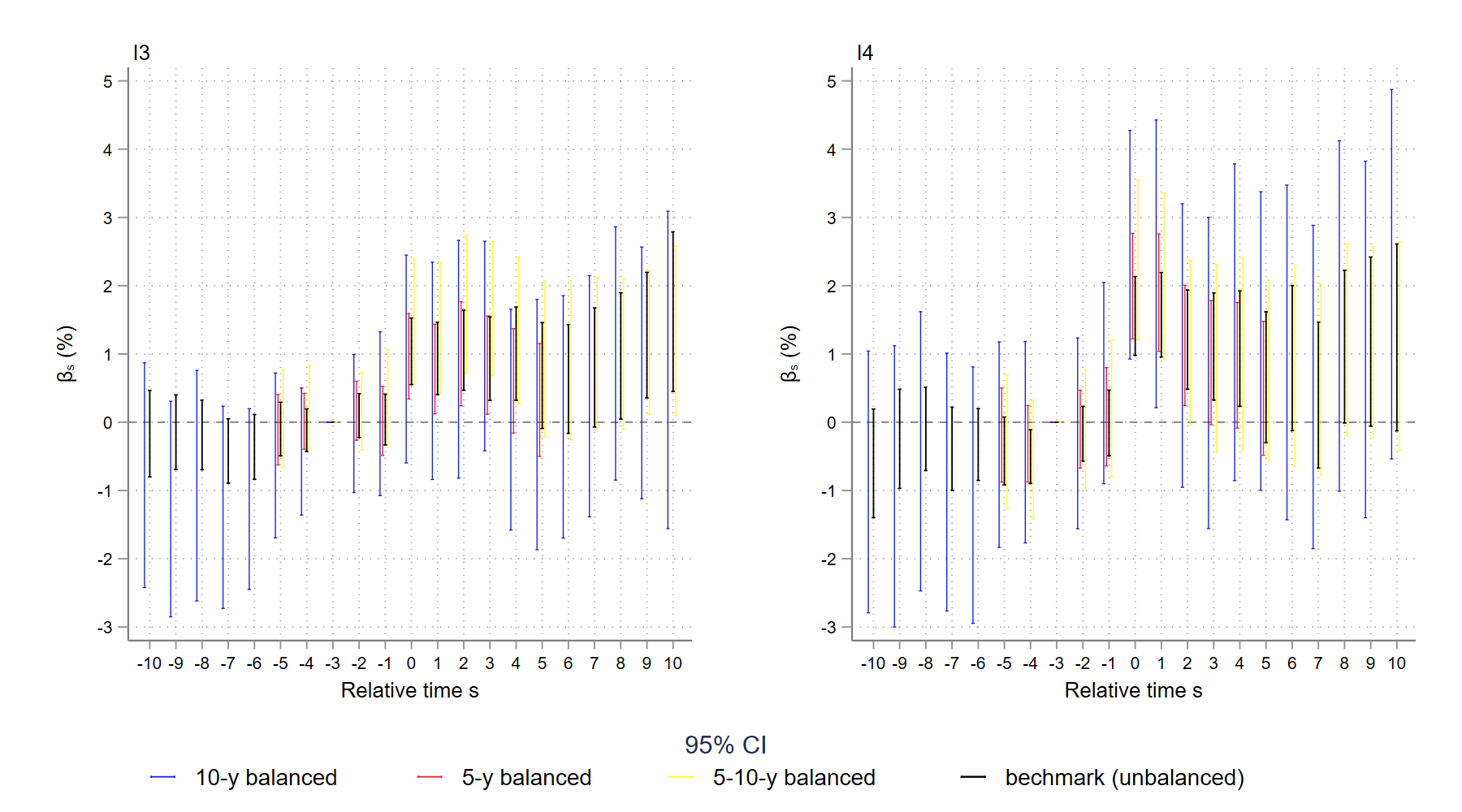}}{\scriptsize Note: $I_3 \in (W^t,2W^t]$, $I_4 \in (2W^t,\infty]$, $W^t$ is the mean annual wage in Norway in year $t$.}
  \caption{$\hat{ATT_s}$ on annual wage income with balanced aggregation schemes}
  \label{fig:combine2_balanced_etpn_p3_bq3}
\end{figure}
\begin{figure}[htbp]
  \centering
  \copyrightbox[b]{\includegraphics[width=0.925\textwidth]{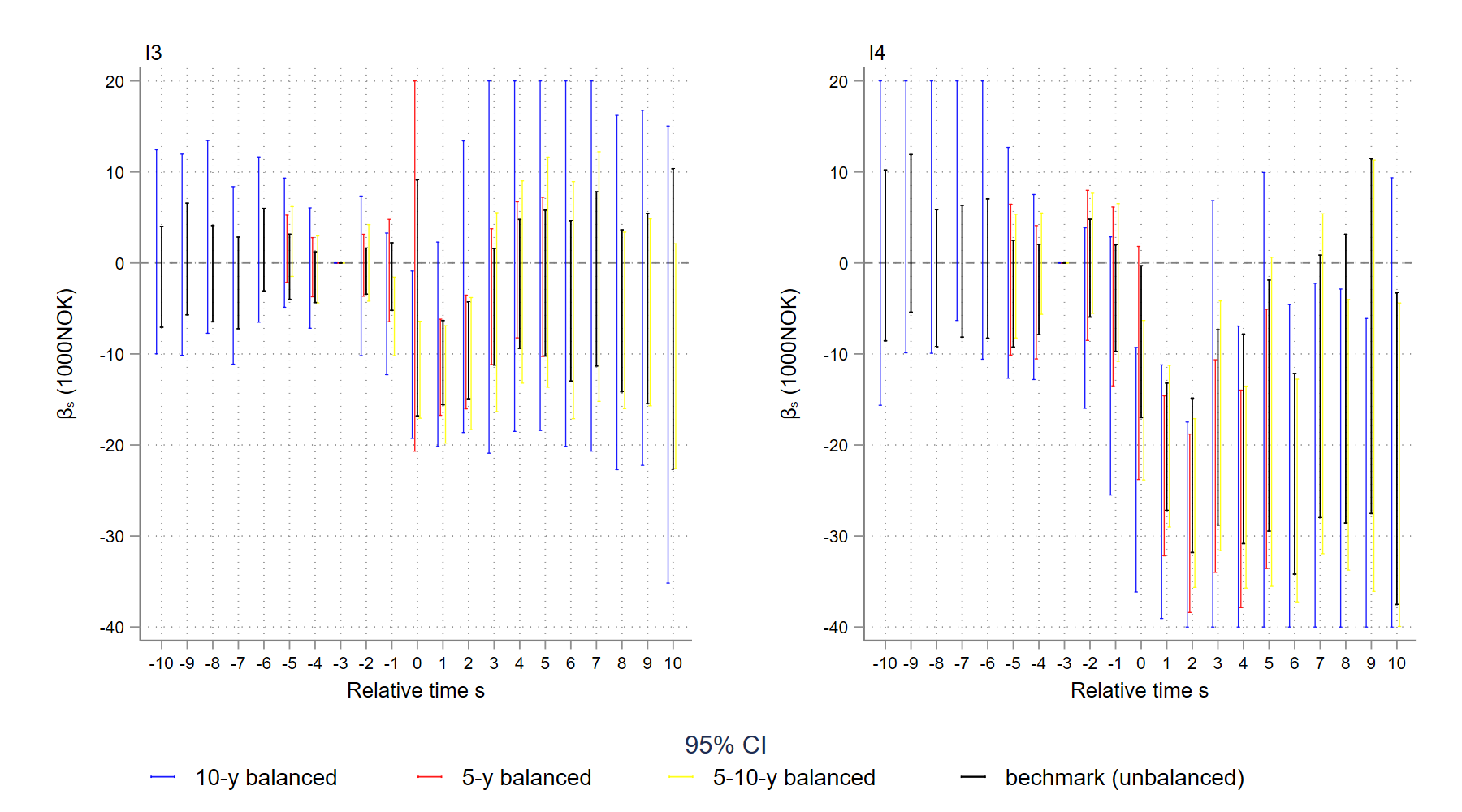}}{\scriptsize Note: $I_3 \in (W^t,2W^t]$, $I_4 \in (2W^t,\infty]$, $W^t$ is the mean annual wage in Norway in year $t$.}
  \caption{$\hat{ATT_s}$ on annual occupational income with balanced aggregation schemes}
  \label{fig:combine2_balanced_yrkinnt_p3_bq3}
\end{figure}

\clearpage

\subsubsection{Birth year constraint}
\label{asubsubsec:birth}


\begin{table}[htbp]
\centering
\def\sym#1{\ifmmode^{#1}\else\(^{#1}\)\fi}
\caption{Description of samples with different birth year constraints}
\label{tab:des_boe}
\vspace{-5pt}

\begin{threeparttable}
\begin{tabular}{l*{11}{c}}
\toprule
            &\multicolumn{2}{c}{Baseline}&&\multicolumn{2}{c}{Boe (unbalanced)}&&\multicolumn{2}{c}{Boe (balanced)}&&\multicolumn{2}{c}{Relaxed}  \\
birth year&\multicolumn{2}{c}{$[1951,1975]$}&&\multicolumn{2}{c}{$[1944,1976]$}&&\multicolumn{2}{c}{$[1944,1976]$}&&\multicolumn{2}{c}{$[1948,1978]$}  \\
cohorts&\multicolumn{2}{c}{$[1996,2014]$}&&\multicolumn{2}{c}{$[1997,2010]$}&&\multicolumn{2}{c}{$[2000,2004]$}&&\multicolumn{2}{c}{$[1999,2014]$}  \\

       \cmidrule{2-3}    \cmidrule{5-6}    \cmidrule{8-9}  \cmidrule{11-12} 

inheritances size & I3   & I4   &&  I3   & I4   &&    I3   & I4   &&    I3   & I4  \\        

\midrule

average cohort &       2008&    2009 && 2007  &  2008    &&    2002   & 2002  &&   2,007   &    2,008  \\        
inheritance amount&     548&    1,691&& 498   &  1,575   &&    462    & 1,251 &&   518     &     1,632 \\       
wage income&            416&      470&& 387   &    432   &&    349    & 378   &&   396     &    443    \\    
occupational income&    447&      513&& 416   &    471   &&    379    & 414   &&   425     &    483    \\   
Pr(self-employment)&   0.13&    0.14 && 0.13  &   0.14   &&    0.15   & 0.18  &&   0.13    &   0.14    \\  
gross wealth&         1,129&    2,369&& 1,213 &  2,408   &&    872    & 1,592 &&   1,211   &   2,423   \\  
-gross debt &           687&   829   && 673   &  799     &&    534    &  634  &&   706     &   833    \\  
-debt rate  &          0.61&  0.35   && 0.55  & 0.33     &&    0.61   & 0.40  &&   0.58    & 0.34   \\  
age         &            48&       49&& 50    &  51      &&    47     & 46    &&   49      &   49      \\  
male        &          0.47&    0.49 && 0.50  &  0.51    &&    0.51   & 0.52  &&   0.49    &   0.51    \\  
education level  &     4.28&    4.69 && 4.2   &  4.7     &&    4.2    & 4.6   &&   4.3     &  4.7      \\

 \midrule
\multicolumn{1}{l}{Number of individuals} & 34,177 &23,924  && 9,293 & 6,515 &&45,424 &40,517 &&40,947 & 35,502     \\
\bottomrule
\end{tabular}
\begin{tablenotes}
      \small
      \item Note: (1) Sample ``Boe (balanced)'' are individuals born between $[1944,1976]$ and inheriting between $[2000,2004]$, which is the same as the literature \citep{Boe2019}. (2) $I_3 \in (W^t,2W^t]$, $I_4 \in (2W^t,\infty]$, $W^t$ is the mean annual wage in Norway in year $t$. (3) Income and wealth are in 1,000 Norwegian kroner (NOK) at 2015 price. (4) ``occupational income'' = wage income + business income; entrepreneurship=1 if business income $\ne$ 0; ``gross wealth'' includes debt; ``education level'' is an ordered 0-8 categorical variable as defined by SSB: \href{https://www.ssb.no/klass/klassifikasjoner/36/}{https://www.ssb.no/klass/klassifikasjoner/36/}, e.g. education=4 for upper secondary education, education=5 for Post-secondary non-tertiary education; ``number of kids'' is the number of underage children; parental death indicators equal to 1 if the corresponding parents are dead.
    \end{tablenotes}
\end{threeparttable}
\end{table}


\begin{figure}[htbp]
  \centering
  \copyrightbox[b]{\includegraphics[width=1\textwidth]{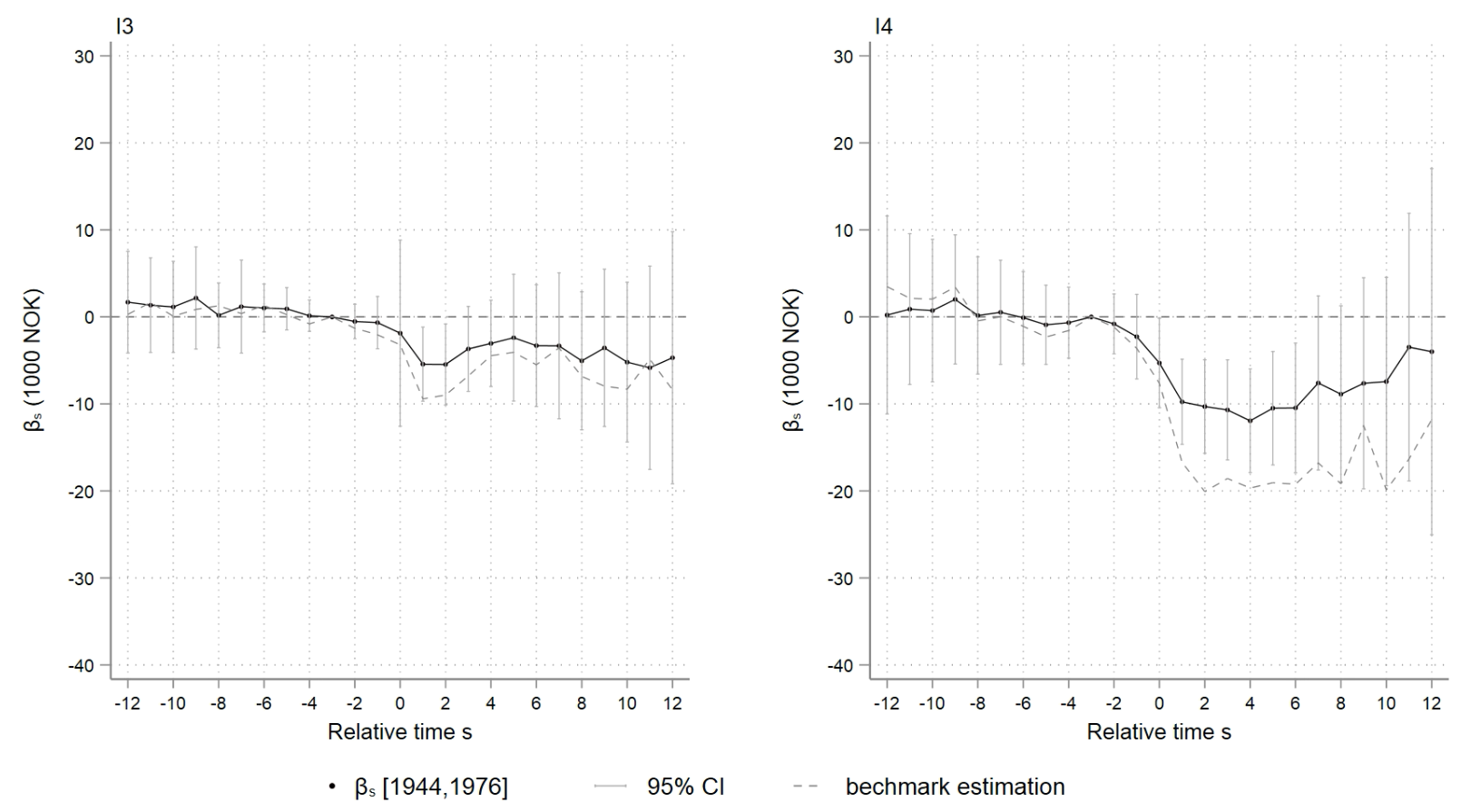}}{\scriptsize Note: $I_3 \in (W^t,2W^t]$, $I_4 \in (2W^t,\infty]$, $W^t$ is the mean annual wage in Norway in year $t$.}
  \caption{$\hat{ATT_s}$ on annual wage income: born in $[1944,1976]$}
  \label{fig:combine2_basic_wage_p3_44_76_bq3}
\end{figure}
\begin{figure}[htbp]
  \centering
  \copyrightbox[b]{\includegraphics[width=1\textwidth]{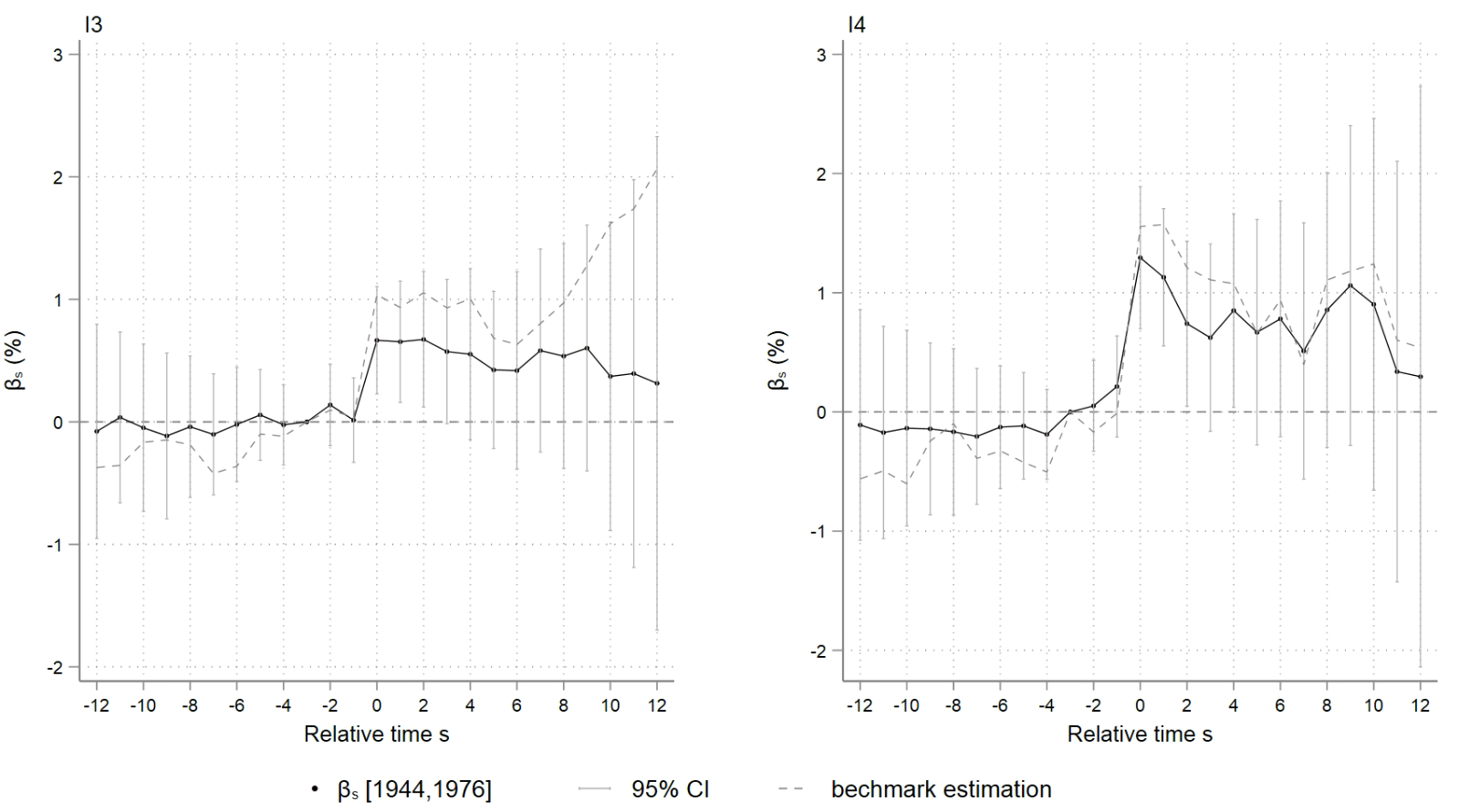}}{\scriptsize Note: $I_3 \in (W^t,2W^t]$, $I_4 \in (2W^t,\infty]$, $W^t$ is the mean annual wage in Norway in year $t$.}
  \caption{$\hat{ATT_s}$ on Pr(self-employment): born in $[1944,1976]$}
  \label{fig:combine2_basic_etpn_p3_44_76_bq3}
\end{figure}
\begin{figure}[htbp]
  \centering
  \copyrightbox[b]{\includegraphics[width=1\textwidth]{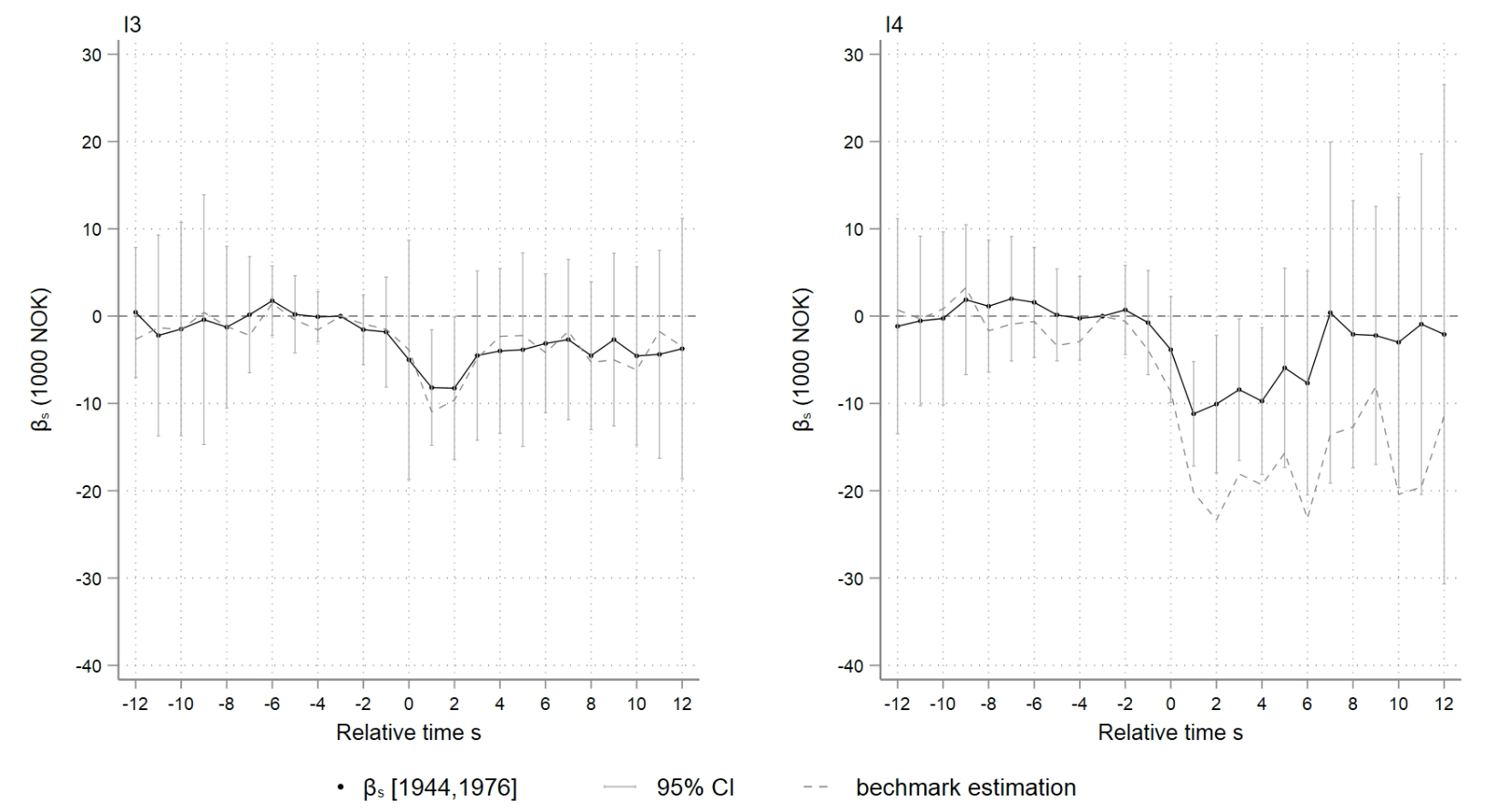}}{\scriptsize Note: $I_3 \in (W^t,2W^t]$, $I_4 \in (2W^t,\infty]$, $W^t$ is the mean annual wage in Norway in year $t$.}
  \caption{$\hat{ATT_s}$ on annual occupational income: born in $[1944,1976]$}
  \label{fig:combine2_basic_yrkinnt_p3_44_76_bq3}
\end{figure}
\begin{figure}[htbp]
  \centering
  \copyrightbox[b]{\includegraphics[width=1\textwidth]{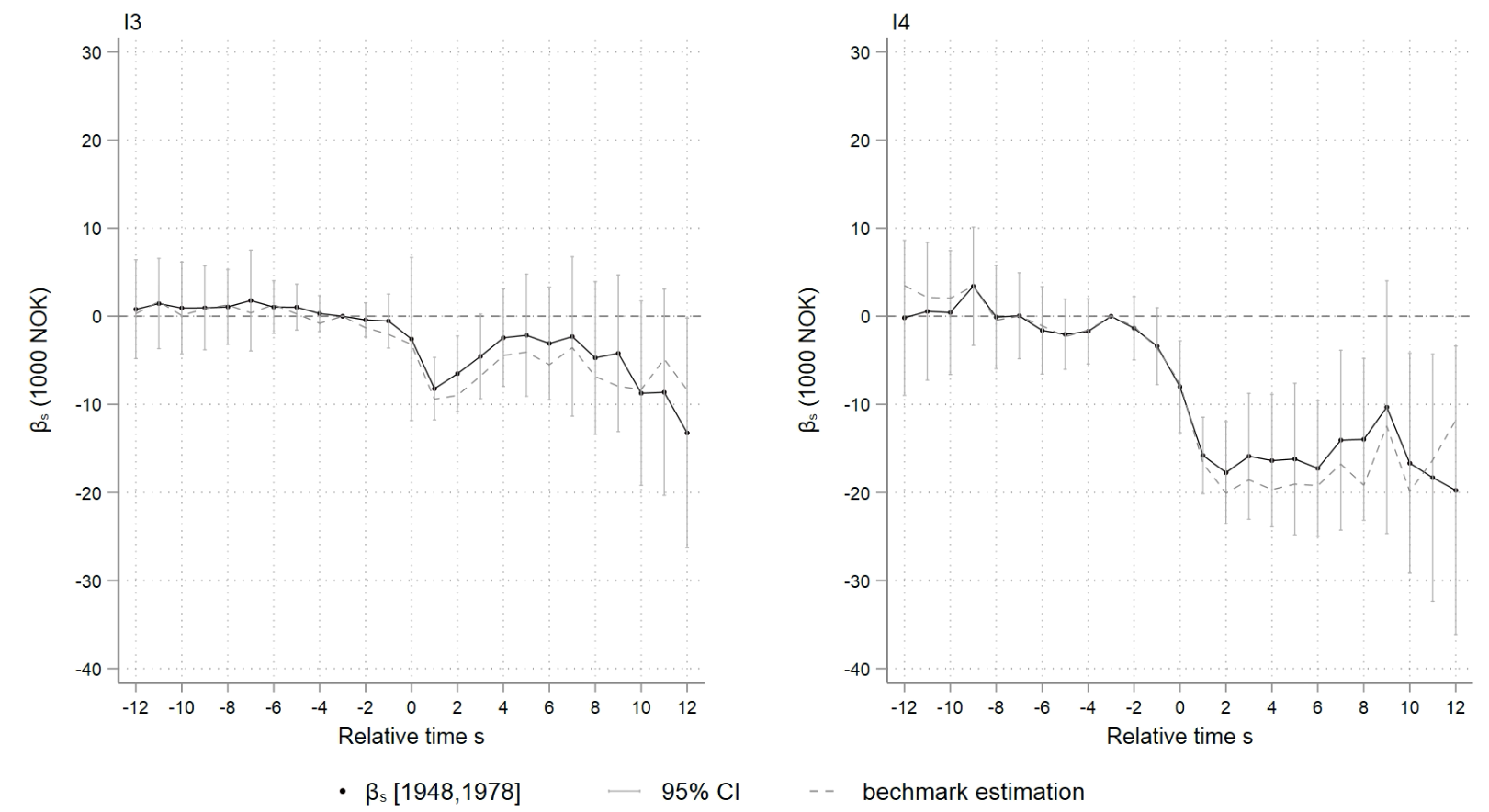}}{\scriptsize Note: $I_3 \in (W^t,2W^t]$, $I_4 \in (2W^t,\infty]$, $W^t$ is the mean annual wage in Norway in year $t$.}
  \caption{$\hat{ATT_s}$ on annual wage income: born in $[1948,1978]$}
  \label{fig:combine2_basic_wage_p3_48_78_bq3}
\end{figure}
\begin{figure}[htbp]
  \centering
  \copyrightbox[b]{\includegraphics[width=1\textwidth]{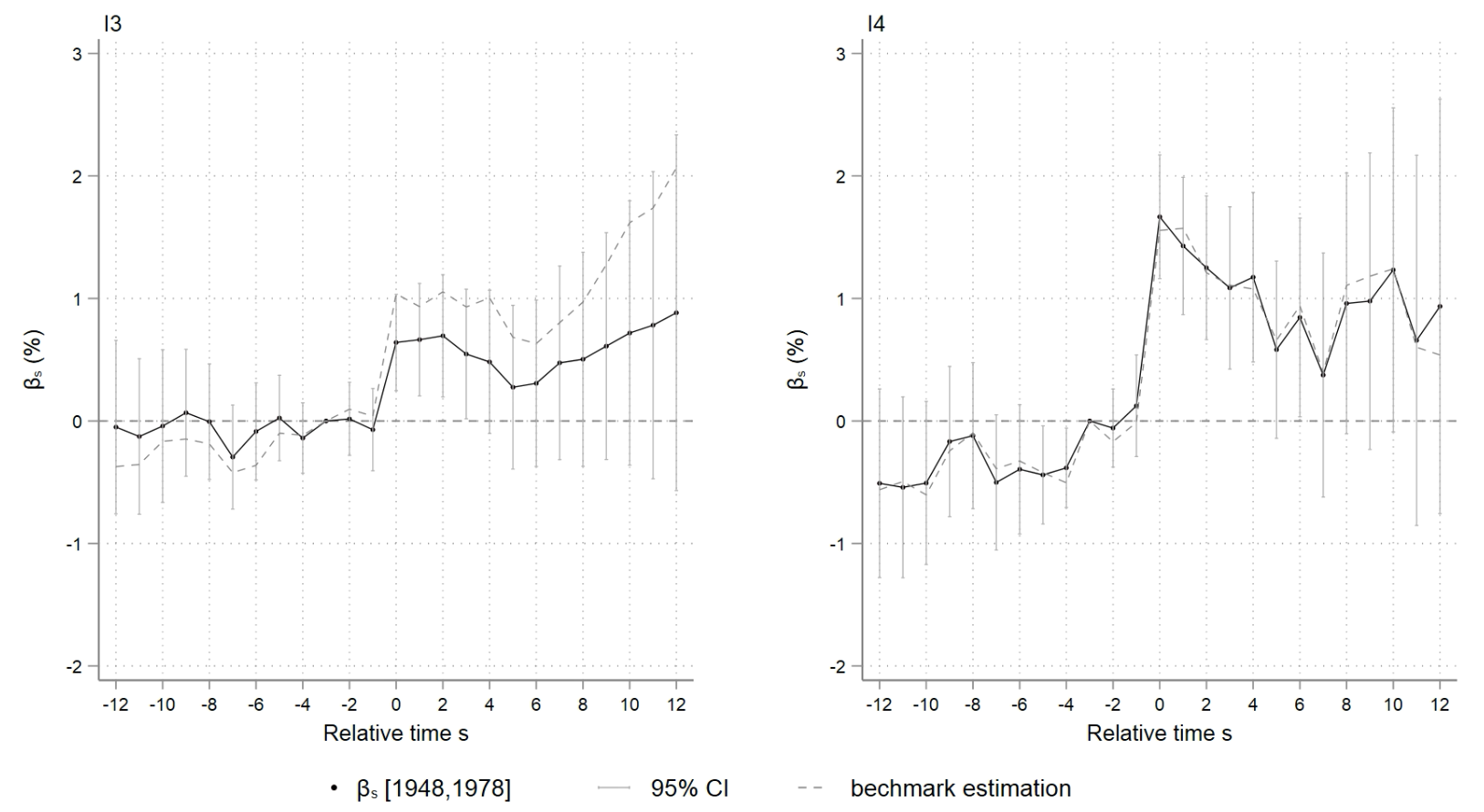}}{\scriptsize Note: $I_3 \in (W^t,2W^t]$, $I_4 \in (2W^t,\infty]$, $W^t$ is the mean annual wage in Norway in year $t$.}
  \caption{$\hat{ATT_s}$ on Pr(self-employment): born in $[1948,1978]$}
  \label{fig:combine2_basic_etpn_p3_48_78_bq3}
\end{figure}
\begin{figure}[htbp]
  \centering
  \copyrightbox[b]{\includegraphics[width=1\textwidth]{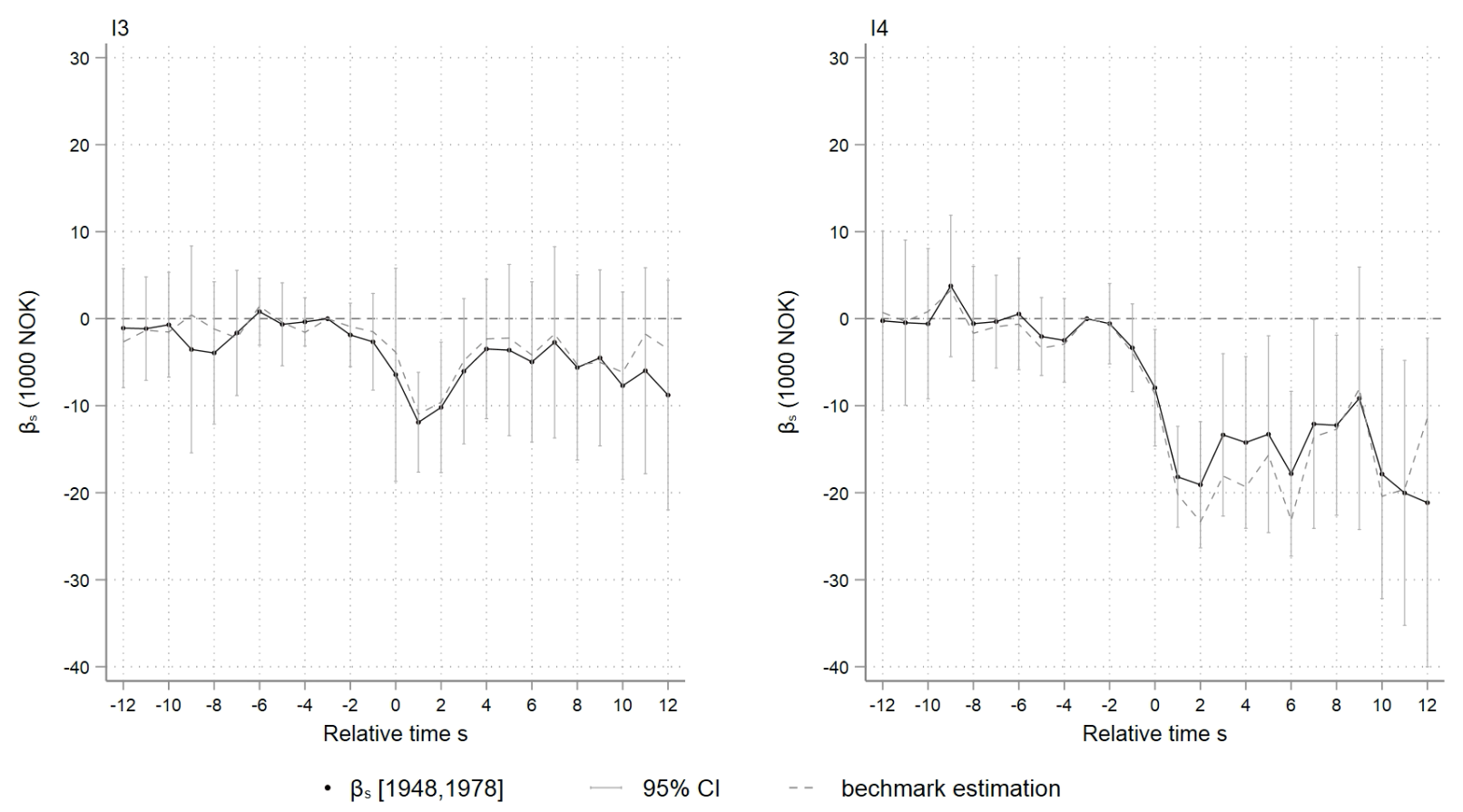}}{\scriptsize Note: $I_3 \in (W^t,2W^t]$, $I_4 \in (2W^t,\infty]$, $W^t$ is the mean annual wage in Norway in year $t$.}
  \caption{$\hat{ATT_s}$ on annual occupational income: born in $[1948,1978]$}
  \label{fig:combine2_basic_yrkinnt_p3_48_78_bq3}
\end{figure}
\begin{figure}[htbp]
  \centering
  \copyrightbox[b]{ \includegraphics[width=1\textwidth]{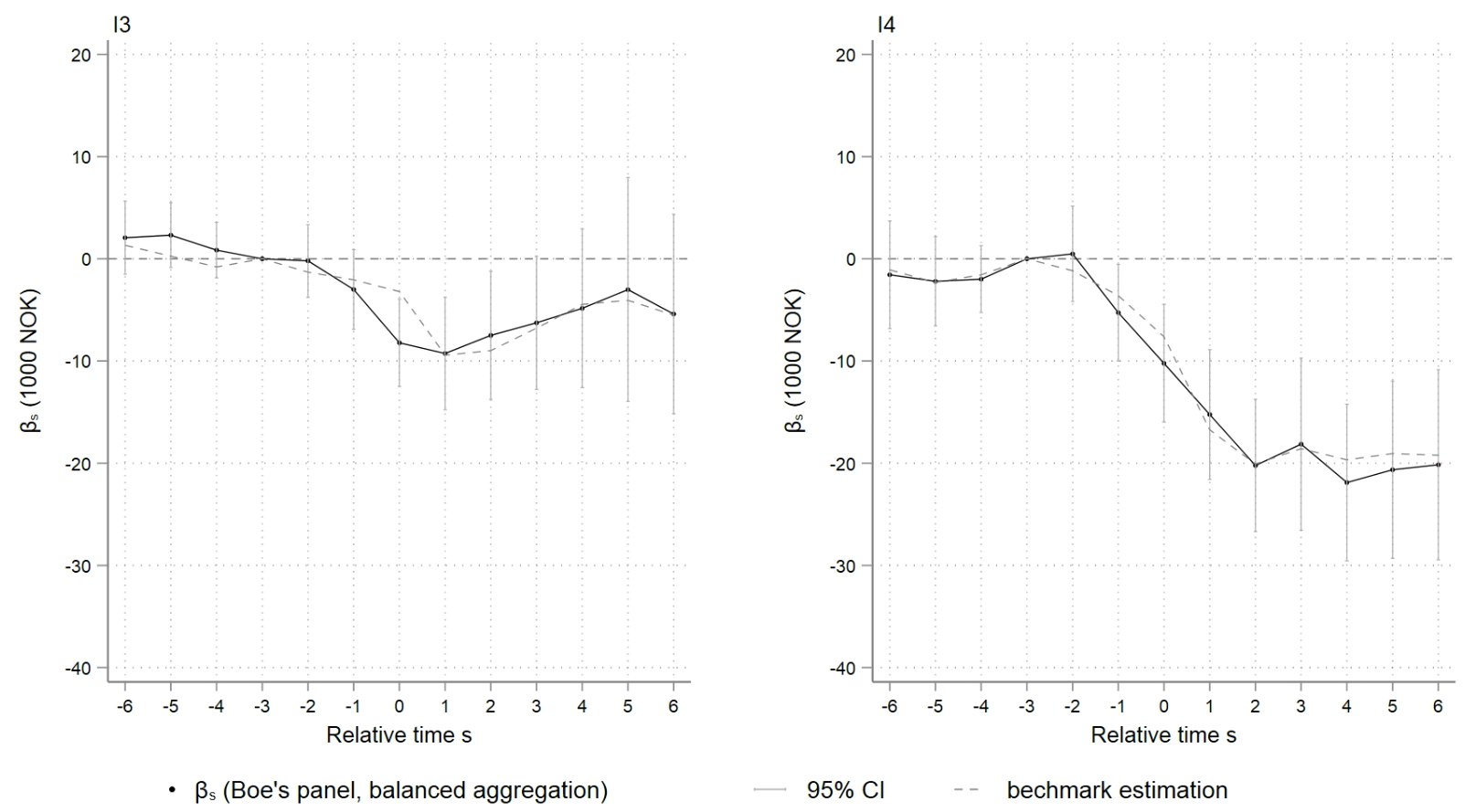}}{\scriptsize Note: $I_3 \in (W^t,2W^t]$, $I_4 \in (2W^t,\infty]$, $W^t$ is the mean annual wage in Norway in year $t$.}
  \caption{$\hat{ATT_s}$ on annual wage income: Boe's cohorts and balanced horizon}
  \label{fig:combine2_boe_wage_p3_44_76_bq3}
\end{figure}
\begin{figure}[htbp]
  \centering
  \copyrightbox[b]{\includegraphics[width=1\textwidth]{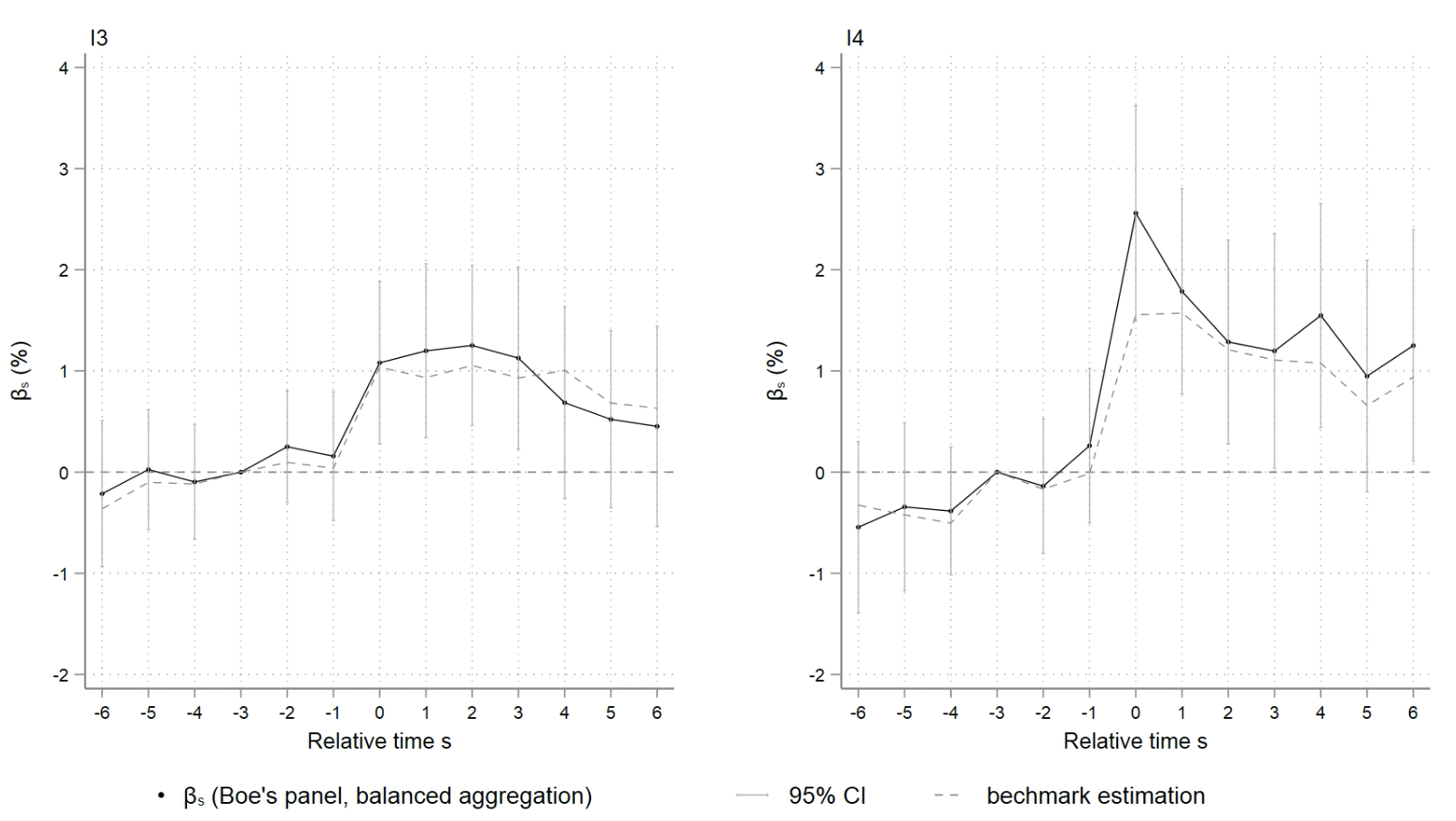}}{\scriptsize Note: $I_3 \in (W^t,2W^t]$, $I_4 \in (2W^t,\infty]$, $W^t$ is the mean annual wage in Norway in year $t$.}
  \caption{$\hat{ATT_s}$ on Pr(self-employment): Boe’s cohorts and balanced horizon}
  \label{fig:combine2_boe_etpn_p3_44_76_bq3}
\end{figure}
\begin{figure}[htbp]
  \centering
  \copyrightbox[b]{\includegraphics[width=1\textwidth]{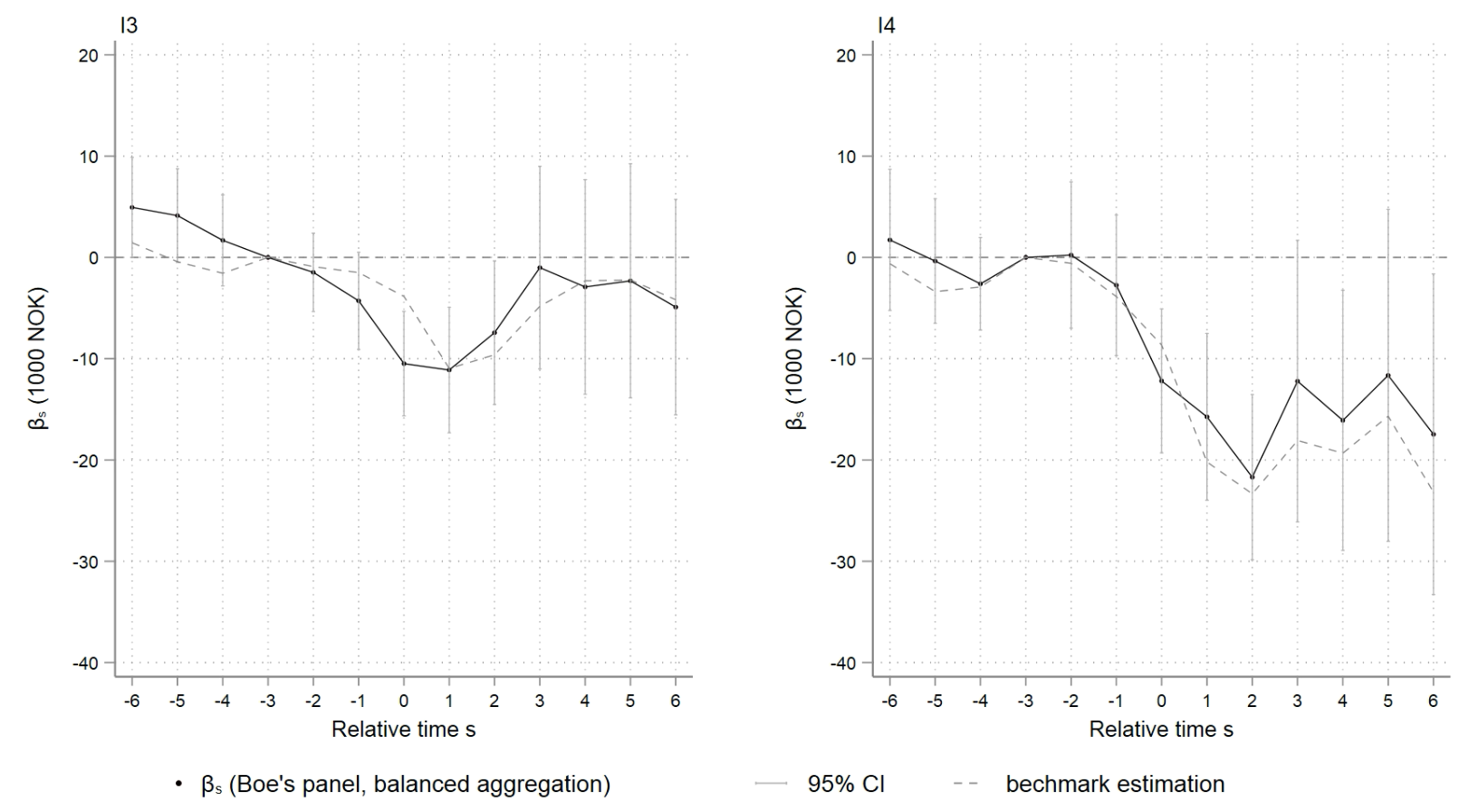}}{\scriptsize Note: $I_3 \in (W^t,2W^t]$, $I_4 \in (2W^t,\infty]$, $W^t$ is the mean annual wage in Norway in year $t$.}
  \caption{$\hat{ATT_s}$ on annual occupational income: Boe’s cohorts and balanced horizon}
  \label{fig:combine2_boe_yrkinnt_p3_44_76_bq3}
\end{figure}

\clearpage

\subsubsection{Nearest-n-cohorts as control}
\label{asubsubsec:nn}

\begin{figure}[htbp]
  \centering
  \copyrightbox[b]{ \includegraphics[width=0.925\textwidth]{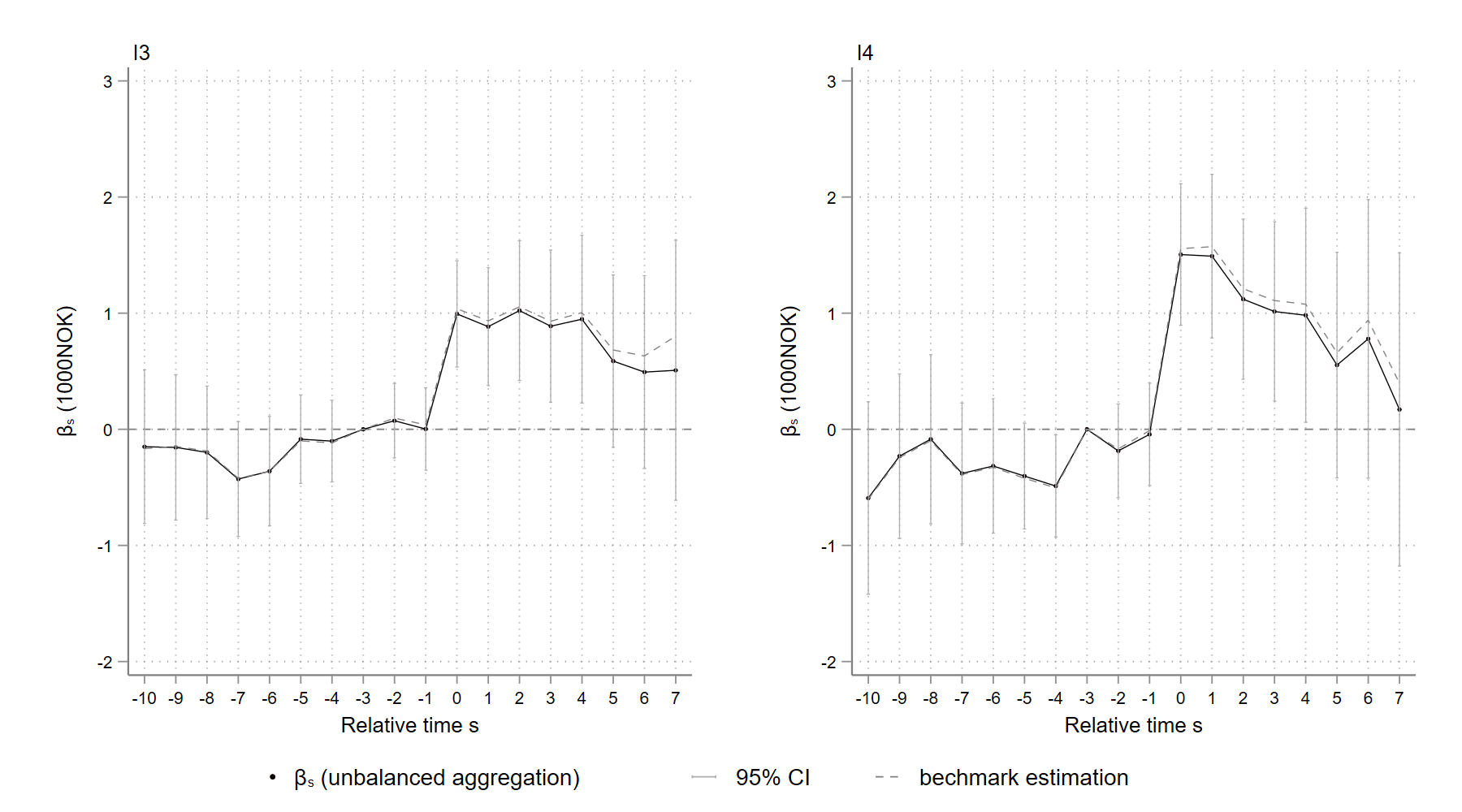}}{\scriptsize Note: $I_3 \in (W^t,2W^t]$, $I_4 \in (2W^t,\infty]$, $W^t$ is the mean annual wage in Norway in year $t$.}
  \caption{$\hat{ATT_s}$ on Pr(self-employment): nearest-10 not-yet-treated cohorts as control}
  \label{fig:combine2_robust_etpn_n10_p3_bq3}
\end{figure}
\begin{figure}[htbp]
  \centering
  \copyrightbox[b]{ \includegraphics[width=0.925\textwidth]{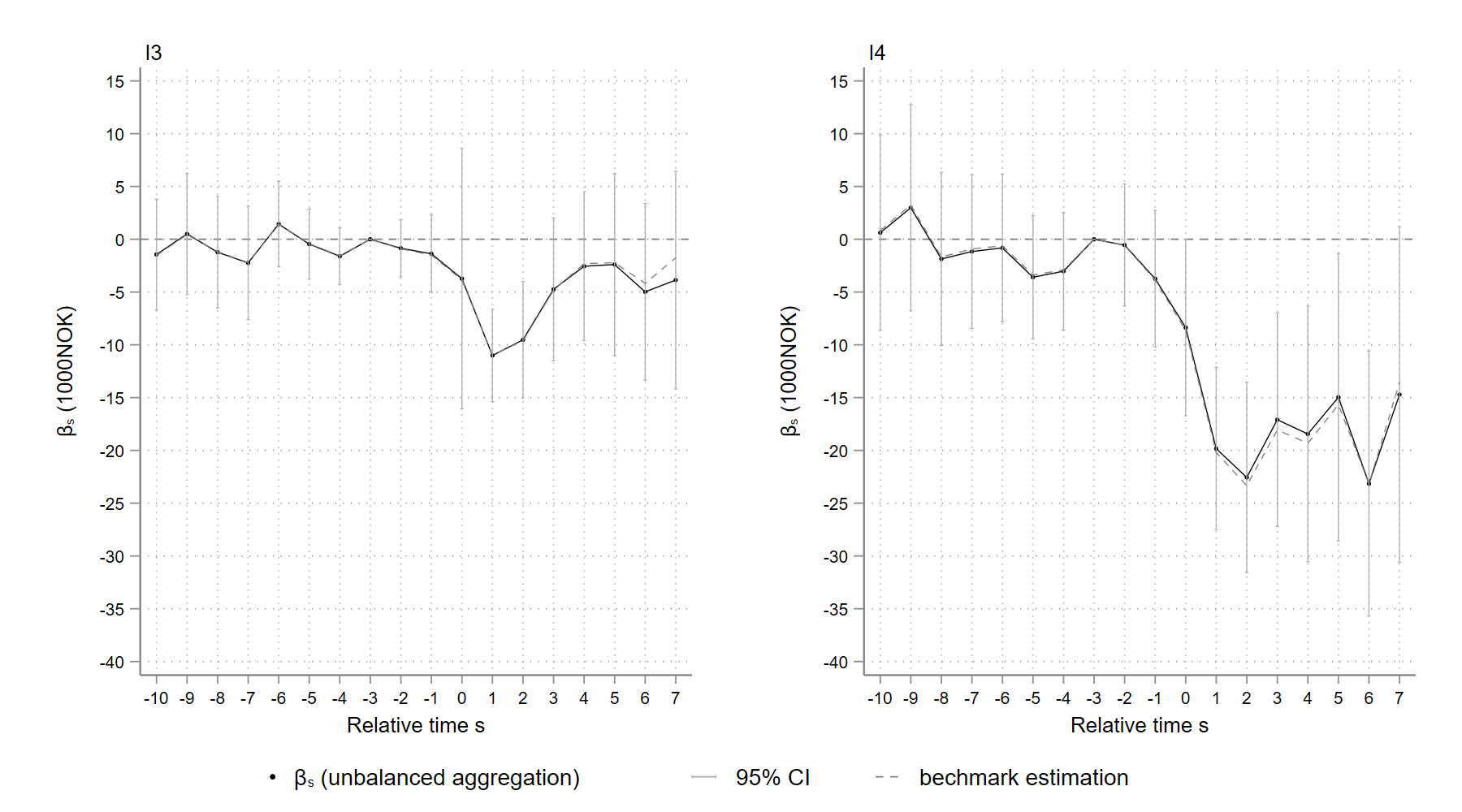}}{\scriptsize Note: $I_3 \in (W^t,2W^t]$, $I_4 \in (2W^t,\infty]$, $W^t$ is the mean annual wage in Norway in year $t$.}
  \caption{$\hat{ATT_s}$ on annual occupational income: nearest-10 not-yet-treated cohorts as control}
  \label{fig:combine2_robust_yrkinnt_n10_p3_bq3}
\end{figure}

\clearpage

\subsubsection{Limited anticipation}
\label{asubsubsec:anti}

\begin{figure}[htbp]
  \centering
  \copyrightbox[b]{ \includegraphics[width=0.925\textwidth]{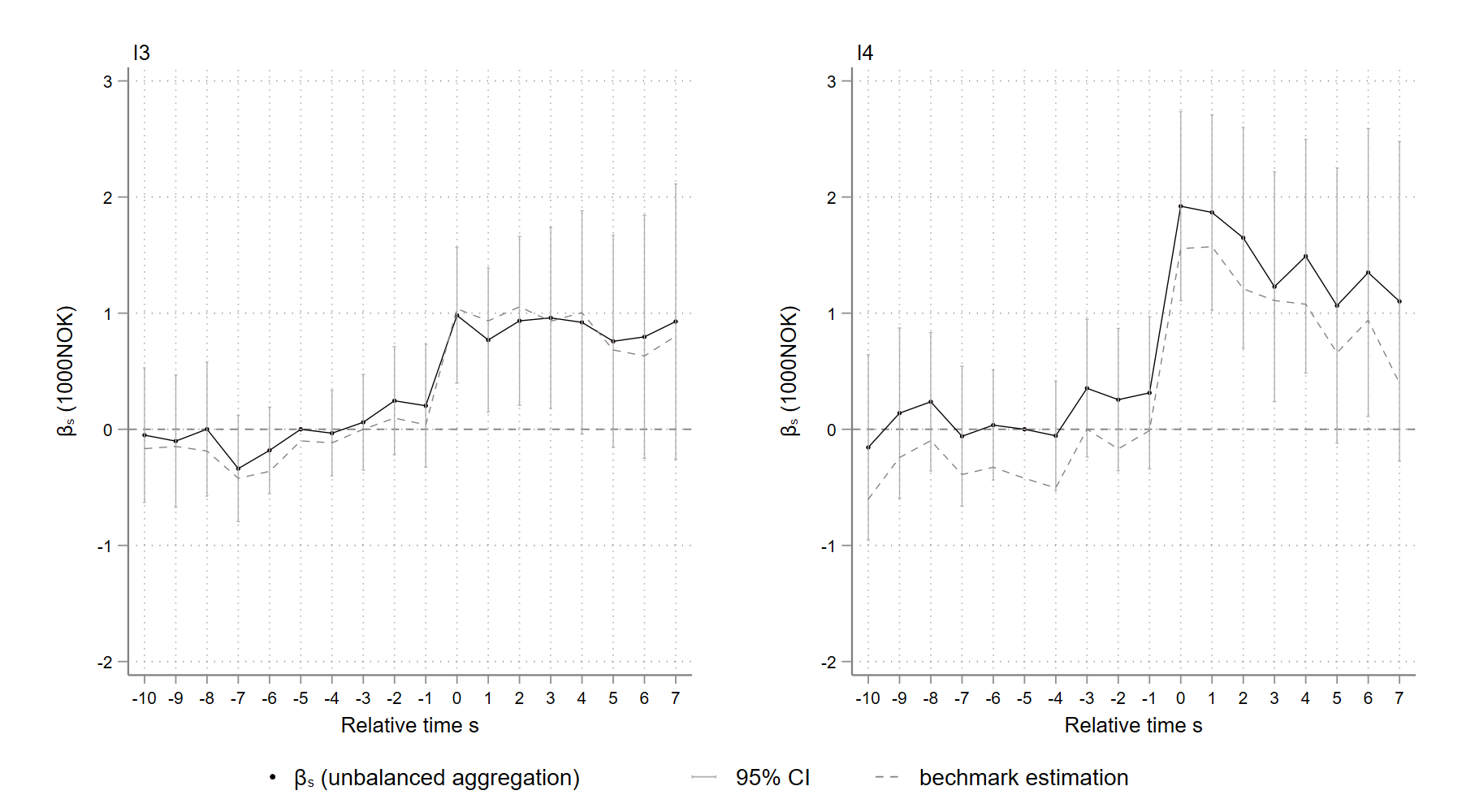}}{\scriptsize Note: $I_3 \in (W^t,2W^t]$, $I_4 \in (2W^t,\infty]$, $W^t$ is the mean annual wage in Norway in year $t$.}
  \caption{$\hat{ATT_s}$ on Pr(self-employment) with $s=-5$ as reference}
  \label{fig:combine2_robust_etpn_p5_bq3}
\end{figure}
\begin{figure}[htbp]
  \centering
  \copyrightbox[b]{ \includegraphics[width=0.925\textwidth]{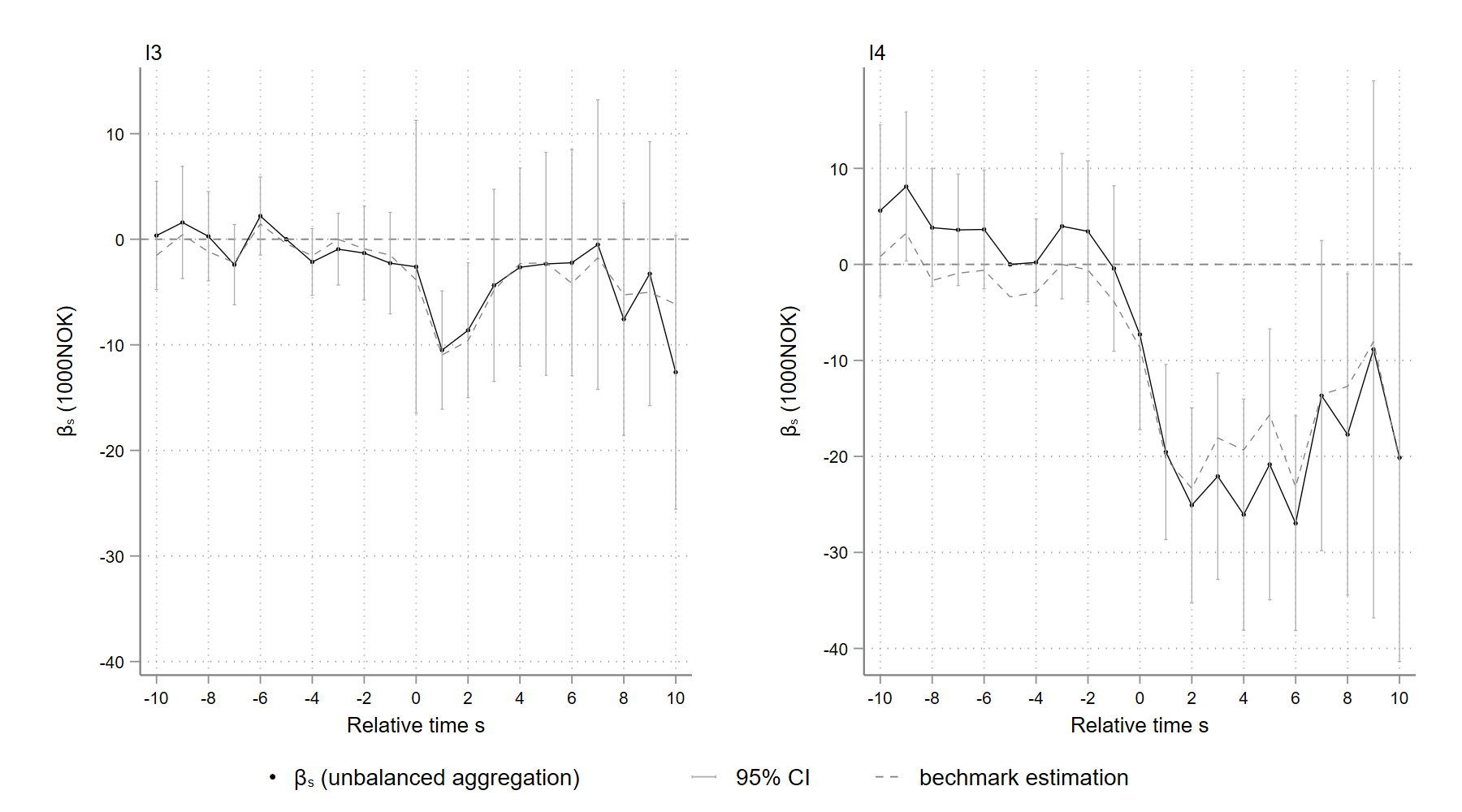}}{\scriptsize Note: $I_3 \in (W^t,2W^t]$, $I_4 \in (2W^t,\infty]$, $W^t$ is the mean annual wage in Norway in year $t$.}
  \caption{$\hat{ATT_s}$ on annual occupational income with $s=-5$ as reference}
  \label{fig:combine2_robust_yrkinnt_p5_bq3}
\end{figure}

\clearpage

\subsubsection{Multiple small gifts}
\label{asubsubsec:gift}

\begin{figure}[htbp]
  \centering
  \copyrightbox[b]{ \includegraphics[width=0.925\textwidth]{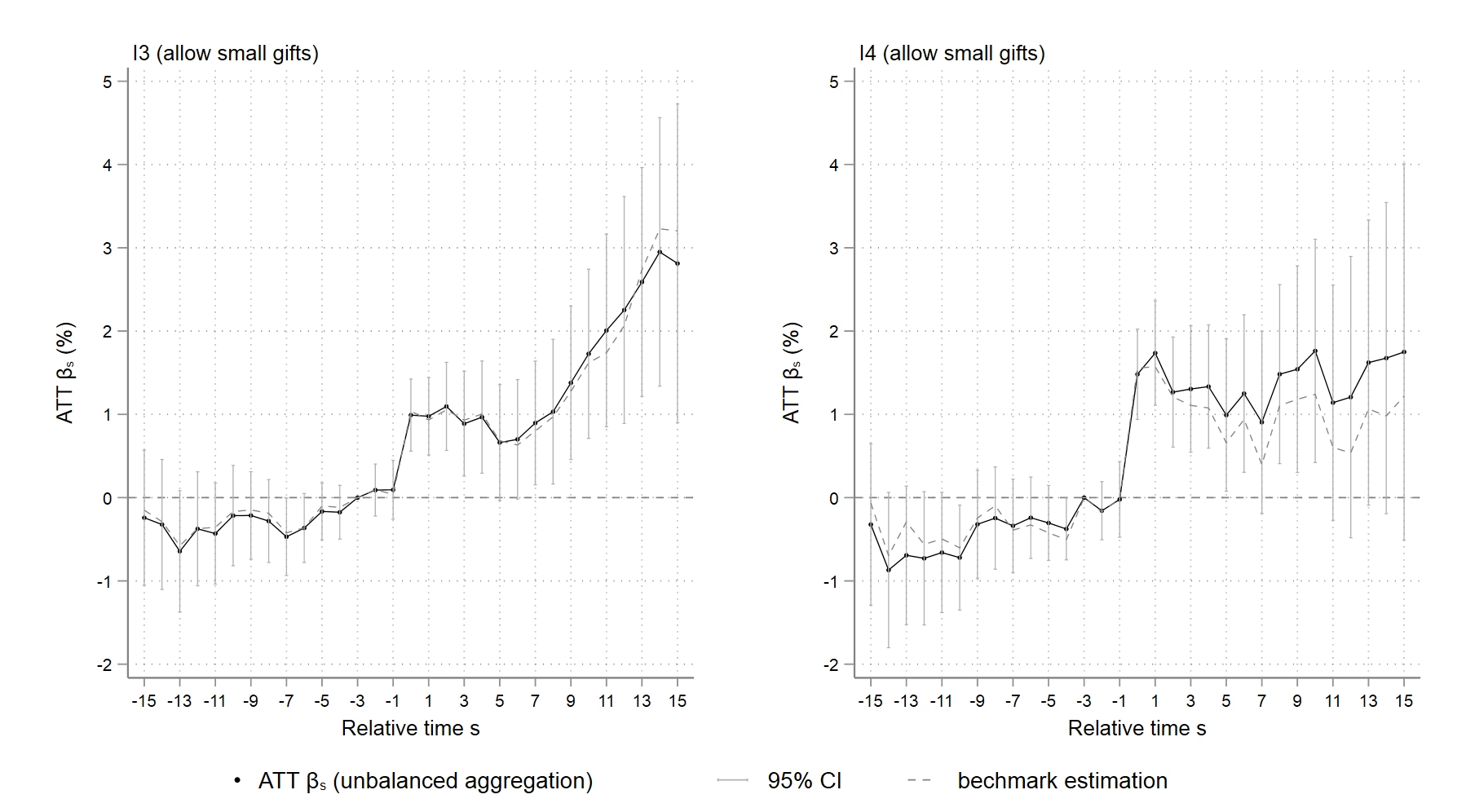}}{\scriptsize Note: $I_3 \in (W^t,2W^t]$, $I_4 \in (2W^t,\infty]$, $W^t$ is the mean annual wage in Norway. The notation $ig$ means small gifts are allowed in the sample.}
  \caption{$\hat{ATT_s}$ on Pr(self-employment) allowing small gifts}
  \label{fig:combine2_ignore_etpn_p3_bq3}
\end{figure}
\begin{figure}[htbp]
  \centering
  \copyrightbox[b]{ \includegraphics[width=0.925\textwidth]{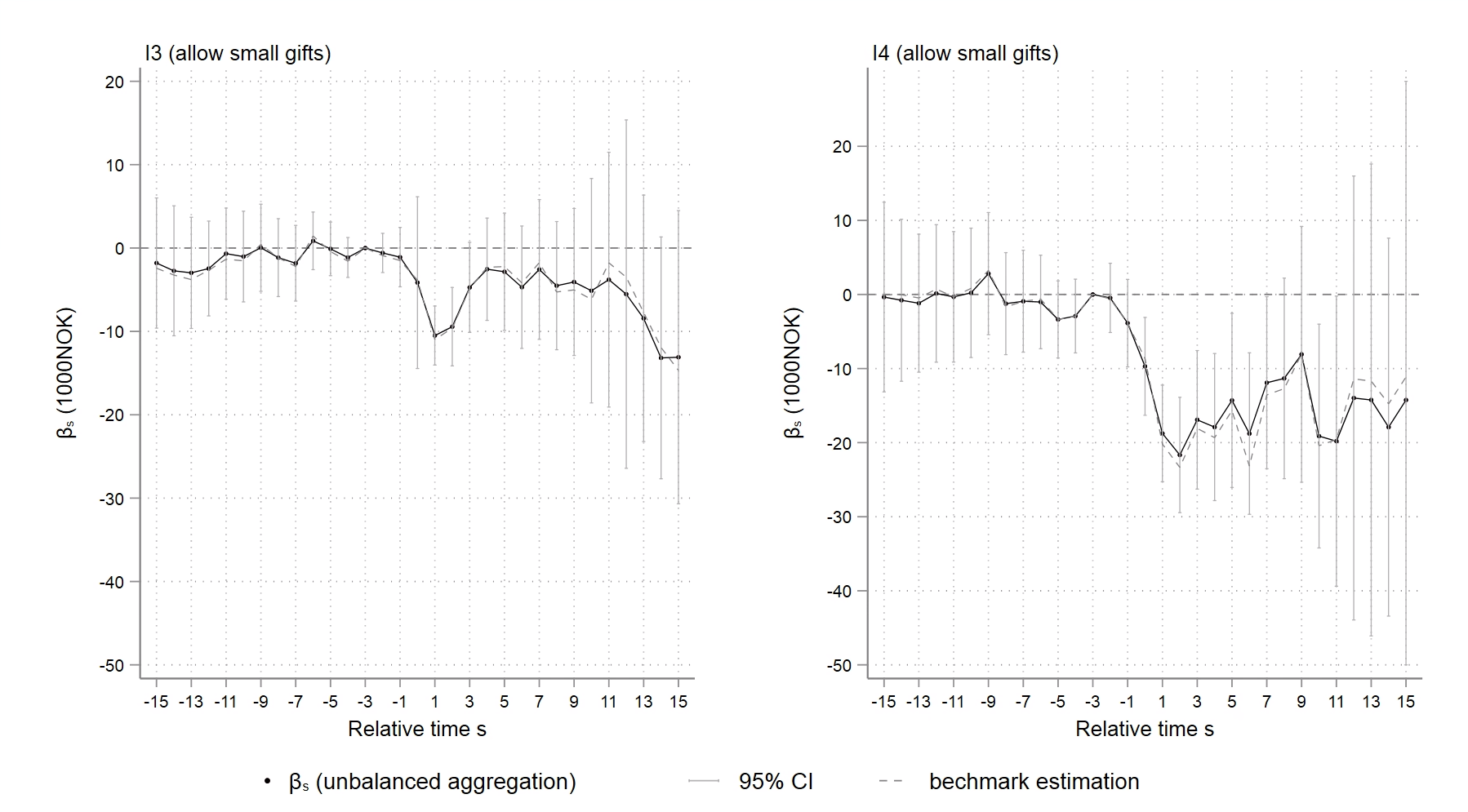}}{\scriptsize Note: $I_3 \in (W^t,2W^t]$, $I_4 \in (2W^t,\infty]$, $W^t$ is the mean annual wage in Norway. The notation $ig$ means small gifts are allowed in the sample.}
  \caption{$\hat{ATT_s}$ on annual occupational income allowing small gifts}
  \label{fig:combine2_ignore_yrkinnt_p3_bq3}
\end{figure}

\end{document}